\documentclass[numberedappendix]{emulateapj}
\usepackage{amsmath}
\usepackage{float}
\usepackage{units}
\usepackage[titletoc,title]{appendix}
\usepackage{mathtools}
\usepackage[dvipsnames]{xcolor}
\usepackage[normalem]{ulem}
\newcommand{\braket}[1] {\left< #1 \right>}
\newcommand{\dvg}{\delta v_{gk}}
\newcommand{\dvp}{\delta v_{pk}}
\newcommand{\dvrel}{\delta v_{pg}}
\newcommand{\avvg}{\boldsymbol{\bar{v}}_{gk}}
\newcommand{\avvp}{\boldsymbol{\bar{v}}_{pk}}
\newcommand{\avvrel}{\boldsymbol{\bar{v}}_{pg}}
\newcommand{\vg} {\boldsymbol{v}_{gk}}
\newcommand{\vp} {\boldsymbol{v}_{pk}}
\newcommand{\vrel} {\boldsymbol{v}_{pg}}
\graphicspath{{figures/}{../../Thesis_Code/gas_assist_plots}{revised_figures/}}
\usepackage[breaklinks,colorlinks,urlcolor=blue,citecolor=blue,linkcolor=blue]{hyperref}
\definecolor{mygreen}{HTML}{139703}

\usepackage[pdf]{pstricks}
\usepackage{color}

\begin{document}

	\title{Gas-Assisted Growth of Protoplanets in a Turbulent Medium} 
    \author{M. M. Rosenthal}
    \author{R. A. Murray-Clay}
    \affil{Department of Astronomy and Astrophysics, University of California, Santa Cruz, CA 95064, USA}
    \author{H. B. Perets}
    \affil{Physics Department, Technion - Israel Institute of Technology, Haifa, Israel 3200003}
    \author{N. Wolansky}
    \affil{Computer Science Department, Columbia University, 500 West 120 Street, Room 450, New York, New York, 10027, USA}

\begin{abstract}
Pebble accretion is a promising process for decreasing growth timescales of planetary cores, allowing gas giants to form at wide orbital separations. However, nebular turbulence can reduce the efficiency of this gas-assisted growth. We present an order of magnitude model of pebble accretion, which calculates the impact of turbulence on the average velocity of small bodies, the radius for binary capture, and the sizes of the small bodies that can be accreted. We also include the effect of turbulence on the particle scale height, which has been studied in previous works. We find that turbulence does not prevent rapid growth in the high-mass regime: the last doubling time to the critical mass to trigger runaway gas accretion ($M \sim 10 \, M_\oplus$) is well within the disk lifetime even for strong ($\alpha \gtrsim 10^{-2}$)  turbulence. We find that while the growth timescale is quite sensitive to the local properties of the protoplanetary disk, over large regimes of parameter space large cores grow in less than the disk lifetime if appropriately-sized small bodies are present. Instead, the effects of turbulence are most pronounced for low planetary masses. For strong turbulence the growth timescale is longer than the gas disk lifetime until the core reaches masses $\gtrsim 10^{-2}-10^{-1} M_\oplus$.  A ``Flow Isolation Mass," at which binary capture ceases, emerges naturally from our model framework. We comment that the dependence of this mass on orbital separation is similar to the semi-major axis distribution of solar-system cores.
\end{abstract}

\section{Introduction}

In the core accretion model of gas giant formation, the growth of a gas giant is constrained by two main factors: the growth timescale of the planet relative to the lifetime of the gas disk, and the amount of solid material available for a growing planet to accrete. Because early stages of formation were not well understood, many classic models of planet formation focus on later stages of growth, beginning with solid ``planetesimals" of size $\gtrsim \, \text{km}$. In these models, growth is too slow to produce gas giants at wide orbital separations. In contrast, close to the host star there is insufficient material locally available to produce a solid core massive enough to grow a gas giant. While these processes can produce architectures similar to the solar system, they are not sufficient to explain the diverse system architectures that are observed around other stars. In particular, recent theoretical work has pointed to the possibility that accretion of ``pebble" sized bodies may be important in both determining the growth timescale of cores and providing a reservoir of solid material through radial drift  (\citealt{OK10}, \citealt{pmc11}, \citealt{OK12}, \citealt{lj12}, \citealt{ljm_2014}, \citealt{lkd_2015}, \citealt{mljb15}, \citealt{vo_2016}, \citealt{igm16}, \citealt{xbmc_2017}). In this paper we will introduce an order of magnitude model of protoplanetary growth by pebble accretion, focusing on the regime in which the core is sufficiently massive that the gravity of the core is non-negligible. In what follows we will use the term ``protoplanet" to refer to cores in this regime.  We will focus on incorporating the effects of local disk turbulence into the various length and velocity scales that set the growth timescale.

Before proceeding, we briefly define a number of standard terms that will be used throughout this work. Within the context of a bottom up formation model, the growth of gas giant planets proceeds by ``core accretion" -- a gas giant core grows until it reaches a large enough mass, $M_{\rm{crit}}$, that its atmospheric mass is comparable to its core mass. At this point the core rapidly accretes gas from the nebula, culminating in a gas giant (see e.g. \citealt{pollack_gas_giants}). In this work we will not address the physics that set $M_{\rm{crit}}$ (see e.g. \citealt{raf06}, \citealt{pymc_2015}), and will instead consider growth timescales as a function of core mass. In the absence of some dissipative mechanism, the largest enhancement to the collision cross section comes from ``gravitational focusing" by the large cores. Gravitational focusing refers to the effect where large bodies can accrete material with impact parameters far outside their physical radius through the influence of gravity. This effect is significant for particles with velocity dispersions smaller than the large body escape velocity. In what follows we will refer to models of core accretion where gravitational focusing is the largest enhancement to the accretion cross section as ``canonical core accretion" or ``planetesimal accretion." 

Two particular challenges to planetesimal accretion stem from the existence of directly imaged planets at wide orbital separations and ``Super Earths" close in to their host stars. The star HR8799 has a system of four gas giants orbiting at $a \approx 15-70\,\text{AU}$ (\citealt{HR8799_orig}, \citealt{HR8799_fourth}). \textit{N}-body integrations show that it is unlikely that this system was formed by scattering  (\citealt{dodson-robinson_gas_giants}), indicating that these planets likely formed \textit{in situ}. Gravitational instability may be an alternative way to form the HR8799 planets, as reviewed by \cite{kl_2016}. However, \cite{kratter_gas_giants} argue that, if formed by gravitational instability, the wide orbital separation gas giants should represent the low-mass tail of a distribution of stellar companions. Thus far, observations do not clearly connect the population of wide orbital separation gas giants to the Brown Dwarf population (\citealt{bowler_DI_review}). There exist a number of other wide separation gas giants, but whether they formed \textit{in situ} is less well constrained. Super Earths are difficult to explain through local isolation mass due to their large masses and proximity to the central star. At such small orbital separations, there is not enough material locally to grow such massive planets without causing the protoplanetary disk to become unstable to collapse (\citealt{hilke_iso_mass}), indicating that radial drift of particles may be an important factor in planet formation. More massive particles can also move radially due to Type I migration and/or gas dynamical friction (e.g. \citealt{gp_2015}).

These difficulties can be amended through a more detailed consideration of the interaction between the gas present in the disk and the material accreted by the growing cores. While the effect of gas drag on smaller  ($\lesssim 0.1-1\,\text{km}$) planetesimals can be substantial (\citealt{raf}), even more striking is the effect of drag on smaller, mm---cm sized particles. For these bodies gas drag can enhance accretion rates by dissipating the relative kinetic energy between the small bodies and growing cores during their interaction. Due to the sizes of bodies for which this is possible, this processes is commonly referred to as ``pebble accretion."  We will alternatively refer to this process as ``gas-assisted growth," to highlight the idea that enhancements to growth come from the constructive effect of collisions between the small bodies being accreted and the gas particles, and to avoid confusion with the geological use of the term ``pebble." We also note that ``pebbles" need not necessarily be particles of small sizes, but could also include ``fluffy" aggregates of low density that have similar aerodynamic properties to rocky mm--cm sized particles. Because the term ``pebble accretion" is well established, we use these two terms interchangeably.

Models of gas-assisted growth in the context of planet formation find that gas drag acting on pebble sized particles can lead to substantially higher growth rates than models that rely on growth by planetesimal accretion. For a wide range of disk parameters, massive ($M \gtrsim 10^{-3} M_\oplus$) cores can accrete pebble sized particles at impact parameters comparable to the core's Hill radius (\citealt{OK10, lj12}). If particles of these sizes are present, cores accreting at these rates can easily grow a gas giant at wide orbital separation, as opposed to cores undergoing planetesimal accretion. The presence of these smaller pebbles is supported by observations of protoplanetary disks. Matching observations of the spectral energy distribution of disks requires a dust size distribution where most of the mass is in $\sim 1\,\text{mm}$ sized particles (\citealt{dal_disk_models}), while sub-mm images of disks find solid surface densities in this size range which are comparable to the minimum mass solar nebula (\citealt{andrews_09}, \citealt{andrews}). 

While these rapid growth timescales can solve some of the issues present in growing wide orbital separation gas giants, they present issues of their own. Chief among these is that pebble accretion is, in some respects, \textit{too} efficient. Because the growth rate at large masses is so fast, the last doubling timescale to $M_{\rm{crit}}$ is extremely short in pebble accretion. Thus gas-assisted growth seems to predict that growth of gas giants should be a ubiquitous phenomenon. Direct imaging surveys, however, place severe constraints on the existence of gas giants at wide orbital separation (e.g. \citealt{brandt_di}, \citealt{chauvin_di}, \citealt{bowler_DI_review}, \citealt{gal_gas_giant_freq}). Pebble accretion must therefore be inhibited in some manner from what current models naively predict.

One commonly neglected effect in models of the pebble accretion is the effect of turbulent gas velocities on planetary accretion efficiency. Turbulence can increase the velocity of the gaseous component of the disk. This in turn has a number of ramifications for gas-assisted growth: pebble velocities are now higher as well, accretion cross sections can shrink substantially, and the scale height of particles can increase. These effects can greatly decrease the efficiency of accretion. 

The effects of turbulence on pebble accretion have been discussed in a number of different regimes. The majority of works including turbulence discuss the growth of lower mass planetesimals -- in these models the growing body is assumed to be of low enough mass that its gravity is negligible, i.e. these models discuss the effect of turbulence for accretion where the cross section is comparable to the body's geometric cross section (e.g. \citealt{hgb_2016}). Previous models of pebble accretion for higher mass protoplanets generally neglect the effects of turbulence, or include it only by modifying the scale height of small bodies. A few works do modify the particle velocities or impact parameters due to the influence of turbulence. \cite{OK12} examine growth over a large range of large body sizes, and include a turbulent component to the velocity through ``turbulent stirring" on the random velocities of small bodies (\citealt{OK12}). \cite{gio_2014} include the effect of turbulence on the  radial motion of small bodies. \cite{c_2014} employ a methodology more similar to our own, using asymptotic expressions from \cite{oc07} for the relative velocity between the small bodies and gas due to turbulence, and extending the \cite{OK10} expressions for impact parameter and accreted particle sizes to include this turbulent component. In this formulation however, the impact parameter, as well as the sizes of small bodies that can be accreted, are functions solely of the relative velocity between the small body and the core at infinity. Our approach, which separately calculates the parameters relevant to growth, as well as the velocity preceding and during the encounter between the small body and the core, can be more naturally extended to include turbulence, and captures facets of the problem not covered by the \citeauthor{c_2014} results. Furthermore, the focus of our study is distinct from those described above: these papers are concerned with holistically studying growth of planets at a few points in parameter space by including a wide variety of processes and modeling the problem numerically. Our methodology instead focuses on studying the effects of turbulence over a broad parameter space, and understanding the conditions under which turbulence is important to pebble accretion.

With these considerations in mind, in this paper we present an order of magnitude model of pebble accretion. We approach the problem in a different manner than past theories have, allowing separate changes to the different parameters that set the growth timescale, as opposed to grouping growth timescales into a few regimes. This allows us to more fully take into account the effects of turbulence than previous studies, including the effects of turbulence on not just the particle scale height, but also the velocity dispersion of the small bodies as well as the impact parameters for accretion. This model can be applied over a wide range of parameter space to give results accurate to order of magnitude, and can accurately describe the trends present in gas-assisted growth. We use this model to discuss the overarching features of pebble accretion, as well as to investigate how turbulence modifies these features. We also discuss how pebble accretion operates in different regions of parameter space, particularly at wide orbital separations and low core masses. In these regimes growth at intermediate masses may dominate the timescales for gas giant growth, which we will discuss in more detail in our Paper II (Rosenthal and Murray-Clay, submitted).

In Section \ref{overview} we give an overview of how growth operates in the presence of nebular gas, and discuss broadly how we calculate the growth timescale in our model. In Section \ref {vel} we discuss our choices for modeling the velocities that enter into our calculation. Section \ref{len_scales} details how the length scales relevant to the accretion cross section are calculated. In Section \ref{results}, we give an overview of the output from our model, in particular discussing the broad features of pebble accretion as well as the effects of turbulence. We also give a detailed comparison between our modeling and other works on pebble accretion. Readers that are not concerned with the details of our model can find a summary of our algorithm in Appendix \ref{app:sum}, and may skip directly to Section \ref{results} for our results. In Section \ref{param_space} we discuss how gas-assisted growth operates when various parameters are adjusted. In Section \ref{flow_iso_mass} we note how the relatively simple assumptions on which our model is based lead naturally to a ``Flow Isolation Mass," past which accretion of pebbles ceases. Finally in Section \ref{summary} we summarize our results and discuss future extensions of our model. 

\section{Model Overview} \label{overview}

\subsection{Accretion in the Presence of Gas} \label{energy}
We begin by discussing generally how we model growth of planets in the presence of nebular gas. The details of how specific quantities are calculated are deferred to subsequent sections. 

Our calculation proceeds in an order of magnitude manner -- i.e. the approximations made and the neglected effects mean that quoted values should be correct to within an order of magnitude. In what follows, we consider two types of bodies. The large bodies, or protoplanetary cores, are assumed to be massive enough that they are unaffected by gas drag and thus move at the local Keplerian orbital velocity. This constrains our cores to have radii $\gtrsim 10 \, \text{km}$, in which case their velocity will deviate from Keplerian by at most $10^{-3}$ of the gas velocity, with a weak dependence on stellocentric distance for fiducial disk parameters. In practice we rarely consider cores of such small size. We note also that while these cores are insensitive to aerodynamical gas drag, large bodies with masses in the range $10^{21} \, \text{g} < M < 10^{25} \, \text{g}$ can still be affected by gas dynamical friction (\citealt{gp_2016}). We do not include these effects here. The second type of particles considered are ``small" bodies, which can be substantially affected by gas drag. The growth timescale in pebble accretion is strongly dependent on the size of small body under consideration, unlike canonical core accretion where the size of the planetesimals enters only through its effect on the small body velocity dispersion. Thus, all calculations are performed as a function of small body radius, $r_s$. Quantities of interest can later be averaged over size by assuming a size distribution for the small bodies. Note that quoted values of the growth timescale $t_{\rm{grow}}$ implicitly assume that all of the surface density is contained in particles of the given value of $r_s$. For a size distribution where most of the mass is in the largest sizes of particles present (e.g. a Dohnyani size distribution, \citealt{size_dist}), this value of $t_{\rm{grow}}$ is approximately equal to the growth timescale for a distribution of small body sizes where the maximum size present is the given value of $r_s$. In this paper we will not perform integrations over small body size explicitly; see our Paper II for examples of this process as well as a discussion of the effects of altering the size distribution.

In gas-assisted growth, the interaction between the small body and the nebular gas modifies the accretion process substantially. As the small body approaches the core, gas drag will dissipate the kinetic energy of the small body relative to the large body. This loss of energy can cause small bodies on non-collisional trajectories to become bound to the core and eventually be accreted. This process is similar to the $L^2s$ mechanism identified by \cite{gls_2002} for formation of Kuiper Belt binaries, with gas drag as the source of dissipation in the place of dynamical friction. Gas drag can also stop small bodies from accreting -- if particles couple strongly to the gas as they flow around the core then they will be unable to accrete.

Which of these processes occur depends on the relative size of two different length scales: the stability radius, $R_{\rm{stab}}$ and the Bondi radius, $R_b$. The stability radius is the smallest radius at which stable orbits by the small body about the large body are possible: outside of $R_{\rm{stab}}$ interactions between the small body and either the nebular gas or the central star will shear the small body away from the large body's gravity. Inside of $R_{\rm{stab}}$ the small body can safely inspiral onto the core. The details of how $R_{\rm{stab}}$ is calculated are discussed in Section \ref{acc_cross}. The Bondi radius, on the other hand, is approximately the radius at which the escape velocity from the core is equal to the speed of sound in the gas: 
\begin{equation}\label{eq:r_b}
R_b = \frac{G M}{c_s^2} \;, 
\end{equation}
where $M$ is the mass of the core and $c_s = \sqrt{k T / \mu}$ is the isothermal sound speed of the gas. Here $k$ is Boltzmann's constant and $\mu$ is the mean molecular weight of the nebular gas.  We consider the Bondi radius because it roughly tells us the length scale interior to which the the core can have a stable atmosphere, which has substantial effects on the flow pattern. For the lowest mass cores we consider, the Bondi radius may be less than the physical radius of the core, $R$. This occurs roughly at a core mass of:
\begin{align}
M_a &\equiv \frac{c_s^3}{G}\left(\frac{3}{4 \pi G \rho_p}\right)^{1/2}\\
&\approx 2 \times 10^{-4} M_\oplus \left( \frac{a}{30 \, \rm{AU}} \right)^{-9/14} \left( \frac{\rho_p}{2 \, \rm{g} \, \rm{cm}^{-3}} \right)^{-1/2} \; ,\notag
\end{align}
where $\rho_p$ is the density of the protoplanet (e.g. \citealt{raf06}), and for the expression in the second line we've used our fiducial disk parameters (see Section \ref{params}). If $R_b<R$, then the effects discussed below are unchanged, with $R$ taking the place of $R_b$. In what follows, we will discuss accretion for $R_b < R_H$, where $R_H$ is the core's Hill radius (see Section \ref{r_h}). We discuss accretion in the regime $R_b > R_H$ in Section \ref{flow_iso_mass}.

\begin{figure} [h]
	\centering
\includegraphics[width=\linewidth]{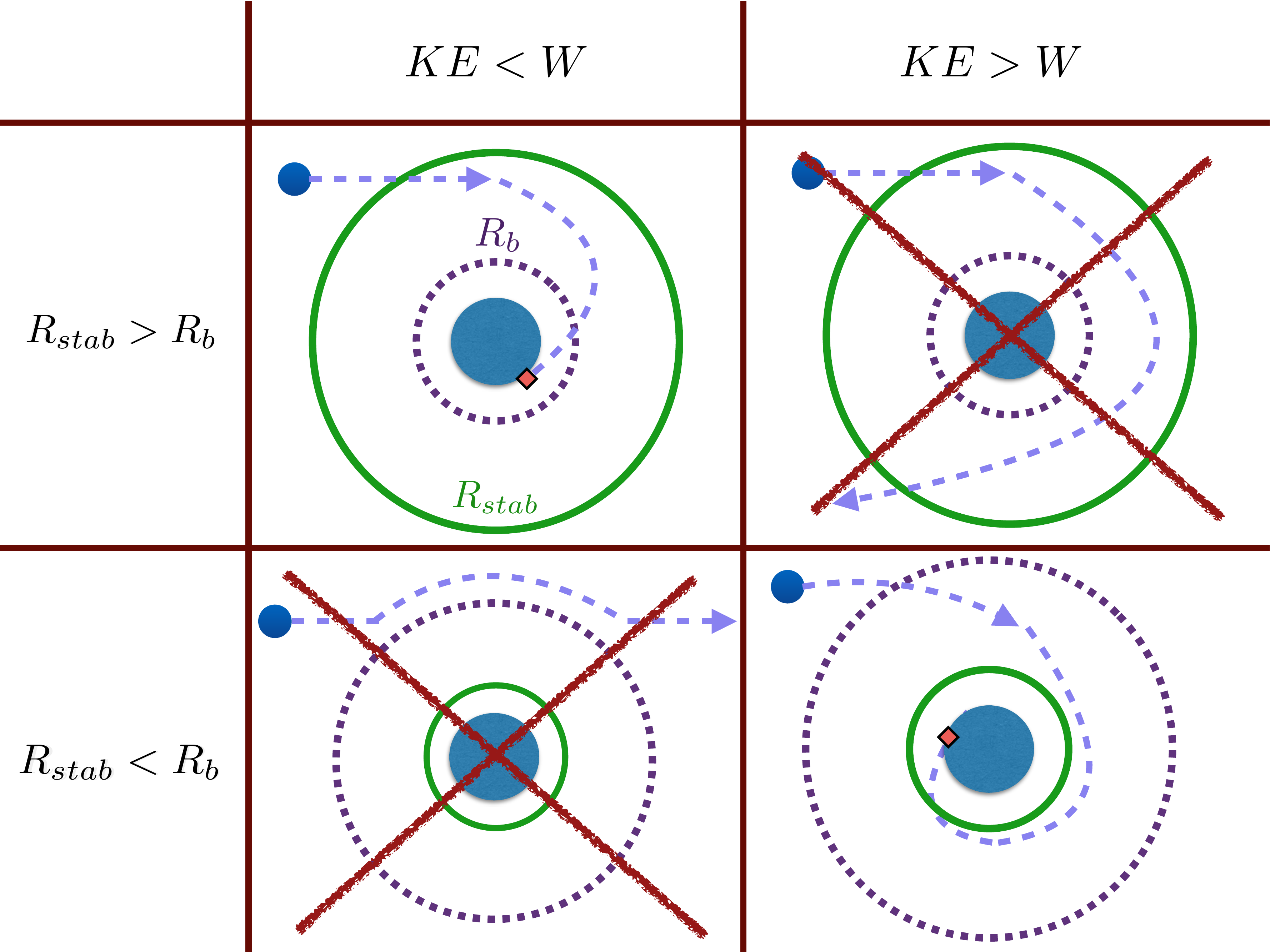}
	\caption{A graphical illustration the energy regimes used to determine whether small bodies are able to accrete. \textit{Upper Left Panel}: Here $R_{\rm{stab}} > R_b$, so particles can inspiral onto the core in a region where the gas flow is not substantially altered by the core's gravity. Particles that deplete their kinetic energy relative to the core via gas drag will inspiral onto the core and be captured. \textit{Upper Right Panel}: If the particle is unable to deplete its kinetic energy during the interaction then it will simply have its trajectory deflected before exiting $R_{\rm{stab}}$, and will not accrete. \textit{Lower Left Panel}: Here $R_b > R_{\rm{stab}}$, so the gas flow is altered substantially when the small body is accreting. The gas will tend to flow around the atmospheric radius, so particles that have $KE < W$, i.e. particles that deplete their kinetic energy, will couple to the gas and flow around the core without accreting. \textit{Lower Right Panel}: Larger particles which do not deplete their kinetic energy and instead penetrate into the atmospheric radius will experience a rapid increase in gas density as they enter the atmosphere of the nascent planet. This increased density may rapidly deplete the kinetic energy of the small body, which will then inspiral and accrete, similar to what occurs in the upper left panel.}
	\label{fig:en_fig}
\end{figure}

Given these considerations, we center our model around two main ideas about accretion in the presence of gas, which are summarized in Figure \ref{fig:en_fig}:
\begin{enumerate}
	\item If the radius for stable orbits exceeds the Bondi radius, i.e. $R_{\rm{stab}} > R_b$, then the flow pattern of gas is not substantially altered in the region where particles can stably orbit the core. In this case any small bodies that deplete their kinetic energy relative to the core within $R_{\rm{stab}}$ will inspiral onto the core and be accreted. On the other hand any particles that are unable to dissipate their kinetic energy in this regime will pass out of $R_{\rm{stab}}$ and will not be accreted.

	\item If instead the Bondi radius exceeds the stable orbit radius, i.e. $R_b > R_{\rm{stab}}$, then particles which \textit{are} able to deplete their kinetic energy relative to the core \textit{will not} accrete. This is due to the fact that in this regime, well coupled particles will tend to flow around the core's atmosphere, which extends up to $R_b$. If instead particles are \textit{unable} to dissipate their kinetic energy, so that they penetrate into the atmospheric radius, the increase in density as the particle enters the growing planet's atmosphere is taken to be so substantial that the particle will now be able to dissipate its kinetic energy and will accrete onto the core.
\end{enumerate}

The first point is supported not only by order of magnitude considerations, but also by detailed numerical simulations of growth of protoplanets in the presence of gas (e.g. \citealt{OK10}, \citealt{lj12}). Analytic calculations also show that even for small bodies several orders of magnitude larger than the sizes we will be concerned with, small bodies inspiral on times shorter than the disk lifetime. Thus we are justified in neglecting this part of the accretion process (\citealt{pmc11}). We also neglect the possibility that the core's envelope is periodically replenished by the protoplanetary disk; see \cite{osk_2015} for a discussion of this possibility, and \cite{a_2017} for an application of this replenishment to pebble accretion.

Our second point invokes the classical solution of flow around an obstacle: for $R_b < R_H$ the flow of the nebular gas is subsonic, meaning the core's atmosphere is approximately incompressible. See \cite{ormel_flows} Figure 5A for an example of this flow pattern in the context of a planet embedded in a protoplanetary disk. We note here that more detailed simulations of the flow show structure that may produce circulation into $R_b$ (\citealt{ormel_flows}, \citealt{faw2015}), which we assume we can neglect for the level of accuracy we desire in this model. In addition, we do not explicitly calculate the work done on particles interior to $R_b$ -- our assumption that particles with $KE>W$ will always be accreted for $R_{\rm{stab}}<R_b$ will eventually be violated for large enough particle sizes. See \cite{ii_2003} for an in depth discussion of accretion in this regime.

\begin{figure} [h]
	\centering
	\includegraphics[width=1.0\linewidth]{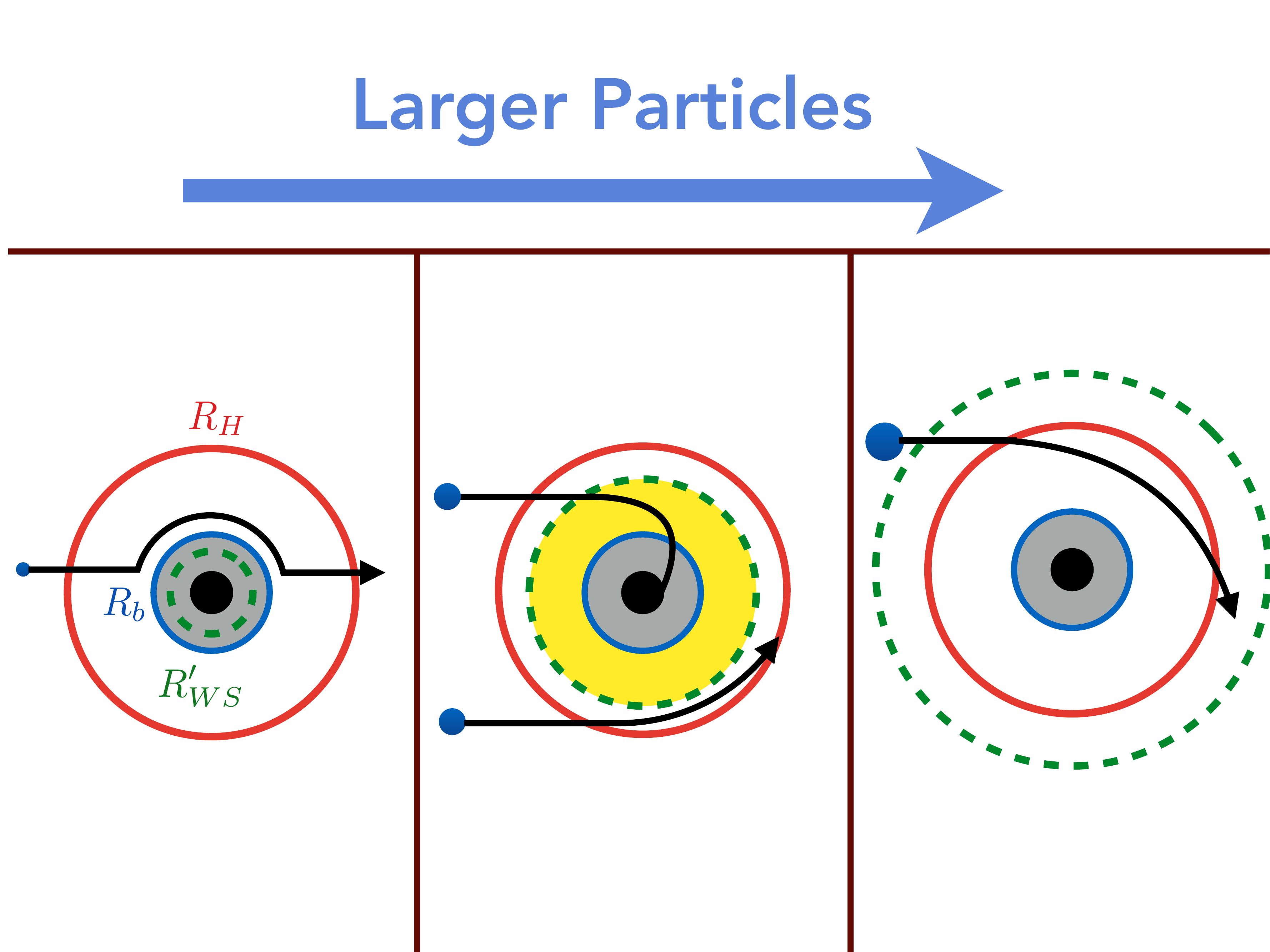}
	\caption{A cartoon illustration of the typical manner in which pebble accretion operates as the small body size is increased. The black circle represents the planet, while the blue circles depict incoming particles. The extent of the planet's atmosphere is denoted by the grey shaded region, and the yellow shaded region shows the region where incoming particles can be accreted. \textit{Left Panel}: For small particles, $R_{WS}^\prime = R_{\rm{stab}} < R_b$, so the core's gravitational sphere of influences lies inside its atmosphere. In this regime particles couple to the local gas flow and flow around the core without being accreted. \textit{Middle Panel}: For intermediate sizes of particles, $R_{WS}^\prime > R_b$, meaning particles can be bound to the core in a region outside the core's atmosphere. For these intermediate sizes of particles the work done by gas drag exceeds the incoming kinetic energy of these small bodies, meaning that  particles that pass interior to $R_{\rm{stab}}$ will be accreted. Particles with impact parameters $>R_{\rm{stab}}$ will be sheared off the core by gas drag. \textit{Right Panel}: Finally, large particles will be so massive that their incoming kinetic energy is too large to be depleted by gas drag. These particles will not be accreted via pebble accretion, regardless of impact parameter. For the case shown here $R_{WS}^\prime$ has grown so much that $R_{\rm{stab}}=R_H$, but this is not always the case for particles with $KE>W$.  }
	\label{fig:rs_ex}
\end{figure}

While the criteria above are used throughout parameter space, in order to better illustrate how we model gas-assisted growth we also provide a simplified ``sketch" of how accretion in our model operates, which accurately describes pebble accretion over a large amount of parameter space. \footnote{The main assumption in this description is that particles that have $R_b > R_{\rm{stab}}$ will always have $KE<W$. To see this, note that $R_{\rm{stab}} < R_b$ implies that $F_D(v_{cg}) > G M m/R_b^2$, where $v_{cg} = \max(v_{\rm{gas}},v_{\rm{shear}})$ is the relative velocity of the gas and the core at the Bondi radius (see Section \ref{vel} and Equation \ref{eq:rws_imp}). This can be rewritten as $m/R_b < F_d(v_{cg})/c_s^2$. Taking the ratio $KE/W=m v_\infty^2/(4 F_d(v_{\rm{enc}}) R_b)$ (see Equations \ref{eq:ke} and \ref{eq:work}) and inserting this inequality implies that $KE/W < \left(v_\infty^2/c_s^2 \right) F_d(v_{cg})/F_d(v_{\rm{enc}})<1$. For the case $R_b<R$, $c_s \rightarrow v_\infty$ in the last inequality, which still implies $KE/W < 1$ as long as the escape velocity from the planet's surface is larger than the velocity of the gas flow at the planet's surface, which is typically satisfied for planetary radii $R \gtrsim 10-100$ km.} Figure \ref{fig:rs_ex} shows a ``cartoon" of gas-assisted growth for a fixed core mass and semi-major axis in the disk, with small body radius increasing from left to right. Each panel shows the core's Bondi radius, $R_b$ as well as the two radii that are used to determine $R_{\rm{stab}}$ -- the Hill radius $R_H$ (see Section \ref{r_h}), which is the distance past which the stellar gravity will pull particles off the core, and the radius $R_{WS}^\prime$ (see Section \ref{WISH}) beyond which gas drag pulls particles off the core. Since smaller particles have a higher surface area to volume ratio and therefore experience larger gas drag accelerations, $R_{WS}^\prime$ is smaller for smaller particles. In the far left panel, the particles are low mass, meaning gas drag can easily pull them off of the core, and $R_{WS}^\prime$  lies inside the core's atmosphere. Because these particles have low mass they easily dissipate their kinetic energy during the encounter with the core, meaning they are in the regime in the lower righthand panel of Figure \ref{fig:en_fig} and do not accrete. As particle size grows, $R_{WS}^\prime$ increases as well, until it exceeds $R_b$, the scale of the core's atmosphere, i.e. there now exists a region exterior to the core's atmosphere where particles can stably orbit the core. Because these particles are still of relatively low mass they are able dissipate their kinetic energy through gas drag. We therefore fall into the upper lefthand panel of Figure \ref{fig:en_fig}, which signals the onset of pebble accretion (middle panel of Figure \ref{fig:rs_ex}). Finally, as particle size continues to increase we eventually reach a point where the particles are so massive that they no longer dissipate their kinetic energy. As particle size increases, $R_{WS}^\prime$ will continue to increase, meaning we clearly still have $R_{\rm{stab}} > R_b$. These particles are therefore in the upper righthand panel of Figure \ref{fig:en_ratio} and will not accrete (righthand panel of Figure \ref{fig:rs_ex}). 

Given this formalism, for the purposes of discussing whether a given small body will accrete, we simply need to compare the magnitude of the kinetic energy of the particle relative to the core and the work done on the particle during its encounter with the core. The kinetic energy of the particle relative to the core before the encounter is
\begin{align} \label{eq:ke}
KE=\frac{1}{2}mv_{\infty}^2 \; ,
\end{align}
 while we take the work done by gas drag to be simply
 \begin{align} \label{eq:work}
 W = 2 R_{\rm{acc}} F_D(v_{\rm{enc}}) \; ,
 \end{align}
where $R_{\rm{acc}} = \max(R_{\rm{stab}},R_b)$, and $F_D(v_{\rm{enc}})$ is the drag force on a small body moving at a velocity $v_{\rm{enc}}$, which is the relative velocity between the small body and the large body during the encounter. Calculation of $v_{\rm{enc}}$ and discussion why $R_{\rm{acc}}$ is used for determining the work done by gas drag are located in section \ref{work}.

Our consideration of the ranges of particle sizes that can be accreted is an important aspect to our modeling that is often not present in other works. See section \ref{comp} for a detailed discussion.

Given the uncertainties in the size distribution of small bodies present in protoplanetary disks, understanding the extent of particle radii for which gas-assisted growth is possible is an important facet to studying the role of pebble accretion in planet formation. Furthermore, an important effect of nebular turbulence is to change the range of small body sizes that can be accreted (see Section \ref{var_a_M}), which makes a detailed consideration of the small body sizes where pebble accretion can operation an important facet of our model.

In summary, a particle will be able to accrete if one of the following criteria are met:
\begin{enumerate}
	\item $R_{\rm{stab}} > R_b$ and $2 F_D(v_{\rm{enc}}) R_{\rm{acc}} > \frac{1}{2}mv_{\infty}^2$
	\item  $R_{\rm{stab}} < R_b$ and $2 F_D(v_{\rm{enc}}) R_{\rm{acc}} < \frac{1}{2}mv_{\infty}^2$
\end{enumerate}

Small bodies which do not satisfy either of these criteria will not be able to accrete, that is we set $t_{\rm{grow}} = \infty$ for these particles. We emphasize that setting $t_{\rm{grow}} = \infty$ refers \textit{only} to the timescale for growth by pebble accretion: it is possible these particles could still be accreted via less efficient processes, such as capture by gravitational focusing unassisted by gas drag or by collisions unaffected by gravity (see Equation \ref{eq:r_focus}). In particular, having $t_{\rm{grow}} = \infty$ does not imply that growth literally halts. See Section \ref{comp} for further discussion. While in principle our model could be extended to include these effects, in this work we are concerned primarily with gas-assisted growth, and therefore we do not explicitly include these other timescales.

\subsection{Growth Timescale for Protoplanets} \label{t_grow}
\begin{figure} [h]
	\centering
	\includegraphics[width=0.9\linewidth]{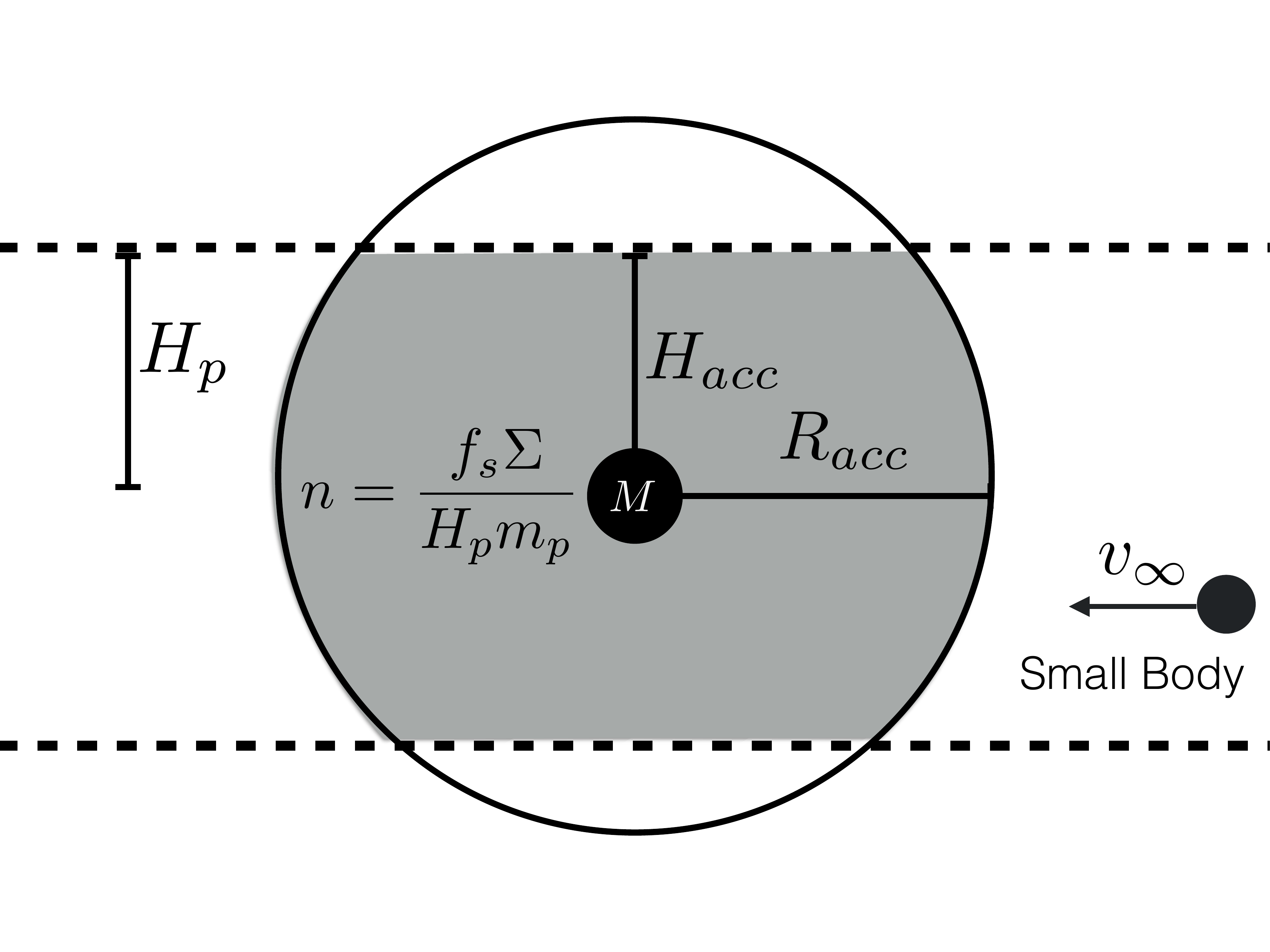}
	\caption{A graphical illustration of the quantities used to determine the growth timescale for the planetary core.}
	\label{fig:growth_fig}
\end{figure}

We now discuss in more detail how the growth timescales for cores are computed. In order to calculate the growth timescale for the large bodies for a given core mass $M$, we use the usual expression (see e.g. \citealt{gold}, hereafter GLS)
\begin{align} \label{accRate}
t_{\rm{grow}} \equiv \left(\frac{1}{M} \frac{dM}{dt}\right)^{-1} \; .
\end{align}
The rate that small bodies encounter the core is given by $n \sigma_{\rm{acc}} v_{\infty}$, where $n$ is the volumetric number density of small bodies of a given size, $\sigma_{\rm{acc}}$ is the cross section for accretion of the small body by the large body, and $v_{\infty}$ is the velocity at which small bodies encounter the large body. Note that $v_\infty$ is not necessarily the velocity of the small body during its encounter with the core, since this encounter can change the relative velocity; $v_\infty$ is the relative velocity between the two bodies at large separations.  The number density of solids is simply $n=\rho_s / m=f_s \Sigma / (2 m  H_p)$ where $\rho_s$ is the mass density of the small bodies, $\Sigma$ is the surface density of the gaseous component of the disk, $f_s$ is the solid to gas mass ratio, $m$ is the mass of the small body, and $H_p$ is the scale height of solids in the disk. Since each accretion of a small body increases the mass of the large body by $m$, the growth timescale is given by
\begin{align} \label{t_grow_full}
t_{\rm{grow}} = \frac{M H_p}{2 f_s\Sigma v_{\infty} R_{\rm{acc}} H_{\rm{acc}}} \; ,
\end{align}
where we've decomposed $\sigma_{\rm{acc}}$ into the product of lengthscales in the plane of the disk and perpendicular to it: $\sigma_{\rm{acc}} = (2R_{\rm{acc}})(2H_{\rm{acc}})$. Thus, the aim of our calculation is to determine the quantities $R_{\rm{acc}}$, $H_{\rm{acc}}$, $H_p$, and $v_{\infty}$. Once these quantities are known we can immediately determine the growth timescale. The role each of these quantities plays in the growth timescale is illustrated graphically in Figure \ref{fig:growth_fig}.

In the next few sections we will discuss in detail how each of the above quantities are calculated in the context of pebble accretion. Our algorithm is summarized in Appendix \ref{app:sum}.

\section{Velocity of the Small Bodies through the Disk} \label{vel}
As small bodies move through the disk their velocities are affected by drag from the nebular gas as well as gravitational interactions with both the central star and the core. In order to calculate $t_{\rm{grow}}$ we need to understand what sets $v_\infty$, the velocity of a small body relative to the big body, and $v_{\rm{enc}}$, the velocity of the small body during its encounter with the core. Calculating these velocities requires us to treat not only the gas drag force on the small bodies, but also to understand how drag from the laminar and turbulent components of the gas velocity each contribute to the velocity of the small bodies. Furthermore the gas velocity influences the size of both $R_{\rm{acc}}$ and $H_{\rm{acc}}$, in ways that can have substantial impact on $t_{\rm{grow}}$.

We begin by reviewing our choices for modeling gas drag regimes and introduce the stopping time to parameterize the coupling between the small bodies and the gas (Section \ref{stoppings}). The gas is taken have both a bulk, laminar component that is independent of time, and a fluctuating, turbulent component that time averages to zero. Both of these components can have an effect on the small bodies' velocity, and we discuss each separately (Sections \ref{lam_vel} and \ref{turb_vel}). These two components can be combined to give the average velocity between the small body and the large one due to gas drag, $v_{pk}$. (Section \ref{rey_avg}). Gas drag is not the only source of relative velocity in the disk -- the shear present in the disk due to the dependence of the Keplerian orbital frequency on semi-major axis can also affect these relative velocities (Section \ref{v_shear}). Gas drag can also have a strong effect on the relative velocity between the small body and the core during their encounter, $v_{\rm{enc}}$, as can the gravitational force from the core (Section \ref{work}). These sections synthesize results from many works, which we present in detail so that the framework and assumptions our model is based on are clearly laid out. Readers who only wish to review our choices for modeling the velocity may consult the summary of our calculation in Appendix \ref{app:sum}.

\subsection{Gas Drag and Stopping Time} \label{stoppings}
Gas drag is generally treated by breaking the drag force, $F_D$, into a number of different regimes. We summarize our choices below; for a more in depth discussion, see \cite{Bat}. First, we distinguish between the ``diffuse regime," which applies for particles with $r_s < 9\lambda/4$, and the ``fluid regime," $r_s > 9\lambda/4$. Here $r_s$ is the radius of the small body and $\lambda$ is the mean free path of the gas particles. In the diffuse, non-supersonic,\footnote{
	It is easy to see we can in general neglect the super-sonic regime ($v_{\rm{rel}} > c_s$): since $v_{\rm{rel}} \lesssim v_{\rm{gas}} \approx \eta v_k$ (see section \ref{lam_vel} for a discussion of the notation), we have $c_s <  \eta v_k \Rightarrow \, H_g/a > 1$, where $H_g \sim c_s / \Omega$ is the gas scale height, and $\Omega$ is the local Keplerian orbital frequency. Thus for the super-sonic case the protoplanetary disk has an aspect ratio greater than 1, in strong opposition to observations of protoplanetary disks.
	} regime, the drag force on the particle is given by the Epstein drag law
\begin{align} \label{eq:epstein}
F_D = \frac{4}{3}\pi\rho_gv_{th} v_{\rm{rel}} r_s^2 \; ,
\end{align}
where $\rho_g = \Sigma/(2H_g)$ is the density of the gas, $v_{th}=\sqrt{8/\pi} \, c_s$ is the thermal velocity of the gas, and $v_{\rm{rel}}$ is the relative velocity between the gas and the object. For a particle in the fluid regime, we must consider an additional parameter -- the Reynolds number of the particle, given by $Re = 2 r_s v_{\rm{rel}} / \left(0.5 \, v_{th} \lambda \right)$. For $Re<1$ the particle is in the Stokes drag regime, and the drag force is given by
\begin{align} \label{eq:stokes}
F_D = 3 \pi \rho_g v_{th} v_{\rm{rel}} \lambda r_s \; ,
\end{align}
where $\lambda$ is the mean free path of the gas particles. For $Re>1$, the particle is in the Ram pressure regime, and 
\begin{align} \label{eq:RAM}
F_D = \frac{1}{2} \rho_g \pi r_s^2 v_{\rm{rel}}^2 \; .
\end{align}

To mitigate discontinuities in the drag force we use a smoothed drag force law in the fluid regime given by \cite{cheng_drag}:

\begin{align} \label{eq:cheng_drag}
F_D = \frac{1}{2} C_D  (Re) \pi r_s^2 \rho_g v_{\rm{rel}}^2 \; ,
\end{align}
where 

\begin{align}
C_D(Re) =& \frac{24}{Re} \left ( 1 + 0.27 Re \right)^{0.43} \\ \quad +& 0.47 \left [ 1 - \exp \left( - 0.04 Re^{0.38} \right) \right] \; . \notag
\end{align}
As $Re \rightarrow 0$ we have $C_D \rightarrow 24/Re$, so $F_D$ reduces to the Stokes drag law given in \eqref{eq:stokes}. As $Re \rightarrow \infty$, $C_D \rightarrow 0.47$, in which case $F_D$ becomes the Ram drag force given in \eqref{eq:RAM} (with a slightly different prefactor).

We can define a timescale from the drag force, known as the ``stopping time" of the particle
\begin{align} \label{eq:t_s_full}
t_s \equiv \frac{m v_{\rm{rel}}}{F_D} \; .
\end{align}
The Epstein and Stokes drag laws are both linear in velocity -- in these linear drag regimes, it is straightforward to show from Equations \eqref{eq:epstein} and \eqref{eq:stokes} that for spherical particles of uniform density $\rho_s$:

\begin{align} \label{t_s_exp}
t_{s}=\begin{dcases*}
\frac{\rho_{s}}{\rho_{g}}\frac{r_{s}}{v_{th}}, & \text{Epstein}\\
\frac{4}{9}\frac{\rho_{s}}{\rho_{g}}\frac{r_{s}^{2}}{v_{th}\lambda}, & \text{Stokes}
\end{dcases*}
\end{align}
In these regimes the stopping time of the particle is a function only of the properties of the particle and the gas, and in particular is independent of the particle's velocity. Hence $t_s$ is often used as a parameterization of the particle's size, $r_s$, in terms of how well the small body is coupled to the gas flow (e.g. \citealt{cy10}). If the drag law is instead quadratic in velocity, as in Equation \eqref{eq:cheng_drag}, we numerically solve for the stopping time using Equations \eqref{eq:v_lam_gas}, \eqref{RMSturb_gas} and \eqref{turb_lam}. In practice we solve these equations iteratively to calculate a self-consistent solution.

\vspace{4mm}

\subsection{Laminar Velocity of Small Particles} \label{lam_vel}
Due to the internal pressure of the gas, the gas component of the protoplanetary disk will move at a sub-Keplerian orbital velocity. \cite{weiden_lam_drift} gives the difference in velocity $\Delta v$ as $\Delta v \approx \eta v_{k}$, where $v_{k}$ is local Keplerian orbital velocity, $v_{k} = a\Omega$, and $\eta$ is a measure of the local gas pressure support, with approximate value $\eta \approx c_{s}^2 / \left( 2 v_{k}^2 \right)$. Due to this sub-Keplerian rotation, small bodies experience a ``headwind" from the gas, which produces a drag force on the small bodies. If we use a polar coordinate system such that $\boldsymbol{\hat{r}}$ denotes the direction pointing away from the central star and $\boldsymbol{\hat{\phi}}$ denotes the direction of the disk's rotation, then the drag force causes the small bodies to move with a sub-Keplerian velocity in the  $\boldsymbol{\hat{\phi}}$ direction, and to drift in the $-\boldsymbol{\hat{r}}$ direction. In the above notation, the particle acquires a laminar velocity relative to the gas given by
\begin{align} 
v_{r,\rm{gas}}&=-2\eta v_{k} \left[ \frac{\tau_s}{1 + \tau_s^2} \right] \; , \label{vr}\\ 
v_{\phi,\rm{gas}}&=-\eta v_{k} \left[ \frac{1}{1+\tau_s^2} - 1 \right ] \; , \label{vphi}
\end{align} 
where $\tau_s = t_s \Omega$. See \cite{nag} for further details. \footnote{There will also be turbulent component to $v_r$, which stems from the inward diffusion of the gas due to the turbulent viscosity $\nu = \alpha c_s H_g$ (see Section \ref{turb_vel} for a discussion of the notation.) \cite{gio_2014} give this velocity as $v_{r,\rm{turb}} = v_\nu / (1 +\tau_s^2)$, where $v_\nu \sim \alpha c_s H_g / a$ is the radial velocity of the gas due to turbulence. Following \citeauthor{gio_2014}, the turbulent component $v_r$ dominates for $\tau_s < \tau_{s,\nu} = \alpha c_s^2 / (2 \eta v_k) \approx \alpha$. \cite{LJ14} show that, for the parameters they consider, this velocity is negligible compared to $v_{r,k}$ because the diffusive lengthscale $\ell$ is always less than the global scale of the disk $a$. However, their expression for $\ell$ can be rewritten as $\ell/r \approx \alpha (H_g/a)^2/(2 \tau_s \eta) \approx \alpha/\tau_s$, which leads to the same conclusion as above for the size at which turbulence dominates. For our purposes we neglect this effect, since the small particles for which $v_\nu > v_{r,k}$ move at velocities comparable to the fluctuating turbulent velocity $v_t$, which dominates: $v_\nu / v_t \sim \sqrt{\alpha} H_g /a \ll 1$.}

Since the laminar gas velocity relative to Keplerian is simply $v_{gas,k}  = - \eta v_k \boldsymbol{\hat{\phi}}$, the velocity of the particle relative to Keplerian is
\begin{align} 
v_{r,k}&=-2\eta v_{k} \left[ \frac{\tau_s}{1 + \tau_s^2} \right] \label{vr_kep} \; ,\\ 
v_{\phi,k}&=-\eta v_{k} \left[ \frac{1}{1+\tau_s^2} \right ] \label{vphi_kep} \; .
\end{align} 
It is straightforward to show that the magnitude of the laminar component of the particle's velocity is
\begin{align} \label{eq:v_lam_gas}
v_{pg,\ell} = \eta v_k \tau_s \frac{ \sqrt{4+\tau_s^2} }{1+\tau_s^2} \; ,
\end{align}
relative to the gas, and
\begin{align} \label{eq:v_lam_kep}
v_{pk,\ell} = \eta v_k \frac{ \sqrt{1+4\tau_s^2} }{1+\tau_s^2} \; ,
\end{align}
relative to Keplerian.

\subsection{Turbulent Velocity of Small Particles} \label{turb_vel}
In order to describe the strength of the turbulence in the disk, we use the standard Shakura-Sunyaev $\alpha$ parameterization of the effective kinematic viscosity. We employ $\alpha$ simply as a convenient parameterization of the strength of turbulence; for our purposes $\alpha$ is fundamentally a local quantity, and is not necessarily connected with the accretion rate onto the star.  In terms of $\alpha$, the effective kinematic viscosity of the turbulent gas, $\nu_t$, is given by (\citealt{ss_alpha}):

\begin{align}
\nu_t = \alpha c_s H_g \; ,
\end{align}
where $H_g = c_s/\Omega$ is the scale height of the gas. If we write the viscosity as the product of the local turbulent velocity and the largest scale turbulent eddies, $\nu_t = v_t \ell_t$, and have $l_t \approx v_t/\Omega$, then $v_{t}=\sqrt{\alpha}c_{s}$. 

We use the Kolmogorov theory of turbulence to determine the turbulent energy spectrum. As described in \cite{ch03}, in the Kolmogorov theory turbulence exists over a range of scales or ``eddies," which are characterized by their wave number $k_\ell = 1/\ell$. The largest scale eddies, which occur on the lengthscale $l_t$, are the scale on which energy is supplied by the turbulence; these large scale eddies ``turn over" and transfer their energy to smaller scale eddies, until energy is finally dissipated by kinematic viscosity on some smallest scale $\tilde{\eta}$. If we assume that the rate of energy transfer between scales is independent of eddy size, we can show that the energy spectrum of the turbulence is given by $E(k) \propto k^{-5/3}$, where $E(k)$ has units of energy per mass per wavenumber. The velocity associated with $k$ is then $v(k) = (2kE(k))^{1/2} \propto k^{-1/3}$, and the overturn time for an eddy of wavenumber $k$ is $t_k = 1/(k v(k)) \propto k^{-2/3}$. Thus the larger scale eddies overturn more slowly and contain more of the turbulent kinetic energy. The size of the smallest scale eddies can be determined by setting the rate of energy loss from molecular viscosity equal to the eddy turnover time, which gives $\tilde{\eta }= \left (\nu^3 / \epsilon \right)^{1/4}$, where $\nu \sim v_{th} \lambda $ is the molecular viscosity and $\epsilon$ the rate of energy dissipation. The length $\tilde{\eta}$ is known as the ``Kolmogorov microscale" of length. We can determine the relation between the smallest and largest scale eddies by setting the rate that energy is supplied at the largest scales, $v_t^2 / t_L \sim v_t^3/l_t$ equal to $\epsilon$, the rate that energy is dissipated at small scales. Plugging in our expression for $\epsilon$ in terms of $\tilde{\eta}$ and $\nu$ gives $\tilde{\eta} / l_t \sim  \left (v_t l_t / \nu \right)^{-3/4} = Re_t^{-3/4}$, where we've defined the Reynolds number of the turbulence, $Re_t \equiv \nu_t / \nu$. In terms of $\alpha$ we have $Re_t = \alpha c_s H_g / (v_{th} \lambda)$.

Qualitatively, the behavior of a small body in response to an eddy of wavenumber $k$ depends on the ratio of the particle's stopping time to the eddy turnover time $t_k$. Particles with $t_s < t_k$ will come to equilibrium with the eddy before it turns over, and will follow the large scale motion of the eddy. On the other hand particles with $t_s > t_k$ will not come to equilibrium with the eddy before it turns over, and will therefore only receive a small perturbative kick from the eddy. To characterize the small body's response to the turbulence, most authors use the Stokes number, $St \equiv t_s / t_L$, where $t_L$ is the overturn time of the largest eddies. We make the usual assumption that $t_L = \Omega^{-1}$, in which case $St = \tau_s$, the parameter used in describing the particle's interaction with the laminar gas flow. This permits the use of one parameter, which we will hereafter refer to as $St$, to characterize the particle's interaction with both the laminar and turbulent components of the nebular gas flow. See \cite{orb} for a more in depth discussion of the effect of varying $t_L$.

Due to their interaction with the turbulent gas, small bodies will move relative to inertial space with some non-zero root-mean-square (RMS) velocity, $v_{pi,t}$. The RMS velocity can be calculated through order of magnitude means as follows (\citealt{orb}): For $St \gg 1$ we have $t_s \gg t_L$; in this regime particles receive many uncorrelated ``kicks" from the largest scale eddies over a single stopping time, causing the particle to random walk in velocity. Since the particles' velocity damps out over a time $t_s$, the particle receives approximately $N \sim t_s / t_L \sim St$ kicks of amplitude $v \sim v_t \, t_L/t_s = v_t/St$ resulting in a velocity $v_{pi,t}\sim v_t / \sqrt{St}$. On the other hand, for $St\ll1$, we expect $v_{pi,t} \sim v_t$. The simplest expression that has the correct behavior in each of these limits is given by
\begin{align} \label{RMSsimple}
v_{pi,t}=\frac{v_{t}}{\sqrt{1+St}} \; .
\end{align}

In order to better calculate the velocities of smaller particles, i.e. particles with $St<1$,  we employ expressions from \cite{oc07} (hereafter OC07), who give closed form equations for the RMS velocity of solid particles suspended in a turbulent medium. By following the methodology of \cite{volk}, but also using results which take into account the finite inner scale of eddies at $k_\eta = 1/\tilde{\eta}$, \footnote{
	The OC07 results also use a different auto-correlation function (ACF) for the gas velocity of an eddy with turnover time $t_k$. \cite{mmv} encapsulate the \cite{volk} ACF in the more general form
	\begin{align*}
	R(t,t^\prime;k) = \frac{E(k)}{2\pi k^2} \left(  1 + \frac{\left| t -t^\prime \right|  }{t_k} \right)^n e^{- \left | t-t^\prime \right|/t_k} \; ,
	\end{align*}
	The \citeauthor{volk} results use $n=0$, but \citeauthor{mmv} instead use $n=1$ because of the zero slope behavior at $t=t^\prime$. Fits to the results of numerical simulations by \cite{ch03} further validate the choice of $n=1$ over $n=0$.
	} OC07 arrive at an analytic expression for the RMS inertial space velocity of particles suspended in a turbulent medium
\begin{align} \label{RMSturb}
v_{pi,t}^2=v_{t}^2\left(1-\frac{St^2(1-Re_t^{-\frac{1}{2}})}{(St+1)(St+Re_t^{-\frac{1}{2}})}\right) \; .
\end{align}
Relative to the gas, the velocity is:
\begin{align} \label{RMSturb_gas}
v_{pg,t}^2=v_{t}^2\left(\frac{St^2(1-Re_t^{-\frac{1}{2}})}{(St+1)(St+Re_t^{-\frac{1}{2}})}\right) \; .
\end{align}

 Equation \eqref{RMSturb_gas} was originally derived by \cite{ch03} for particles with $St\ll1$. OC07 however, argue from the results of numerical calculations and comparison with simulations, that Equations \eqref{RMSturb} and \eqref{RMSturb_gas} hold to order unity for particles of arbitrary Stokes number. We are therefore justified in applying this equation to determine the RMS turbulent velocity of arbitrary sized accreting particles, with one caveat. We first note that, for large Stokes numbers, we have $v_{pi,t}\rightarrow v_{t}/Re_t^{1/4}$ which cannot be completely correct, as in the limit as $St\rightarrow \infty$ we expect for the particles to be so massive that they are completely unaffected by turbulence, and thus $v_{pi,t}\rightarrow 0$. In addition, we note that, for $St\ll Re_t$, as long as we do not have $St \ll 1$, Equation \eqref{RMSturb} reduces to \eqref{RMSsimple}. Thus we can use \eqref{RMSsimple} to determine the velocity of larger $St$ particles; in practice we use \eqref{RMSsimple} for $St \geq 10$. This form should connect smoothly to the more precise form given in equation \eqref{RMSturb} as long as $Re_t \gg 10$. Using the fiducial values given in Section \ref{params}, the Reynolds number of the turbulence is given by
 \begin{align} \label{eq:rey_fid}
 Re_t=4.07\times10^{10} \alpha \left(\frac{a}{\text{AU}}\right)^{-1} \; .
 \end{align}   
 Thus, unless we are at extremely large orbital separations with weak turbulence, we can model the full range of Stokes numbers in this manner with little error.

\subsection{Combining Velocities and Changing Frames} \label{rey_avg}
We have previously discussed how to calculate the laminar and turbulent components of the particle's velocity. However, in order to calculate the total RMS velocity we need to understand how to combine these two components. Furthermore, calculation of drag forces requires knowledge of the velocity of the particle relative to the gas, whereas calculation of $v_\infty$ will require the velocity of the particle relative to the local Keplerian velocity, since this is the velocity at which small bodies approach the core. We therefore also need to understand how to convert our velocities from one frame to another. The methodology necessary for our calculation is summarized here; for a more detailed derivation see Appendix B.

We let $\boldsymbol{\delta v}$ represent the ``fluctuating" or turbulent component of the gas velocity and $\boldsymbol{\bar{v}}$ represent the laminar component -- i.e. if we write $\boldsymbol{v} = \boldsymbol{\bar{v}} + \boldsymbol{\delta{v}}$ then $\braket{\boldsymbol{v}} = \boldsymbol{\bar{v}}$. We can then show that

\begin{align}
v^2=  \bar{v}^2 + \braket{\delta v^2} \label{turb_lam} \; ,
\end{align}
i.e. the turbulent and laminar components combine in quadrature.  

If the subscripts $p,g,$ and $k$ denoted the velocity of the small bodies, the gas, and the local Keplerian velocity respectively, then 

\begin{align} \label{rel_v_avg}
\avvp = \avvrel + \avvg \; .
\end{align}
So the laminar component of the particle's velocity can be changed from one frame to another in the usual manner (i.e. the $\boldsymbol{\hat{\phi}}$ component is altered by $\eta v_k$ while the $\boldsymbol{\hat{r}}$ component is unchanged) independent of the turbulent velocity.

For the turbulent component of the velocity, one can show
\begin{align} \label{turb_frames}
\braket{\dvg^{2}}=\braket{\dvp^{2}}+\braket{\dvrel^{2}} \; ,
\end{align}
which is given in a number of works (e.g. \citealt{csanady}, \citealt{cuzzi93}, and OC07), and is usually derived by considering the Fourier components of the turbulent velocities in frequency space. An alternate derivation is given in Appendix B.

Neglecting the effects from the fact that the Keplerian velocity is not truly an inertial reference frame (see \citealt{orb} for a discussion of non-inertial effects), Equations \eqref{turb_lam}, \eqref{rel_v_avg} and \eqref{turb_frames} fully specify how to calculate all the velocities relevant to the problem from the input given by Equations \eqref{vr}, \eqref{vphi} and \eqref{RMSturb}.

\subsection{$v_{\rm{shear}}$} \label{v_shear}
Because the Keplerian velocity of the disk varies as $v_k \propto a^{-1/2}$, bodies that are separated substantially in the radial direction will move relative to one another in the azimuthal direction even in the absence of other effects. For our purposes we approximate this ``shear" velocity as

\begin{align} \label{eq:v_shear}
v_{\rm{shear}} = \Omega r \; ,
\end{align}
where $r$ is the separation between the two bodies in the radial direction. For small bodies encountering growing protoplanets we have $r\approx R_{\rm{acc}}$. If this velocity is larger than the drift-dispersion velocity of small bodies, $v_{pk}$, then $v_{\rm{shear}}$ will set $v_\infty$, the velocity at which small bodies encounter cores. That is, we set

\begin{align} \label{eq:v_infty}
v_\infty = \max (v_{pk}, v_{\rm{shear}} ) \; .
\end{align}
We will refer to particles with $v_\infty = v_{pk}$ as begin ``drift-dispersion'' dominated, and particles with $v_\infty = v_{\rm{shear}}$ as being ``shear dominated.''

\subsection{$v_{\rm{enc}}$ and Calculating Work} \label{work}

\begin{figure} [h]
	\centering
	\includegraphics[width=1\linewidth]{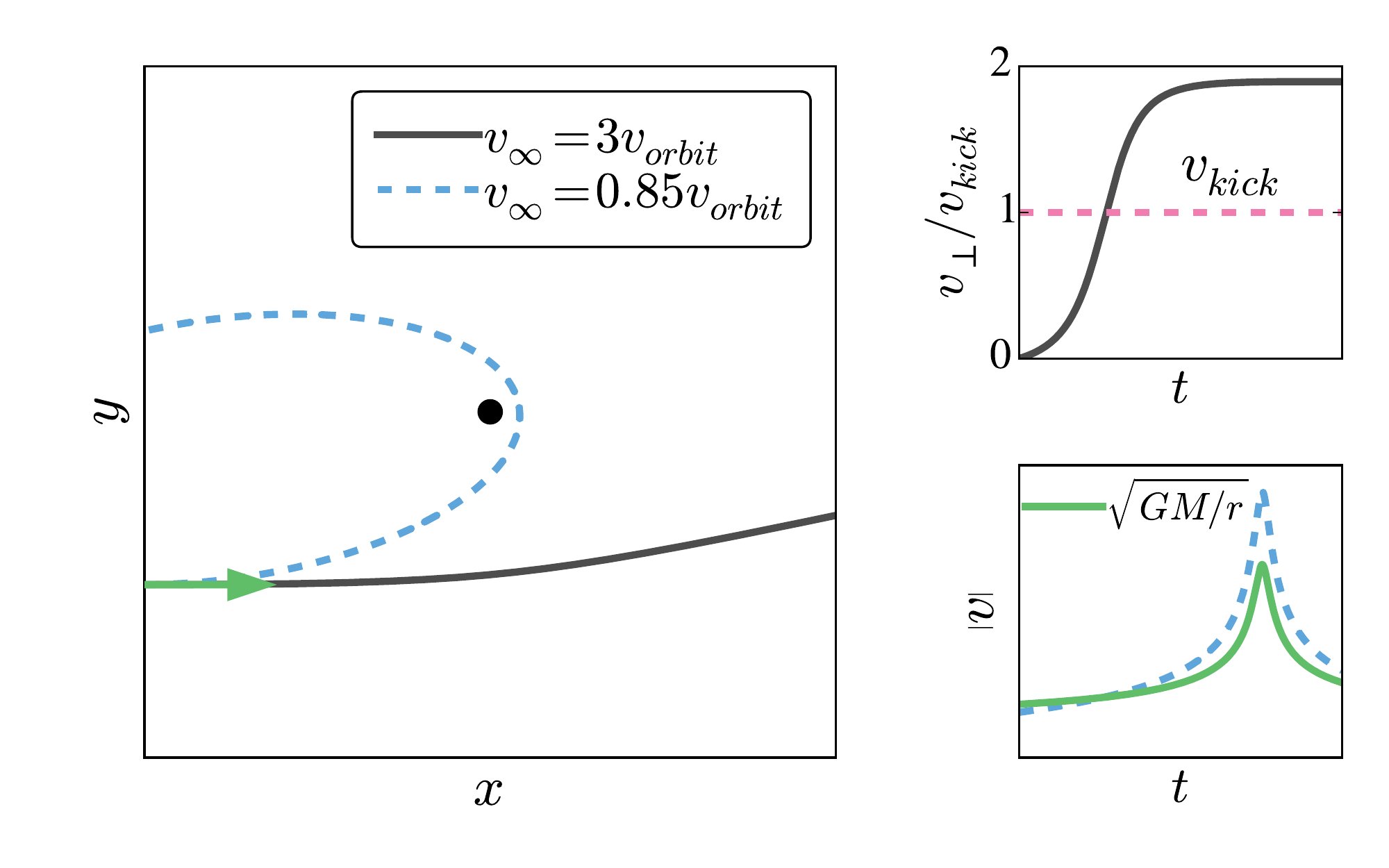}
	\caption{An illustration of the effects of the gravitational interactions between the small body and the core on the velocity between the two bodies during their encounter, $v_{\rm{enc}}$. \textit{Left Panel}: Both particles enter from the lower left, as indicated by the arrow. The upper particle (dashed curve) has $v_{\infty} < v_{\rm{orbit}}$ while the lower particle (solid curve) has $v_\infty > v_{\rm{orbit}}$. The upper, slow moving particle has the direction of its velocity substantially changed during the interaction, while the lower, fast moving particle only receives a small perturbation to its velocity in the direction perpendicular to its motion. \textit{Lower Right Panel}: The slow moving particle is excited to a velocity comparable to $v_{\rm{orbit}}(r) = \sqrt{GM/r}$ during its interaction with the core. \textit{Upper Right Panel}: The fast moving particle receives a small perturbation to its velocity, of order $v_{\rm{kick}} = GM/(R_{\rm{acc}} v_\infty)$. As can be seen in the figure, the actual perturbation is approximately a factor of 2 larger than $v_{\rm{kick}}$. Including this factor of 2 does not have a substantial effect on our results.}
	\label{v_enc_orb}
\end{figure}

It remains to determine the relevant velocity for calculating the drag force. Note that this velocity is relative to the gas (in contrast to $v_\infty$, which is relative to the local Keplerian velocity) and cannot be set by shear, since the gas and the particles will have the same shear velocity.\footnote{Particles that are in the regime depicted in the lower lefthand panel of Figure \ref{fig:en_fig} may experience the full shear velocity over an extended distance as they are turned relative to the core. Because these particles already deplete their kinetic energy this effect does not affect our conclusions about the relative size of $KE$ and $W$.} For particles that approach the core with velocities faster than the local circular orbit velocity about the large body, i.e. for particles with $v_\infty > v_{\rm{orbit}} = \sqrt{G M/R_{\rm{acc}}}$, the dominant effect of the gravitational force from the core will be to give the particle a small ``kick" in the direction perpendicular to its motion. By using the impulse approximation, one can show that the magnitude of the resultant perpendicular velocity is of order $v_{\rm{kick}} = G M / (R_{\rm{acc}} v_\infty)$ (see e.g. \citealt{binney_tremaine}). This effect is illustrated in Figure \ref{v_enc_orb} (solid curve). As also shown in Figure \ref{v_enc_orb}, particles that approach the core with velocities such that $v_\infty < v_{\rm{orbit}}$ will experience a substantial change in their total velocity during their interaction with the core (dashed curve). During the encounter, the magnitude of the particle's velocity when the particle is a distance $r$ from the core is approximately $|v| \approx \sqrt{G M /r}$ (see the lower righthand panel of Figure \ref{v_enc_orb}). If the magnitude of the work done on the particle can be expressed as $W(r) = F_D(r) r$, then for a particle in a linear drag regime we have $W(r) \propto r^{1/2}$, i.e. the majority of the work done during encounter occurs when the particle is at the largest scales. If the particle is in a quadratic (Ram) drag regime, then $F_D$ is independent of $r$, so the work done at all scales is approximately equal. In either regime we are therefore justified in approximating the work done during the encounter as $W = 2 F_D(v_{\rm{enc}}) R_{\rm{acc}}$, where $v_{\rm{enc}} = v_{\rm{orbit}} (R_{\rm{acc}}) = \sqrt{G M / R_{\rm{acc}}}$. 

 Finally, if the particle's drift velocity relative to the gas, $v_{pg}$, is larger than the velocity from the gravitational influence of the core, then $v_{pg}$ will set the velocity during the encounter. 
 
 In summary, we define the relevant velocity for the work calculation, $v_{\rm{enc}}$ to be
\begin{align}
v_{\rm{enc}} = \begin{dcases}
\max\left(v_{\rm{orbit}}, v_{pg}\right), & v_\infty < v_{\rm{orbit}} \\
\max\left(v_{\rm{kick}}, v_{pg}\right), & v_\infty > v_{\rm{orbit}}
\end{dcases} 
\end{align} 
\section{Calculation of Accretion Cross Section} \label{len_scales}
We now turn to how the particle velocities discussed in the previous section are used to calculate the growth timescale of the core. We discuss how the length and width of the accretion cross section, $R_{\rm{acc}}$ and $H_{\rm{acc}}$, are calculated, as well as how the scale height of the small bodies, $H_p$, is determined.

\subsection{Determining $R_{\rm{acc}}$} \label{acc_cross}
The capture radius, $R_{\rm{acc}}$, is the radius interior to which a small body will accrete if certain energy criteria are met, as discussed in Section \ref{energy}. The scale $R_{\rm{acc}}$ is also used to determine both the incoming kinetic energy of a small body and the work done by gas drag during the encounter. In this work we assume that particles that cannot dissipate their kinetic energy at $R_{\rm{acc}}$ will not be able to accrete on a smaller length scale. To see this, we note that, for a general impact parameter $b$, we have
\begin{align}
\frac{KE}{W} = \frac{m v_\infty^2}{4 F_D(v_{\rm{enc}}) b} \; .
\end{align}
If the small body's velocity is shear dominated, so that $v_\infty = v_{\rm{shear}}$, then it can be shown analytically that the only particles that may have $KE/W>1$ are those shearing into $R_H$. In this case particles with substantially smaller impact parameters will not penetrate into $R_H$ due to the nature of three-body trajectories (see e.g. \citealt{hill_enc}). If instead $v_\infty = v_{pk}$, then we have
\begin{align} \label{eq:ke_w_b}
\frac{KE}{W} \propto \frac{1}{F_D(v_{\rm{enc}}) b} \propto \begin{dcases*} \left(v_{\rm{enc}} b \right)^{-1},& \rm{Epstein, Stokes}\\
\left(v_{\rm{enc}}^2 b \right)^{-1},& \rm{Ram}
\end{dcases*}
\end{align}
The maximal possible scaling of $v_{\rm{enc}}$ with impact parameter is $v_{\rm{enc}}=v_{\rm{kick}}\propto b^{-1}$. Therefore, Equation \eqref{eq:ke_w_b} implies that $KE/W$ is always minimized at large $b$ when gas drag is in the Epstein or Stokes regime, and our calculation of the energy criterion at the largest possible accretion scale is sufficient.  The only case for which gas-assisted growth may be ruled out by the energy criterion at a large scale and yet possible at a smaller impact parameter is in the Ram pressure drag regime. We do not include this possibility in our calculation of the accretion rate. This cases holds over small amount of parameter space, as it requires particles to have both $r_s > 9 \lambda/4$ and $Re \gg 1$. Our modeling may rule out accretion of large particles in the inner disk (where the Ram regime is most important) that would be able to accrete at smaller scales.

To determine $R_{\rm{acc}}$, we note that capture is in principle possible either within the planet's atmosphere, where the gas density increases substantially, or within the radius at which a small body can stably orbit the core. Thus the capture radius is set by the relative size of the atmospheric radius, which extends up to $R_b$ (for $R_b < R_H$; see Section \ref{flow_iso_mass} for $R_b > R_H$), and the stability radius, $R_{\rm{stab}}$:
\begin{align}
R_{\rm{acc}} = \max \left( R_{\rm{stab}}, R_b \right) \; .
\end{align}
The stability radius is determined by requiring the that the force on the small body is dominated by the gravity of the core. We consider two other forces that can disrupt accretion, which leads to two different length scales that can set $R_{\rm{stab}}$.
\subsubsection{The Hill Radius} \label{r_h}
The first scale is set by demanding that the small body not be sheared off by the gravity from the host star; this leads to the traditional measure of stability, the Hill radius, where the gravitational acceleration from the core is balanced by the tidal gravity from the star (\citealt{hill}). For $M_* \gg M$, where $M_*$ is the mass of the central star, 
\begin{align} \label{eq:hill}
R_H \approx a\left(\frac{M}{3M_*}\right)^{1/3} \; ,
\end{align}
where $a$ is the semi-major axis of the core's orbit. 

The fact that accretion can occur for impact parameters of order $R_H$ is one of the main enhancements to growth rate that comes from pebble accretion. In the absence of gas, the maximal impact parameter for accretion is smaller than $R_H$ by a factor $(R/R_H)^{1/2}$, where $R$ is the radius of the planet (see Appendix \ref{GF_sec}). Both \cite{lj12} (hereafter LJ12) and \cite{OK10} (hereafter OK10) and ) find regimes where $R_{\rm{acc}} \sim R_H$ in their numerical calculations of pebble accretion. In LJ12's framework, accretion at $R_H$ occurs for core masses greater than $M_t = \sqrt{1/3} v_{\rm{gas}}^3 / (G \Omega)$ and particle sizes  $St\gtrsim 1$.\footnote{LJ12 do not explicitly detail the Stokes number range where accretion at $R_H$ occurs -- because they run their simulations only for particles of size $St = 10^{-2}, 10^{-1}, 10^0$, they merely note that only particles with $St=10^{-2}$ appear to accrete at an impact parameter less than the Hill radius (see below). Following \cite{xbmc_2017}, we take their regime where $R_{\rm{acc}} = R_H$ to hold only for $St \gtrsim 1$.}

In OK10's framework, impact parameters comparable to $R_H$ occur in what they refer to as the  ``three body regime", which occurs for $St>1$, and $St>\zeta_w \equiv \eta v_k/v_H$, where $v_H \equiv R_H \Omega$ is the Hill velocity. In this regime, OK10 give the impact parameter as $b_{3b}/R_H = 1.7 p^{1/2} + 1.0/St$, where $p \equiv R/R_H$. This agrees generally with our results that $R_{\rm{acc}} \approx R_H$ in the Hill accretion regime.
	
The $1/St$ dependence given by OK10 appears to encapsulate the chaotic nature of trajectories for particles that shear into the Hill sphere: often particles will orbit the core many times before exiting the Hill sphere. This effect can cause particles that would not accrete according to our energy criterion to be captured, since in our modeling we only include particles that dissipate their energy over one orbital crossing. If we use the result from \cite{gls_2002}, who studied binary capture of Kuiper Belt objects by dynamical friction, that the probability of capture, $P_{cap}$, for a particle that shears into the Hill radius is equal to the fraction of energy it dissipates over one orbit, then this leads to an effective accretion rate of
\begin{align}
\frac{dM}{dt} \sim R_H v_H \Sigma_p \frac{W}{KE} \; .
\end{align}
For high mass cores, the gravitational force from the core dominates the velocity of the small body during the encounter, and the encounter velocity is $v_{\rm{enc}} \sim v_H$. This in turn implies that $W/KE \propto 1/St$.  For 2D accretion ($H_p < R_{\rm{acc}}$), which is the regime modeled by OK10, our expression therefore agrees with their results. For low mass cores however, the particle-gas relative velocity will instead dominate the velocity during the encounter, which will lead to scaling that differs from $1/St$ in this regime.
	
In summary, particles that have $R_{\rm{stab}} = R_H$ and $v_\infty = v_H$ accrete with timescales given by
\begin{align} \label{eq:r_h_eff}
t_{\rm{grow}}^\prime = \frac{t_{\rm{grow}}}{\min(1,W/KE)} 
\end{align}
where $t_{\rm{grow}}$ is given by Equation \eqref{t_grow_full}. In what follows Equation \eqref{eq:r_h_eff} is used to modify $t_{\rm{grow}}$ for particles that shear into $R_H$.

\subsubsection{The WISH Radius and Shearing Radius} \label{WISH}
In the presence of gas, even if the gravity from the core dominates over the tidal gravity from the star, it is possible that the relative acceleration of the core and the small body will be strong enough to strip the small body away from the core. In this regime, $R_{\rm{stab}}$ is set not by the Hill radius, but by what \citeauthor{pmc11} refer to as the wind-shearing (WISH) radius \citep*{pmc11}. At the WISH radius, the differential acceleration between the large and the small body due to gas drag is balanced by the gravitational acceleration. That is
\begin{align} \label{eq:r_ws}
R_{WS}^\prime=\sqrt{\frac{G(M+m)}{\Delta a_{WS}}} \; ,
\end{align}
where $m$ is the mass of the small body, and $\Delta a_{WS}$ is the differential acceleration between the small and the big body, given by
\begin{align} \label{DiffAcc}
\Delta a _{WS} = \left| \frac{F_D(M)}{M} - \frac{F_D(m)}{m} \right| \; ,
\end{align}
where $F_D$ is the force exerted on the particle due to gas drag. For the mass ratios considered here the first term in Equation \eqref{DiffAcc} is generally negligible. To determine the relevant velocity for calculating the drag force $F_D$, we note that particles that are accreted by the core will have their velocity modified by the core's gravity, increasing their velocity relative to the gas during the encounter. The most extreme case occurs when the velocity of particles match the local Keplerian velocity at which the core is moving. These particles temporarily experience the full gas velocity with respect to the core, which is a combination of the sub-Keplerian and turbulent velocities, as well as the Keplerian shear in the disk. We can encompass this behavior, in the limit $M\gg m$ by writing $R_{WS}^\prime$ as
\begin{align} \label{eq:rws_imp}
R_{WS}^\prime \approx \sqrt{\frac{G M m}{F_d(\max[v_{\rm{gas}},v_{\rm{shear}}])}}
\end{align}
where  $v_{\rm{gas}} = \sqrt{\eta^2 v_k^2 + \alpha c_s^2}$ is the RMS velocity of the nebular gas, and we use the prime to differentiate this radius from the definition of $R_{WS}$ used below. The case $v_{\rm{gas}} > v_{\rm{shear}}$ is discussed in detail by \cite{pmc11}. In the three drag regimes discussed in Section \ref{stoppings}, \cite{pmc11} give $R_{WS}$ as

\begin{align} \label{r_ws_exp} 
R_{WS} = 
\left(G M \rho_s\right)^{1/2} \times \begin{dcases*}
\left(\frac{1}{\rho_g v_{th} v_{\rm{gas}}}\right)^{1/2} r_s^{1/2},& \text{Epstein} \\
\left( \frac{4}{9} \frac{\sigma}{\mu v_{th} v_{\rm{gas}}}\right)^{1/2} r_s, & \text{Stokes} \\
\left( \frac{1}{0.165} \frac{1}{\rho_g v_{\rm{gas}}^2}\right)^{1/2} r_s^{1/2}, & \text{Ram}
\end{dcases*}
\end{align}
where $\rho_s$ is the internal density of the small body. We will use the term WISH radius and the symbol $R_{WS}$ to refer solely to the case $v_{\rm{gas}} > v_{\rm{shear}}$. For the case $v_{\rm{shear}} > v_{\rm{gas}}$ on the other hand, the righthand side of Equation \eqref{eq:rws_imp} now depends on impact parameter. In this regime we will refer to the impact parameter as the ``shearing radius", $R_{\rm{shear}}$. In general, $R_{\rm{shear}}$ is determined by numerically solving the equation
\begin{align}
R_{\rm{shear}} = \sqrt{\frac{G M m}{F_d(R_{\rm{shear}} \Omega)}}
\end{align}
For a particle in a linear drag regime, the particle's Stokes number is independent of velocity, which allows us to solve for $R_{\rm{shear}}$ analytically:
\begin{align} \label{eq:r_shear}
R_{\rm{shear}} = (3 St)^{1/3} R_H \; .
\end{align}

In Section \ref{comp} we will compare our results to OK10 and LJ12 in the laminar regime -- for now we translate their impact parameters into our notation. 
	
The WISH radius is equivalent, in the laminar regime, to the effective accretion radius, $r_d$, used by LJ12, as well as to the settling radius $b_{\rm{set}}$, used by OK10. LJ12 give the effective accretion radius as $r_d = (t_B/t_s)^{-1/2} r_B$. Here $t_B = r_B /  v_\infty$ is the drift time across the ``Bondi radius" -- $r_B = G M /v_\infty^2$.\footnote{As LJ12 note, their definition of the Bondi radius differs from the definition used in other works (including this one), which uses $c_s$ in the place of $v_\infty$.}  As LJ12 note,  $r_d$ is equal to the WISH radius, as can be seen from that fact that
\begin{align} \label{eq:lj12_rd}
r_d = \sqrt{ t_s \eta v_k \frac{G M }{(\eta v_k)^2}} = R_{WS} \; .
\end{align}
LJ12 also give a radius equal to $R_{\rm{shear}}$ (up to a factor of $3^{1/3}$) for particles with $St=10^{-2}$ and $M>M_t$. 

In their ``settling" regime, OK10 determine the impact parameter $b_{\rm{set}} = R_{\rm{acc}}/R_H$ by solving the cubic equation
\begin{align}\label{eq:cubic}
b_{\rm{set}}^3 + \frac{2 \zeta_w}{3} b_{\rm{set}}^2 - 8 St = 0
\end{align}
If the cubic term in the above equation is negligible, then we get the solution 
\begin{align}\label{eq:bset1}
b_{\rm{set}} = \sqrt{12 St / \zeta_w} \;. 
\end{align}
In terms of $\zeta_w$, we can rewrite $R_{WS}$ as
\begin{align}  \label{eq:ok10_r_ws}
R_{WS} = \sqrt{3} \left( \frac{St}{\zeta_w} \right)^{1/2} R_H \; ,
\end{align}
so $b_{\rm{set}} = 2 R_{WS}$ in the laminar regime. If the middle term is negligible on the other hand, the solution is
\begin{align}\label{eq:bset2}
b_{\rm{set}} = 2 St^{1/3}
\end{align}
which is $\sim R_{\rm{shear}}$.  One can verify by inspection that the solution to Equation (\ref{eq:cubic}) is approximately the minimum of Equations (\ref{eq:bset1}) and (\ref{eq:bset2}).

\subsubsection{Combining Length Scales}
Once $R_H$, $R_{WS}$, and $R_{\rm{shear}}$ are calculated, we take the stability radius to simply be the smallest of these three radii:
\begin{align} \label{eq:r_stab}
R_{\rm{stab}} = \min(R_{WS},R_{\rm{shear}},R_H)
\end{align}

We now have the pieces necessary to determine $R_{\rm{acc}}$ -- we first compute $R_{WS}$ and $R_{\rm{shear}}$ -- the minimum of these two length scales determines the scale on which gas drag will pull small bodies off the core before they can be accreted. We then calculate $R_H$ -- the scale on which the stellar gravity disrupts accretion -- and calculate $R_{\rm{stab}}$ by taking in the minimum of these scales. Finally, we take $R_{\rm{acc}}$ to be:
\begin{align} \label{eq:r_acc}
R_{\rm{acc}} = \max(R_{\rm{stab}},R_b) \; .
\end{align} 
Note that accretion within $R_{\rm{acc}}$ only occurs when additional energy criterion are met, as discussed in Section \ref{energy}.

\subsection{Determining $H_{\rm{acc}}$}
We now turn to the height of the accretion rectangle. The height of this rectangle will be the minimum of the dust scale height and the capture radius, since any particle outside the capture radius will not be able to accrete on to the core:
\begin{align} \label{eq:hacc}
H_{\rm{acc}} = \min\left(R_{\rm{acc}},H_{p}\right) \; .
\end{align}
In general, the solid particles will tend to the settle to the midplane of the disk due to the vertical influence of gravity and the lack of pressure support. However, several processes will tend to oppose this settling, giving the particles some finite vertical extent. In our model, we include two different processes that drive particles vertically --  the first is turbulent diffusion due to the isotropy of the small body's turbulent velocities.

For $St>\alpha$, the scale height of the particles due to their interaction with turbulence is given by (\citealt{orb}):
\begin{align} \label{eq:H_p}
H_t = \sqrt{\frac{\alpha}{St}} H_g \; .
\end{align} 
We can understand this expression by separately considering larger particles with $St \gg 1$ and small, well coupled particles with $St \ll 1$. For the larger particles, as discussed in Section \ref{turb_vel}, we can think of the particle's velocity as resulting from the large number of uncorrelated kicks it receives from the eddies. In this case we have $v_{pk,t} \sim v_t/\sqrt{St}$, which gives a scale height of

\begin{align} \label{eq:H_t_oom}
H_t \sim \frac{v_t} {\Omega \sqrt{St}} \sim \sqrt{\frac{\alpha}{St}} H_g  \; ,
\end{align}
since $v_t = \sqrt{\alpha} H_g \Omega$.

As described in the \cite{orb}, for $St \ll 1$ we expect the particles to be well coupled to the gas, and thus expect the particles' diffusion coefficient to be approximately equal to the gas value. We can express the turbulent diffusion coefficient of the gas as $D_g = \alpha_z c_s H_g$. In what follows we take $\alpha_z \approx \alpha$ unless otherwise stated. While $\alpha$ and $\alpha_z$ can differ, since $\alpha$ enters our calculation mainly as a parameterization of the turbulent gas velocity, we are more interested in the relative size of $\alpha_z$ and $v_z$, the vertical turbulent gas velocity. Magnetohydrodynamic simulations find that these two values are similar to order of magnitude (e.g. \citealt{xbmc_2017}). Tightly coupled particles should settle to the midplane at approximately their terminal velocity, so their time to settle to the midplane from a height $z$ is approximately $t_{\rm{set}} \sim z/v_{\rm{term}}$. At terminal velocity the drag force balances the vertical component of gravity, $F_{g,z} \sim (G M m / r^2) (z/r) \sim m \Omega^2 z$. Setting $F_d \sim m \Omega^2 z$ implies that $v_{\rm{term}} \sim St \Omega z$, which gives $t_{\rm{set}} \sim 1 / (St \, \Omega)$. Finally, setting $t_{\rm{set}}$ equal to the particle diffusion time $H_t^2 / D_p$ gives Equation \eqref{eq:H_p}. \cite{carb} derive \eqref{eq:H_p} for $St \gg 1$ by more rigorous means, whereas \cite{dub} do the same for $St \ll 1$.

Note that for $St > \alpha$ we have $H_t > H_g$ which is clearly incorrect; turbulence cannot drive particles to heights larger than the extent of the gas disk. We therefore cap $H_t$ at $H_g$. \cite{dub} derive \eqref{eq:H_p} as the scale height of the dust to gas ratio $\rho_s/\rho_g$, and note that this height $h$ is related to the dust scale height by the equation
\begin{align} \label{H_t_real}
H_t = h \left [1 + \left(\frac{h}{H_g}\right)^2 \right ]^{-1/2} \; ,
\end{align}
which has the expected behavior that $H_t \rightarrow H_g$ as $St \rightarrow \infty$. Equation \eqref{H_t_real} can be used in the place of equation \eqref{eq:H_p}, though this does not substantially change the results. We employ \eqref{eq:H_p} in order to maintain relative simplicity in our analytic results.

Turbulence's modification to the particle scale height is discussed in a number of works, including OK10 and LJ12, though it is often not explicitly included in calculations of growth rate; instead, authors generally note that for $H_t > R_{\rm{acc}}$, the growth timescale is increased by a factor of $H_t/R_{\rm{acc}}$. We also note that equation \eqref{H_t_real} can be written as
\begin{align} \label{H_t_OK}
H_t = H_g\sqrt{\frac{\alpha}{\alpha + St}} \; ,
\end{align} 
which is employed by \cite{OK12} for the dust scale height in the presence of turbulence. Equation \eqref{H_t_OK}  agrees with \cite{orb}, Equation (28) without their factor of $\xi^{-1/2}$, where $\xi \equiv 1+St/\left(1+St\right)$  for the values chosen in this paper. 

Even in the absence of strong turbulence, the Kelvin-Helmholtz shear instability prevents small bodies from settling too close to the midplane. For small particles, which are well-coupled to the gas flow, this leads to a scale height of
\begin{align} \label{HKH}
H_{KH}^{\prime} = \frac{H_g^2}{a} = \frac{2\eta v_k}{\Omega}  \; ,
\end{align} 
(see for example \citealt{kh}). In order to extend this scale height to include larger particles, we assume that even when large bodies dominate the mass distribution, a population of small bodies exists that is substantial enough to drive the Kelvin-Helmholtz shear instability, and therefore turbulence, close to the midplane. We expect the turbulent velocity in this case to be comparable to the vertical shear rate, which is $\sim \eta v_k$. In this case, we can make arguments analogous to the ones preceding Equation \eqref{eq:H_t_oom}, which leads to a scale height $H_p \sim \eta v_k / (\Omega \sqrt{St})$. Thus, we can describe $H_{KH}$ over the full range of particle sizes using the expression
\begin{align} \label{eq:HKH_real}
H_{KH} = \frac{H_g^2}{a}\min(1,St^{-1/2}) \; .
\end{align} 
If only large bodies are present, they may need to settle further in order to excite the Kelvin-Helmholtz instability, which would change the dependence on $St$. We leave a self-consistent calculation to another work. Furthermore, by employing this expression for larger particle sizes, we are neglecting the possibility of dynamical stirring of the large bodies. Mutual scatterings and/or the gravitational force of the core can drive small particles vertically, and may be more important than interactions with the gas for determining the scale height of larger particles, which are decoupled from the gas.

Finally, we take the solid scale height to simply be the maximum of these two heights
\begin{align} \label{eq:H_p_real}
H_p = \max\left(H_t,H_{KH}\right) \; .
\end{align}
The possibility of 3D accretion ($H_p > R_{\rm{acc}}$) in the non-turbulent regime is generally neglected in works on pebble accretion, but we find that it has non-trivial effects on the growth timescale.

\begin{center}
\renewcommand{\arraystretch}{2.25}
\begin{deluxetable*}{cccc}
		\tabletypesize{\footnotesize}
		\tablecaption{Summary of Symbols Used in Text}
		\tablehead{\colhead{Parameter}& \colhead{Symbol}& \colhead{Formula/Value}&\colhead{Equation Number/Section}}
		\startdata
		Mass of large body & $M$ & -- & --\\
		Mass of central star & $M_*$ & -- & --\\
		Keplerian orbital frequency & $\Omega$ & $\sqrt{\frac{G M_*}{a^3}}$& --\\
		Radius of small bodies & $r_s$ & -- & -- \\
		Density of solid bodies & $\rho_s$ & $2\,\text{g}\,\text{cm}^{-3}$ & Section \ref{params} \\
		Mass of small bodies & $m$ & $\frac{4}{3} \pi \rho_s r_s^3$& Section \ref{t_grow} \\
		Stopping time of small bodies & $t_s$ & $\frac{m v_{\rm{rel}}}{F_D}$& Equations \eqref{eq:t_s_full} and \eqref{t_s_exp} \\
		Small body Stokes number/dimensionless stopping time & $St$ & $t_s \Omega$ & Sections \ref{lam_vel} and \ref{turb_vel} \\
		Shakura-Sunyaev $\alpha$ parameter& $\alpha$ & --& Section \ref{turb_vel}\\
		Large body Hill radius & $R_H$ & $a \left(\frac{M}{3 M_*}\right)^{1/3}$& Equation \eqref{eq:hill} \\
		Wind-shearing (WISH) radius & $R_{WS}$ & $\approx \sqrt{\frac{G M t_s}{v_{\rm{gas}}}}$& Equation \eqref{eq:r_ws} \\
        Shearing radius & $R_{\rm{shear}}$ & $\approx R_H \left(3 St \right)^{1/3}$& Equation \eqref{eq:r_shear} \\
		Large body Bondi radius & $R_{b}$& $\frac{G M}{c_s^2}$&  Section \ref{energy} \\
		Largest radius for stable orbits about large body & $R_{\rm{stab}}$ & $\min(R_{WS},R_{\rm{shear}},R_H)$& Equation \eqref{eq:r_stab} \\ 
		Extent of large body's atmosphere& $R_{\rm{atm}}$ & $\min(R_b,R_H)$ & Section \ref{flow_iso_mass}\\
		Impact parameter for accretion & $R_{\rm{acc}}$ & $\max(R_{\rm{stab}},R_{\rm{atm}})$ & Equation \eqref{eq:r_acc} \\
		Scale height of small bodies due to turbulence & $H_t$ & $\sqrt{\frac{\alpha}{St}} H_g$& Equation \eqref{eq:H_p} \\
		Scale height due to Kelvin-Helmholtz shear instability & $H_{KH}$ & $\frac{H_g^2}{a} \min(1,St^{-1/2})$& Equation \eqref{eq:HKH_real} \\
		Scale height of small bodies & $H_p$& $\max\left(H_t,H_{KH}\right)$& Equation \eqref{eq:H_p_real}\\
		Vertical extent of accretion cross section& $H_{\rm{acc}}$& $\min\left(R_{\rm{acc}},H_{p}\right)$& Equation \eqref{eq:hacc}\\
		Velocity of small bodies relative to nebular gas & $v_{pg}$& See text&  Equations \eqref{eq:app_v_pg_ell}, \eqref{eq:app_v_pg_turb}, and \eqref{eq:app_comb_pg}\\
		Velocity of small bodies relative to Keplerian & $v_{pk}$& See text&  Equations \eqref{eq:app_v_pk_ell}, \eqref{eq:app_v_pk_turb}, and \eqref{eq:app_comb_pk}\\
		Keplerian shear velocity& $v_{\rm{shear}}$ & $R_{\rm{acc}} \Omega$& Equation \eqref{v_shear} \\
		Approach velocity of small bodies& $v_\infty$& $\max(v_{pk},v_{\rm{shear}})$& Equation \eqref{eq:v_infty} \\
		Orbital velocity about large body& $v_{\rm{orbit}}$& $\sqrt{\frac{G M}{R_{\rm{acc}}}}$& Section \ref{work}  \\
		Impulse approximation velocity perturbation by large body& $v_{\rm{kick}}$& $\frac{G M }{R_{\rm{acc}} v_\infty}$& Section \ref{work}\\
		Velocity of small body during encounter with large body& $v_{\rm{enc}}$& See text& Equation \eqref{eq:v_enc}
		\label{tab:symbol_table}
\end{deluxetable*}
\end{center}

\section{Overview of Results} \label{results}

Having discussed in detail how our model operates, we can now present the output from our model. For reference, we also provide a table detailing the symbols we use (Table \ref{tab:symbol_table}). In this section we discuss our results broadly in order to introduce how gas-assisted growth operates in the presence of turbulence.

\subsection{Model Parameters} \label{params}
We begin by providing the values of the fiducial parameters used for describing the protoplanetary disk, which are necessary to provide concrete numerical results. A discussion of the effects of varying some of these parameters is given in Section \ref{param_space}.

We assume a surface density profile of
\begin{align} \label{eq:surf_dens}
\Sigma(a)=500\left(\frac{a}{\text{AU}}\right)^{-1} \text{g}\,\text{cm}^{-2} \; ,
\end{align}
where $a$ is the semi-major axis in the disk. This profile is chosen to be consistent with measurements of the surface density in solids of size 0.1-1 mm, taken from sub-mm observations of protoplanetary disks (\citealt{andrews_09}, \citealt{andrews}), which probe these particle sizes. In order to calculate the surface density profile of solid bodies, $\Sigma_p$, we assume a constant dust to gas ratio of $f_s \equiv \Sigma_p / \Sigma = 1/100$. The semi-major axis dependence of the temperature profile is taken from \cite{cy10}
\begin{align}
T=T_0\left(\frac{a}{\text{AU}}\right)^{-3/7} \; .
\end{align}
For most purposes we will take the prefactor to be $T_0 = 200 \, \text{K}$, which would place the Earth outside of the water ice line during its formation (e.g. \citealt{pms_2017}). We also note this prefactor is consistent with assuming the disk is heated primarily by irradiation from a central star with luminosity $L_* \sim 3 L_\odot$ (e.g. \citealt{igm16}), which is appropriate for a solar mass star of age 1 Myr (\citealt{tpd_2011}). The effects of varying $T_0$ are discussed in Section \ref{T_0}.

From here we can then calculate the isothermal sound speed and scale height of the disk in the usual manner: $c_{s}=\sqrt{kT/\mu}$ and $H_g=c_s/\Omega$. Here $\mu$ is the mean molecular weight of particles in the disk; we take $\mu = 2.35 m_H \approx 3.93\times10^{-24}$ g, which assumes a disk composed of 70\% $\text{H}_2$ and 30\% $\text{He}$ by mass. We can also calculate the gas density $\rho_g = \Sigma /(2 H)$, and the mean free path of gas particles, $\lambda = \mu / (\rho_g \sigma)$, where $\sigma$ is the neutral collision cross section, which we take to be $\sigma \sim \left( 3 \text{\AA} \right)^2 \sim 10^{-15}\,\text{cm}^2$. In order convert from Stokes number, which is the relevant parameter for calculating $t_{\rm{grow}}$, to physical size, we need to specify the internal density of the pebbles, $\rho_s$. We take this density to be $\rho_s = 2 \, \text{g}/\text{cm}^3$ unless stated otherwise, which is appropriate for rocky bodies. Note however, that the solids in protoplanetary disks may be fluffy, i.e. their densities may be low, which can have important ramifications for planet formation (see e.g. \citealt{ktow_2013}). While we consider the effect of varying $M_*$ in Section \ref{vary_M_*}, unless otherwise noted all values quoted in the paper are for a solar mass star, $M_* = M_\odot$. As noted in Section \ref{energy}, the radius of the large body needs to be $\gtrsim 10 \, \text{km}$, which corresponds to $M\sim 10^{-9}\,M_\oplus$ for a density of $2 \, \text{g}/\text{cm}^3$. In practice we do not consider cores that are close to this lower limit.

A summary of our fiducial disk parameters is given in Table \ref{tab:param_table}.

\begin{center}
	\renewcommand{\arraystretch}{2.25}
\begin{deluxetable*}{ccc}
	\tabletypesize{\footnotesize}
	\tablecaption{Fiducial Disk Parameters}
	\tablehead{\colhead{Parameter}& \colhead{Symbol}& \colhead{Formula/Value \tablenotemark{a}  }}
	\startdata
	Solid to gas ratio & $f_s$ & 0.01 \\
	Mean molecular weight of gas particles & $\mu$ & $2.35 \, m_H \approx 3.93\times 10^{-24} \, \text{g}$ \\
	Neutral collison cross section & $\sigma$ & $10^{-15} \, \text{cm}^2$\\
	Gas surface density & $\Sigma $ & $\Sigma(a)=500\left(\frac{a}{\text{AU}}\right)^{-1} \, \text{g cm}^{-2} $\\
	Protoplanetary disk temperature & $T$ & $T(a) = 200\left(\frac{a}{\text{AU}}\right)^{-3/7} \, \text{K}$\\
	Isothermal sound speed & $c_s$ & $c_s = \sqrt{\frac{k T}{\mu}}$\\
	Average thermal velocity & $v_{th}$ & $v_{th} = \sqrt{\frac{8}{\pi}} c_s$\\
	Gas scale height & $H_g$ & $H_g = \frac{c_s}{\Omega}$\\
	Gas density & $ \rho_g$ & $\rho_g = \frac{\Sigma}{2 H_g}$\\
	Gas mean free path & $ \lambda$ & $\lambda = \frac{\mu}{\rho_g \sigma}$
	\vspace{1mm}
	\tablenotetext{a}{A value indicates that the quantity is a constant input parameter to the code, whereas a formula indicates that the quantity is calculated in the prescribed manner from input parameters and constants \label{tab:param_table}}
\end{deluxetable*}
\end{center}

\subsection{Basic Model Output} \label{basic}

\begin{figure} [h]
	\centering
	\includegraphics[width=1.1\linewidth]{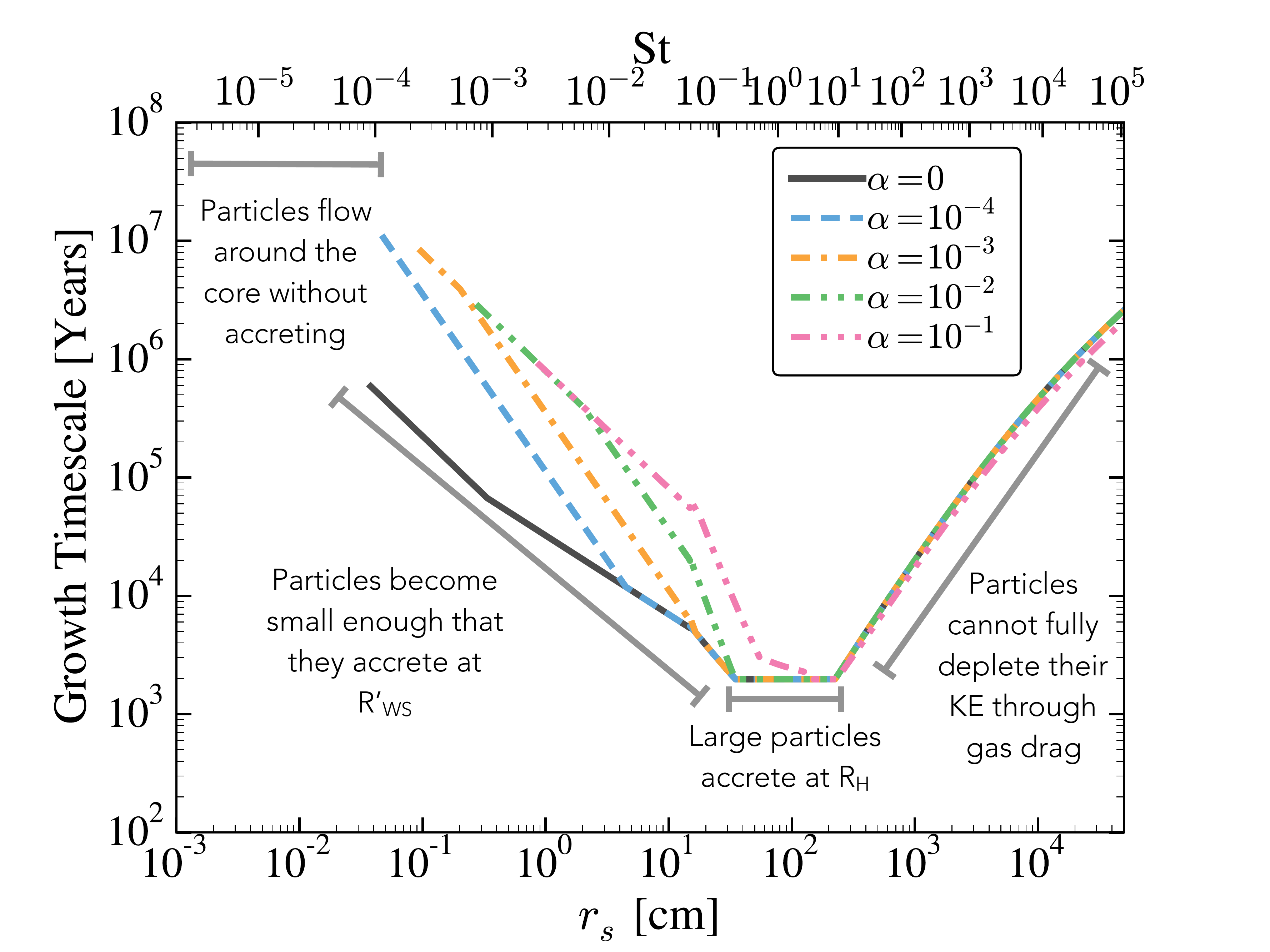}
	\caption{A plot of the timescale for growth of the protoplanetary core as a function of small body radius, plotted for various values of $\alpha$, which measures the strength of turbulence. The values shown are for $a=1 \, \text{AU}$ and $M = 10^{-1} M_\oplus$. Also shown are the Stokes numbers for the plotted values of small body radius. Note that in a non-linear drag regime there is no longer a velocity independent relationship between radius and Stokes number; in this case the given values are calculated for drift velocities in the laminar ($\alpha = 0 $) case.  The timescale is set to $\infty$ for particles that are unable to accrete according to the energy criteria discussed in section \ref{energy}, i.e. the range of radii plotted shows the range of particle sizes the core is able to accrete via pebble accretion. }
	\label{RvsT}
\end{figure}

For the purposes of understanding the broad features of pebble accretion, we present a typical output of our model in Figure \ref{RvsT}. Figure \ref{RvsT} plots the growth timescale (Equation \ref{t_grow_full}) as a function of the small body radius, $r_s$, for a core of mass $M = 10^{-1} M_\oplus$ at $a = 1\, \text{AU}$. The growth timescales for several $\alpha$ values, ranging from purely laminar flow to extremely strong turbulence, are shown. The timescale is set to $\infty$ for particles that do not satisfy the energy criteria discussed in Section \ref{energy}, so the plotted range for each value of $\alpha$ indicates the range of small body radii the core can accrete via pebble accretion. The Stokes numbers corresponding to the given values of radius are shown for reference (the laminar drift velocity is used to calculate $St$ for particles not in a linear drag regime).

We begin by examining the laminar case, where $\alpha = 0$. For large values of $r_s$, particles shear into the Hill radius.  However, growth is slow because particles are large and cannot fully dissipate their kinetic energy relative to the gas, which increases $t_{\rm{grow}}$ by a factor $KE/W$. Once particles are small enough that $KE \sim W$ they can accrete on rapid timescales. In terms of the small body's Stokes number, we can in general write the criterion for particles to be able to deplete their kinetic energy as 
\begin{align} \label{eq:st_en_criterion}
St < \frac{4 v_{\rm{enc}} R_{\rm{acc}} \Omega}{v_\infty^2} \; .
\end{align}
For particles of this size, gravity dominates over gas drag effects -- therefore $R_H$ determines $R_{\rm{acc}}$, as opposed to $R_{WS}$. The large size of these particles also means that their velocity dispersion and drift velocity are low, so the approach velocity is set by shear, i.e. $v_\infty = v_H$. Using these values to calculate $v_{\rm{enc}}$ and plugging the result into Equation \eqref{eq:st_en_criterion} gives
\begin{align} \label{eq:st_crit_num}
St < 4\sqrt{3} \; ,
\end{align}
which will be a general upper limit on Stokes numbers that the core is able to accrete rapidly. Note that if the small body is not in a linear drag regime then the Stokes number is no longer independent of velocity. In this case the Stokes number calculated for the small body moving through the disk (which is relative to $v_{pk}$) will not necessarily be the same as the Stokes number during the encounter with the core (which is relative to $v_{\rm{enc}}$).

Growth for $R_{\rm{acc}} = R_H$ and $v_\infty = v_H$ is extremely rapid in comparison to the planetesimal accretion timescale. As small body size continues to decrease however, eventually gas drag considerations overtake three body effects, and the shear radius shrinks below the Hill radius. This causes the growth timescale to increase for  $r_s \lesssim 20 \, \text{cm}$. From this point the shear radius continues to shrink with small body size, slowing accretion. The first kink in the slope around $r_s \sim 15$ cm comes from small bodies changing from the fluid regime to the diffuse regime, while the second kink around $r_s \sim 0.2$ cm stems from $R_{\rm{shear}}$ shrinking below $H_p$, which dilutes the density of small bodies, in turn slowing accretion. Eventually, the shear radius shrinks below the Bondi radius, which signals the point at which particles are so small that they will couple to the gas and flow around $R_b$. This causes growth to cut off for small values of $r_s$.

As increasingly strong turbulence is taken into account, the picture becomes more complicated. For the $\alpha = 10^{-1}$ case, growth for large particle sizes is quite similar to the laminar case, since the growth parameters are solely determined by the core's mass for high $St$. As the particle size shrinks, we still find a small range of radii where growth occurs at $R_{\rm{acc}} = R_H$, $v_\infty = v_H$. Rather rapidly however, several new features emerge that were not previously present. First, the strong turbulence greatly increases the particle scale height. As particle size decreases this scale height increases until it becomes larger than $R_H$, slowing down growth. As particle size decreases, for $St<1$ the drift-dispersion velocity of particles increases as they couple more strongly to the gas. Due to the rapid turbulent velocity of the gas, $v_{pk}$ actually overtakes $v_H$ around $r_s = 100 \, \text{cm}$, slightly mitigating the increase in growth timescale. The WISH radius is also decreased due to the strong turbulence, which makes it easier to pull particles off of the core. Because of this the radius at which $R_{WS} < R_H$ is much higher than in the laminar case, and around $r_s = 50 \, \text{cm}$ growth is again inhibited as the WISH radius begins determining $R_{\rm{acc}}$. Another kink in the slope appears around $r_s = 20 \, \text{cm}$ as the scale height of particles becomes so large that $H_p = H_g$, at which point the particle scale height stops increasing. The final kink occurs at $r_s \approx 15 \, \text{cm}$, and is again due to the particles switching from the fluid to the diffuse regime. Growth again stops when $R_{WS}  < R_b$, but because of the decreased values of WISH radius relative to the laminar case, this occurs at a larger value of $r_s$. Similar features occur for the smaller values of $\alpha$, but due to the weaker turbulence these features occur at smaller values of $r_s$. The only exception is the domination of the turbulent RMS velocity over $v_H$, which for these values of parameters only occurs for the $\alpha = 10^{-1}$ case. As can be seen in the figure, for lower values of $\alpha$ wider ranges of small body sizes can accrete at what appears to be a minimal value of timescale, which we discuss in the next section. 

\vspace{4mm}

\subsection{$t_{\rm{Hill}}$} \label{min}

The floor to the growth timescale in Figure \ref{RvsT} corresponds to a minimum value of $t_{\rm{grow}}$, which cores reach when they can accrete over their entire Hill sphere. The appearance of this minimum growth timescale, or maximal growth rate, is a consistent thread throughout the parameter space probed by our model. 

We refer to this timescale as the ``Hill timescale," or $t_{\rm{Hill}}$. This timescale is reached when all aspects of the accretion processes are set by the core's gravity -- that is when we have $R_{\rm{acc}} = R_H>H_p$, and $v_\infty = v_H$. From Equations \eqref{t_grow_full} and \eqref{eq:hacc} it is easy to see that if $H_p < R_{\rm{acc}}$ then $t_{\rm{grow}}$ is independent of $H_p$. Using these values for our timescale parameters in \eqref{t_grow_full} gives the value of $t_{\rm{Hill}}$ as
\begin{align}
t_{\rm{Hill}} = \frac{M}{2 f_s \Sigma R_H^2 \Omega} \; .
\end{align}
A similar regime is identified by \cite{lj12}, Section 3.2, who refer to it as ``Hill Accretion." They however, only use this term to differentiate between the drift dominated and shear dominated regimes, which roughly corresponds to where $R_{\rm{acc}} = R_H$ for their model. Within their Hill regime, the growth timescale is not necessarily equal to $t_{\rm{Hill}}$. See Section \ref{comp} for a more in depth comparison of the two models.

In general, $t_{\rm{Hill}}$ represents an extremely rapid timescale for growth. Scaled to fiducial values, the minimum timescale can be expressed as
\begin{align} \label{eq:t_min_fid}
t_{\rm{Hill}} \approx 4300 \left( \frac{a}{\text{AU}}\right)^{1/2} \left( \frac{M}{M_\oplus} \right)^{1/3} \text{years} \; .
\end{align}
This represents a significant enhancement over the timescale achievable via canonical core accretion, as can be seen by comparing equation \eqref{eq:t_min_fid} to \eqref{eqn:t_GF_fid}.

This is the substantial decrease in growth timescale promised by gas-assisted growth. The source of this decrease can be mostly attributed to an increase in $R_{\rm{acc}}$. As previously stated, the minimum timescale indicates accretion in a regime such that essentially all particles that encounter the Hill radius of the growing planet are accreted. In contrast, in the absence of gas the trajectories of particles that enter the Hill radius are chaotic, and whether a collision occurs is extremely sensitive to the value of the small body's impact parameter. See for example \cite{hill_enc}. For this reason encounters with the core in this regime are often treated probabilistically, with the collision rate of small bodies equal to the product of the rate at which small bodies encounter radius and the probability that a particle inside the Hill radius will accrete. Performing such a calculation leads to an effective impact parameter for accretion in the so called ``3-body regime" that is to order of magnitude equal to the geometric mean of the Hill radius and the planetary radius -  $b\sim \sqrt{R R_H}$. See GLS and \cite{OK10} Equation (5). From comparison with Equation \eqref{eqn:t_GF} we see that the maximal increase in accretion rate that can be provided by pebble accretion is of order $R_H/R$.

To see why $t_{\rm{Hill}}$ is in general the fastest rate of growth possible, we first note that since $R_{\rm{stab}} = \min \left( R_H,R_{WS},R_{\rm{shear}} \right)$, it is clear that the maximal accretion cross section is of order $\sigma_{gas,max} \sim R_H^2$. Since turbulence can increase the RMS velocity of small bodies to values substantially larger than $v_H$ however, it is conceivable that strong turbulence could increase the growth rate by increasing $v_\infty$, i.e. by increase the rate at which small bodies encounter the core. However, increasing the velocity of small bodies drives them vertically as well as horizontally, which decreases the density of small bodies and slows accretion. This trade off rules out accretion on timescales faster than $t_{\rm{Hill}}$, which we demonstrate to order of magnitude below. We first note that
\begin{align}
H_{KH} &= \frac{2 \eta v_k}{\Omega} \max(1,St^{-1/2}) \gtrsim \frac{v_{pk,\ell}}{\Omega} \label{KH_vdisp} \; ,
\\H_t  &\approx \frac{v_{pk,t}}{\Omega} \label{Ht_vdisp} \; ,
\end{align}
we also make the approximation that
\begin{align}
v_{pk} &= \max \left( v_{pk,\ell}, v_{pk,t} \right) \label{vdisp_approx} \; .
\end{align}
rather than the quadrature of $v_{pk,\ell}$ and $v_{pk,t}$. We separately consider ``2D accretion," defined by $H_p < R_{\rm{acc}}$, and ``3D accretion," where $H_p > R_{\rm{acc}}$.
 
For 2D accretion, we can write $t_{\rm{grow}}$ as
\begin{align}
t_{\rm{grow}} &= \frac{M}{2 \Sigma_p R_H v_\infty}
\\&=t_{\rm{Hill}} \left(\frac{R_H \Omega}{v_\infty} \right) \;,
\end{align}
where $t_{\rm{Hill}} \equiv M/\left(2 \Sigma_p R_H^2 \Omega \right)$ and $\Sigma_p \equiv f_s \Sigma$. Thus $t_{\rm{grow}} < t_{\rm{Hill}}$ requires $v_\infty > v_H$ while still having $R_H > H_p$. But from Equations \eqref{KH_vdisp} - \eqref{vdisp_approx} we see that $R_H > H_p$ implies that $v_H > v_{pk} $, so $v_\infty = v_H$. Thus the timescale for growth cannot drop below $t_{\rm{Hill}}$ for 2D accretion.

We now consider the 3D accretion regime when the Hill radius sets the geometry, that is $H_p > R_{\rm{acc}} = R_H$. In this case, we can write the growth timescale as
\begin{align}
t_{\rm{grow}} = \frac{M H_p}{2 \Sigma_p R_H^2 v_\infty} = t_{\rm{Hill}} \left( \frac{H_p \Omega}{v_\infty} \right )
\end{align}
If the particle is shear dominated, then $t_{\rm{grow}} > t_{\rm{Hill}}$ since $R_H < H_p$. Furthermore, since Equations \eqref{KH_vdisp} - \eqref{vdisp_approx} imply that $H_p \Omega \gtrsim v_{pk}$, if the particle is drift-dispersion dominated we again have $t_{\rm{grow}} > t_{\rm{Hill}}$. Thus, in the 3D accretion regime $t_{\rm{Hill}}$ also represents the smallest possible timescale.

\subsection{Gas Mitigated Growth}

While growth at $t_{\rm{Hill}}$ is extremely rapid, it is important to realize that in practice the minimum timescale can only be achieved for a certain range of small body sizes. Maximal accretion efficiency occurs for particles where the dominant effect of the gas is to dissipate particles' kinetic energy as they encounter the growing core, and all other aspects of the accretion process are set by the gravitational influence of the core. When the mass of small or large bodies is low enough that gas determines aspects of the accretion process, the timescale for growth will increase, in many cases quite substantially.  

\begin{figure*} [h]
	\centering
	\includegraphics[width=\linewidth]{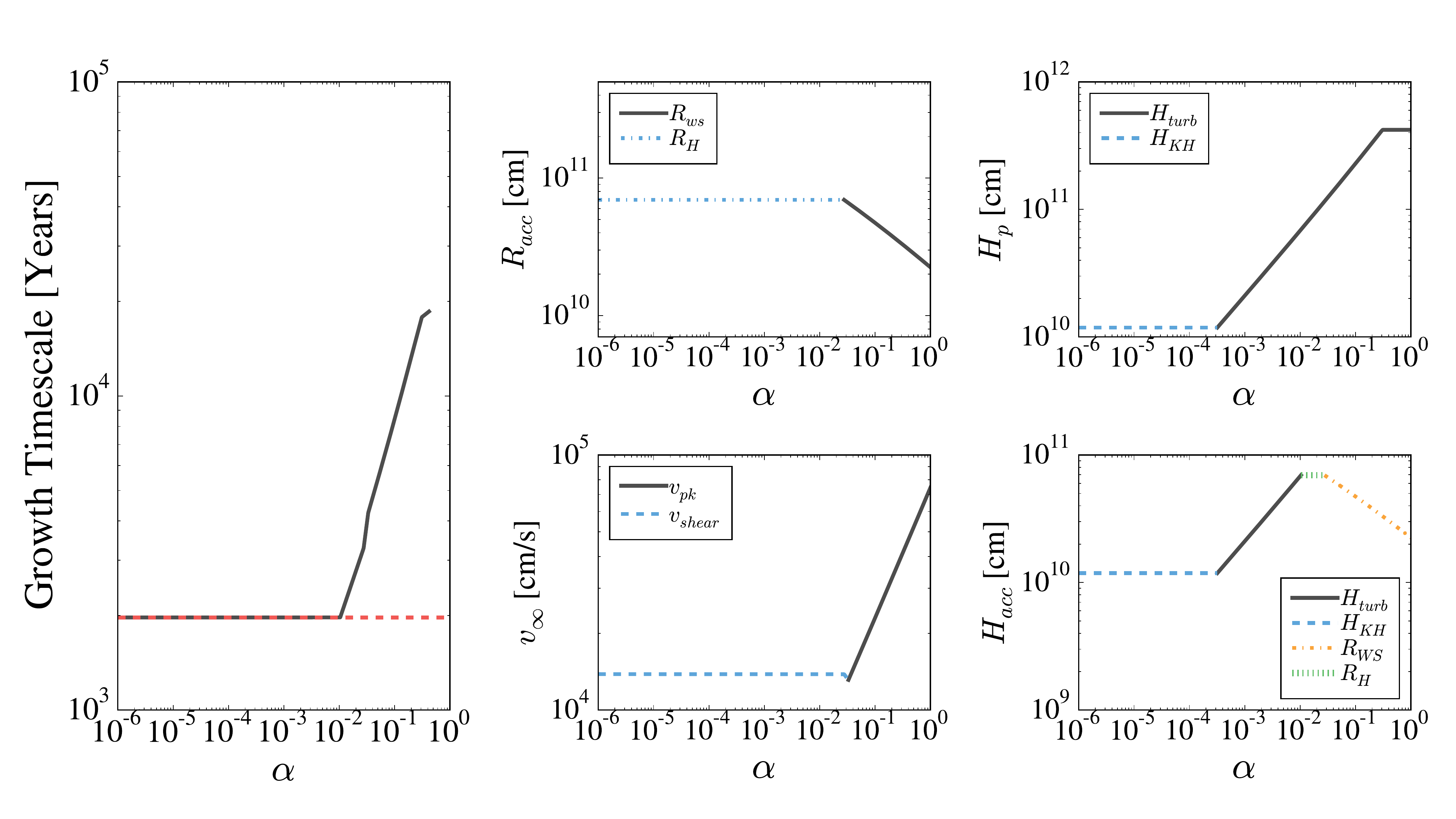}
	\caption{A plot of the timescale for growth of the protoplanetary core as a function of $\alpha$, along with plots of the four quantities used to determine $t_{\rm{grow}}$ (c.f. Equation \ref{t_grow_full}), which are also plotted as a function of $\alpha$. The values shown are for $a=1 \, \text{AU}, M_* = M_\odot$, $M = 10^{-1} M_\oplus$ and $r_s = 35\,\text{cm}$. The minimum timescale $t_{\rm{Hill}}$ is also shown (red dashed line).}
	\label{detail_avsT}
\end{figure*}
	
To examine in more detail what causes the timescale to increase above $t_{\rm{Hill}}$ as the interaction between the gas and the small bodies increases in importance, we plot the four quantities that go into the calculation of $t_{\rm{grow}}$ along with timescale itself in Figure \ref{detail_avsT}. As $\alpha$ increases, the gas-particle interaction will dominate over the gravitational interaction between the small body and the core. For low values of $\alpha$ we have $R_{WS}>R_H>H_{p}$ and $v_\infty = v_{\rm{shear}}$ (as expected for $R_H > H_t$), so the particle is in the 2D accretion regime and is able to accrete at $t_{\rm{Hill}}$. As $\alpha$ increases, the particle scale height begins to increase, until eventually $H_p > R_H$. At this point accretion becomes less efficient, and the timescale correspondingly increases. There are several other kinks in the timescale graph, which are caused by, in order of increasing $\alpha$:
\begin{enumerate}
	\item $R_{WS}$ decreasing to the point that $R_{WS} < R_{H}$. 
	\item $v_\infty$ being set by dispersion instead of shear, i.e. switching to the drift-dispersion dominated regime.
	\item Reaching $H_p = H_g$, at which point $H_p$ stops increasing.
\end{enumerate}

For large $\alpha$ values, the growth timescale is almost an order of magnitude larger than $t_{\rm{Hill}}$, indicating a substantial slow down in growth rate. 

\subsection{Correspondence with Previous Models of Pebble Accretion} \label{comp}

In this Section we compare our model to the analytic results from OK10 and \cite{jl_2017} (hereafter JL17) \footnote{We compare to JL17 as opposed to LJ12 because the JL17 expressions are extensions of LJ12's analytic model, but their framework is more clearly laid out. In generating the comparison we use their Equations (30) and (31) to determine $R_{\rm{acc}}$ for $M<M_t$ and $M > M_t$ respectively.} in the laminar ($\alpha = 0$ regime) as well as the extension of the OK10 model by \cite{c_2014} to include turbulence. We discuss why our framework is more useful for incorporating the results of turbulence.

\begin{figure*} [h]
		\centering
		\includegraphics[width=\linewidth]{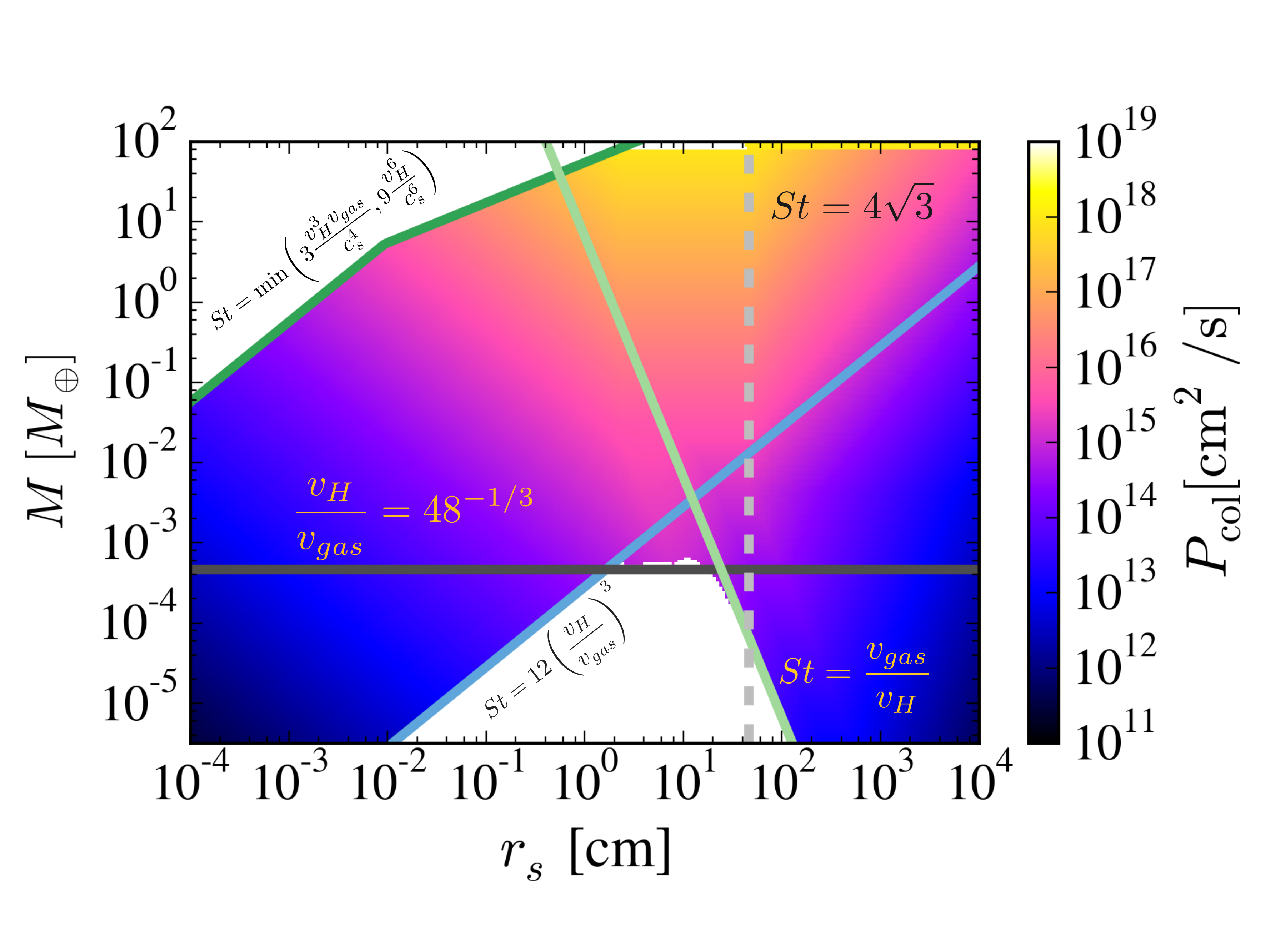}
		\caption{A plot of $P_{col} \equiv 2 R_{\rm{acc}} v_\infty$ as a function of $r_s$ and $M$ for a core at $a = $ 30 AU with $\alpha = 0$. Several important features along with their formulae are marked on the plot. See the text for a discussion of what causes these features to emerge.}
		\label{fig:detail_comp}
	\end{figure*}

In order to facilitate comparison between the various models, we begin by highlighting some important features that emerge in our modeling of the growth rate. Figure \ref{fig:detail_comp} shows a plot of $P_{col} \equiv 2 R_{\rm{acc}} v_\infty$ for $\alpha = 0$ in our model as a function of $r_s$ and $M$. Several features, along with their analytic formulae, are marked on the plot. We discuss these features in detail below. Note that in calculating the analytic expressions for these features we implicitly assume a linear drag regime so that we can employ the analytic expressions for $R_{WS}$ and $R_{\rm{shear}}$ (see Appendix \ref{app:sum}).
	
In the upper lefthand corner of the plot, we see that accretion is cut off once $R_b > R_{\rm{stab}}$. At low core mass, $R_{\rm{stab}} = R_{WS}$ when this cutoff occurs, but at high core masses $R_{\rm{shear}}$ dominates, which causes the kink in the slope seen in the figure. The combination of these two scales causes a cutoff in accretion for
\begin{align} \label{eq:st_low}
St < \min \left(3 \frac{v_H^3 v_{\rm{gas}}}{c_s^4},9 \frac{v_H^6}{c_s^6} \right) \; .
\end{align}
In the bottom of the plot, we see that, at low core masses, accretion shuts off for particles that pass a certain maximum size. In this regime we have $R_{\rm{acc}} = R_{WS}$, $v_\infty \approx v_{\rm{gas}}$, and $v_{\rm{enc}} = v_{\rm{kick}}$, which leads to a maximum size of
\begin{align} \label{eq:st_crit_low}
St = 12 \left( \frac{v_H}{v_{\rm{gas}}} \right)^3 \; .
\end{align}
However, as particle size continues to increase, we see that accretion actually resumes once particles become large enough. This is caused by these large particles decoupling from the gas, which lowers $v_\infty = v_{pk}$ and raises $v_{\rm{enc}} = v_{pg}$ enough to overcome the increased mass of these particles. If we set $R_{\rm{acc}} = R_H$, $v_{pg} \approx v_{\rm{gas}}$ and approximate $v_{pk}$ by its Taylor Expansion at $\infty$ in the laminar regime, $v_{pk} \approx 2 v_{\rm{gas}} / St$, then we can derive an approximate expression for the Stokes number at which accretion commences again:
\begin{align}
St = \frac{v_{\rm{gas}}}{v_H} \; .
\end{align}
Note that as particle size continues to increase, the energy criteria are only satisfied for a small range of particle sizes. However accretion continues probabilistically in this regime in accordance with Equation \eqref{eq:r_h_eff}.

The gap in particle size where accretion is not possible disappears once the core surpasses a certain mass. If particles that have their encounter velocity dominated by the particle-gas relative velocity, $v_{\rm{enc}} = v_{pg}$, become available for accretion, then the range of particle sizes the core can accrete will be extended, since these particles can dissipate more of their kinetic energy due to their larger encounter velocities. See Section \ref{var_a_M} for more discussion of this effect. For these particles to be available, the previously derived upper limit on particle size, $St = 12 v_{H}^3/v_{\rm{gas}}^3$, must occur after the transition where $v_{pg} = v_{\rm{kick}}$. Using the Taylor expansion of $v_{pg}$ about 0 in the laminar regime, $v_{pg} \approx 2 v_{\rm{gas}} St$, we see that the latter transition occurs at $St \approx (3/4)^{1/3} \left(v_H/v_{\rm{gas}}\right) $ Therefore, the mass scale where this transition occurs is given by
\begin{align}
\frac{v_H}{v_{\rm{gas}}} = 48^{-1/3} \; .
\end{align}
Finally, as discussed in Section \ref{basic}, at high core masses the transition where $KE=W$ occurs at a fixed Stokes number of
\begin{align}
St = 4\sqrt{3} \; ,
\end{align}
which is noted on the plot as well.

\begin{figure*} [htbp]
	\centering
\includegraphics[width=7in]{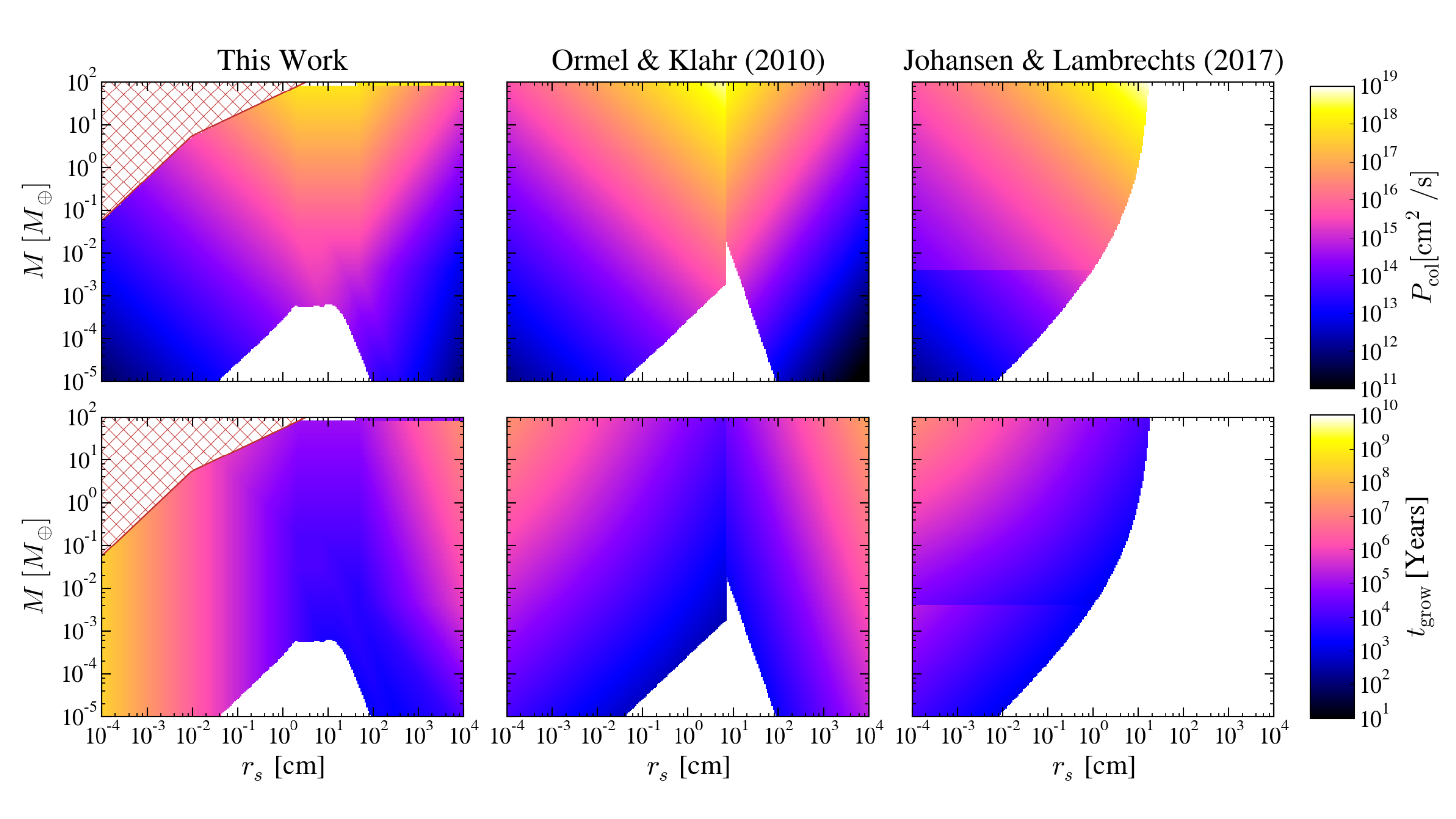}
	\caption{Comparison of results from our model in the laminar regime ($\alpha = 0$) with analytic models of \cite{OK10} and \cite{jl_2017}. The upper row plots $P_{col} = 2 R_{\rm{acc}} v_\infty$ for a core at $a = 30$ AU as a function of small body radius and core mass, while the lower panels instead plot the growth timescale, $t_{\rm{grow}}$. The red hatched region denotes where growth is completely shut off in our modeling, whereas the white regions show places where gas-assisted growth will not operate, but the core can still grow by other means (e.g. gravitational focusing). The upper panels show agreement to order of magnitude between the three models, with exceptions in a few regions (see text for more details). The bottom panels highlight the effect of particle scale height, which is included in our model even in the laminar regime, but not in the other two analytic models.}
	\label{fig:lam_comp}
\end{figure*}

We now contrast the output from our model in the laminar regime with the analytic models of OK10 and JL17. Figure \ref{fig:lam_comp} below shows plots of both $P_{col} \equiv 2 R_{\rm{acc}} v_\infty$ and $t_{\rm{grow}}$ for a core at $30$ AU, using our model, OK10's, and JL17's \footnote{OK10 include a ``Hyperbolic" regime in their modeling, which uses the gravitational focusing cross section for accretion but the laminar terminal small body velocity (our Equation \ref{eq:v_lam_kep}) for $v_\infty$. Because we explicitly do not include gravitational focusing in our model, we similarly do not include this regime when comparing with OK10. We therefore also neglect their exponential term for $R_{\rm{acc}}$ (OK10 Equation 32), which is included in their model to smooth between the settling and hyperbolic regimes, and instead solve their Equation (27) to determine $R_{\rm{acc}}$. JL17 use a similar exponential smoothing term to smooth $R_{\rm{acc}}$ for all Stokes numbers larger than a given value (for fixed core mass). For the same reasons as above, we neglect this term in our comparison.} (for a discussion of how the parameters of the protoplanetary disk are modeled, see Section \ref{params}). We plot $P_{col}$ to disentangle the effects of particle scale height on the growth timescale, as we discuss below.
	
In all three figures, we see the same broad features: while pebble accretion operates for a broad range of particle sizes, there exists an ``optimal" range near $St=1$ ($r_s \approx 10 \, \rm{cm}$) at which accretion reaches its maximal possible rate, and the growth timescale becomes comparable to $t_{\rm{Hill}} = M/(2 R_H^2 \Omega \Sigma_p)$. In all three models, optimum accretion at $St \approx 1$ does not begin until the core has reached a ``large enough" mass. Furthermore, as the core mass increases, the core's growth rate grows as well, in large part due to the increasing size of the core's Hill radius. The growth timescale, however, increases as the core grows, because the core's growth rate scales with $M$ to a power less than one.

In our and OK10's modeling, we see two other notable features as well: firstly, there exists a gap in the range of particle sizes that can be accreted for low core masses, which in our framework cannot accrete because their incoming kinetic energies are too high. Secondly, in the righthand side of the plots ($r_s \gtrsim 100$ m) particles have $R_{H} < R_{WS}, R_{\rm{shear}}$ (i.e. $R_{\rm{acc}} = R_H$), but accretion is inefficient. This is due to the fact that these particles cannot fully dissipate their kinetic energy during a single orbital crossing; but because of the chaotic nature of trajectories for particles with impact parameter $\sim R_H$ their growth timescale is increased by a factor of $KE/W$ (see Section \ref{r_h} for more detail.)

\begin{figure*} [htbp]
	\centering
		\includegraphics[width=7in]{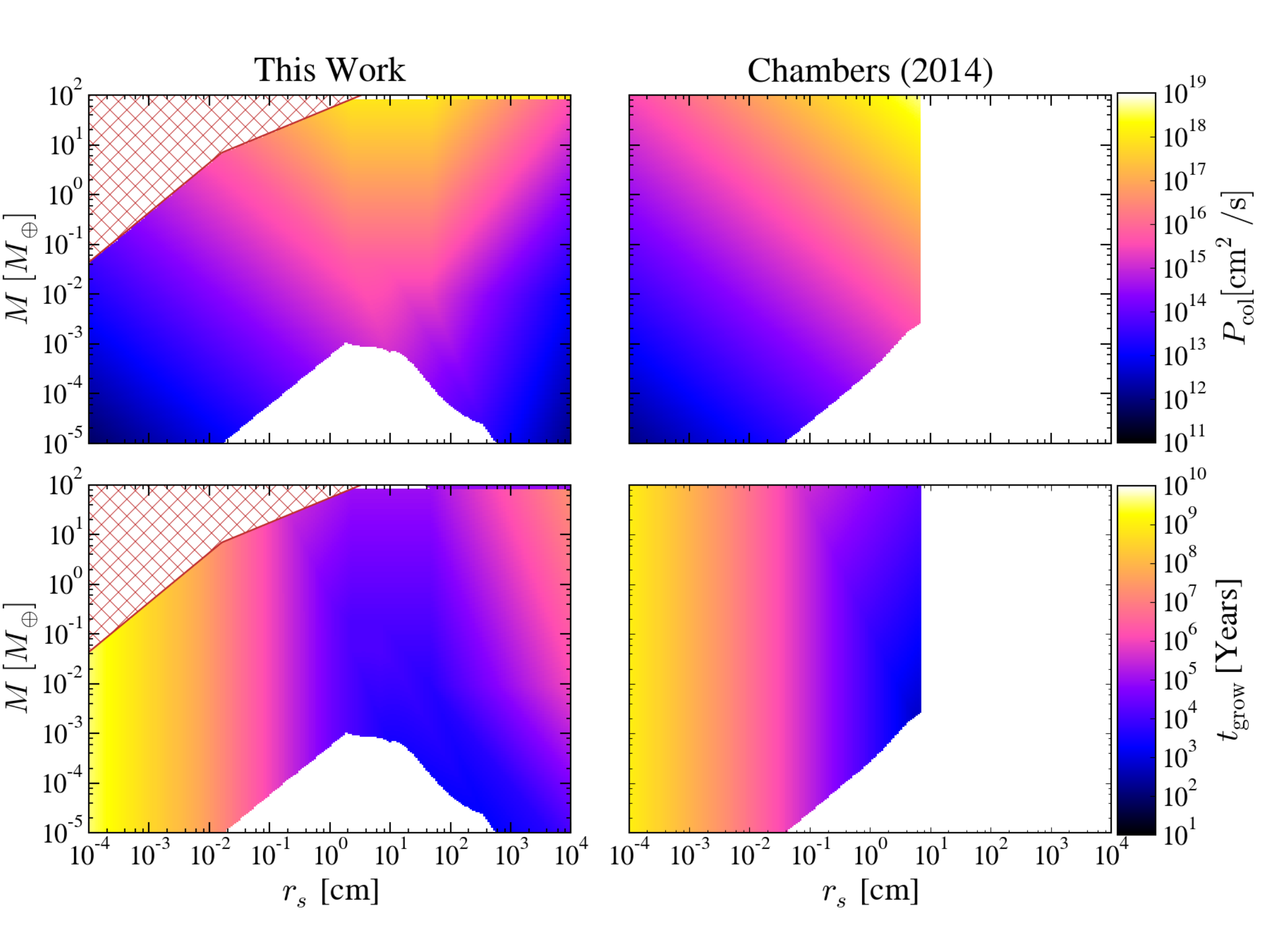}
		\caption{Comparison of results from our model with turbulence included ($\alpha = 10^{-3}$) with modeling of pebble accretion by \cite{c_2014}. The upper row plots $P_{col} = 2 R_{\rm{acc}} v_\infty$ for a core at $a = 30$ AU as a function of small body radius and core mass, while the lower panels instead plot the growth timescale, $t_{\rm{grow}}$. The red hatched region denotes where growth is completely shut off in our modeling, whereas the white regions show places where gas-assisted growth will not operate, but the core can still grow by other means (e.g. gravitational focusing).}
		\label{fig:turb_comp}
	\end{figure*}

As can be verified from the plots, in the regions of parameter space where our model predicts that pebble accretion should operate, our results agree within factors of a few with both of these analytic models. This main disagreements are:
	\begin{enumerate}
		\item The mass scale above which accretion occurs on timescales comparable to $t_{\rm{Hill}}$ is not exactly the same in the models. As discussed previously, in our model there is a transition in behavior past $v_H/v_{\rm{gas}} \approx 48^{-1/3}$. In JL17's model, the parameters that set $t_{\rm{grow}}$ change for a mass $M>M_t$, where $M_t=\sqrt{1/3} v_{\rm{gas}}^3/(G \Omega)$. This is equivalent to $v_H/v_{gas}= \sqrt{1/3}$, which is clearly comparable to the cut in our model. In OK10's modeling, the size of the gap progressively narrows, as the borders between the two regimes -- $St = 12 v_H^3 / v_{\rm{gas}}^3$, and $St = v_{\rm{gas}}/v_H$ -- become comparable. Eventually the gap disappears for $v_H/v_{gas} = 1$.
		\item As discussed above, in our modeling accretion shuts off for particles below a certain radius, which stems from the fact that once $R_{\rm{stab}}$ shrinks below $R_b$ we expect particles to flow around the core's atmosphere (see Figure \ref{fig:en_fig}, in particular the lower righthand panel). The regions where this effect occurs are denoted by red hatching. We note that in these regions we expect growth by all mechanisms to be inhibited, e.g. capture by gravitational focusing or even physical collisions with the core will not occur. (In contrast, in the white regions of the plot we still expect capture mechanisms other than gas-assisted growth to operate.) Because this process is set by the modification of the flow by the core's gravity, we would not expect OK10's integrations to capture it, since they assume a constant headwind velocity in their integrations. LJ12's simulations do not appear to go to high enough core masses with small enough particle sizes to capture this effect.\footnote{LJ12's \texttt{1e-1\_0.01} simulation, which is the highest core mass and smallest particle size they simulate, is actually marginal in the sense that $R_{\rm{shear}}=R_b$ in our notation. While there is a reduction in accretion in this run, it is unclear whether accretion is actually substantially reduced in the way we would expect in our modeling. It is also unclear whether LJ12's resolution is fine enough to resolve the core's atmosphere and therefore the modification of the flow pattern.} This effect does appear to be consistent with the results of \cite{ormel_flows}, who find minimal accretion for small particle sizes at high core masses (see their Figure 12) \footnote{In the bottom panel of Figure 12, \cite{ormel_flows} almost all particle streamlines are unable to accrete for $St = 10^{-4}$ (in our notation). For $St = 10^{-3}$ however, particles over a range of impact parameters are able to accrete. For the parameters used by \cite{ormel_flows} the particle size where $R_{\rm{stab}}$ ($=R_{WS}$ in this region of parameter space) is less than $R_b$ is $St < 5 \times 10^{-4}$.}
		\item JL17 define a ``weak coupling" regime, where the stopping time of the particle exceeds the time to pass the protoplanet, $t_p = GM/\left(v_{\rm{gas}} + v_H\right)^3$. For $v_{\rm{gas}} \gg v_H$ this criterion is equivalent to $St > 3 v_H^3/v_{\rm{gas}}^3$, which is line with our and OK10's criterion $St > 12 v_H^3/v_{\rm{gas}}^3$. For $v_H \gg v_{\rm{gas}}$, the passing time criterion reduces to $St > 3$, which is again similar to our $St > 4\sqrt{3}$ limit. For particle sizes exceeding this critical Stokes number, JL17 exponentially smooth $R_{\rm{acc}}$ to zero, which differs from our and OK10's modeling.
		\item In the lower righthand region, OK10's results and our diverge. In this regime, particles that shear into $R_H$ cannot fully dissipate their kinetic energy over one orbital crossing. This makes them accrete probabilistically, in accordance with Equation \eqref{eq:r_h_eff}, i.e., instead of shutting off accretion because $W<KE$ in this regime, we increase the growth timescale by a factor $KE/W$. For particles with $v_{\rm{enc}} \sim v_H$, which holds for larger cores, this increases the growth timescale by a factor of $St$, which agrees with OK10's expression for $R_{\rm{acc}}$ in this regime -- $R_{\rm{acc}} = R_H/St$. However, for the low mass cores in the bottom righthand corner of the plot, the particle gas relative velocity dominates over the core's gravity, changing the Stokes number dependence.
	\end{enumerate}

Thus while our results are generally in order of magnitude agreement with other modeling in the laminar regime, there are few notable regions where our results diverge. It can be shown analytically that the agreement between our model and the other two essentially stems from the fact that, to order of magnitude, many of the transitions between length and velocity regimes occur in the same region of parameter space. For example, the transitions where $R_H = R_{WS}$ and where $v_{pk,\ell} = v_H$ both occur for $St \sim \zeta_w$. In the laminar case, there are therefore essentially fewer regimes that need to be covered.
	
When turbulence is included however, this is no longer the case. As an example, in Figure \ref{fig:turb_comp} we compare our modeling to the pebble accretion modeling of \cite{c_2014}. The expressions used in \citeauthor{c_2014} are analogous to the OK10 expressions, with the replacement of OK10's $\zeta_w \equiv \eta v_k / v_H$ parameter with $v_{\rm{gas}}/v_H$, where $v_{\rm{gas}} = \max(\eta v_k,\sqrt{\alpha} c_s)$. In our comparison, we neglect the exponential smoothing term they carry over from OK10, as well as terms involving eccentricity and inclination, as they are not included in our model. \footnote{The \cite{c_2014} modeling of the velocity of small bodies induced by turbulence appears to use expressions for the RMS particle-particle relative velocity, as opposed to the velocity of small bodies relative to the local Keplerian orbital velocity. This may be an error in the text, as other expressions used by these authors to model the relative velocity between a small body and an embryo (e.g. their Equation 8), are expressions for a particle relative to the local Keplerian velocity. In particular, their expression tends toward 0 for small Stokes number particles, whereas we would expect that small, well-coupled particles move relative to the embryos at velocities comparable to the RMS turbulent gas velocity (assuming the embryos are decoupled from the gas, which is a good approximation since $M_{\rm{emb}} \geq 5 \times 10^{-6} M_\oplus$ in their work). This appears to underestimate the incoming velocities of small bodies for strong turbulence ($\alpha \gtrsim \eta$). In our comparison with \cite{c_2014} we therefore use the expression $v_{\rm{turb}}^2 = \alpha c_s^2 / (1 + St)$ in place of their Equation (9).}  	
As can be seen from the figure, for low core masses and small body sizes, our models are in order of magnitude agreement. Again, analytic calculations can be used to demonstrate that for low core masses and small particle sizes, the expressions between the two models agree to order of magnitude. For example, the cutoff in Stokes number employed by \citeauthor{c_2014} -- $St_{\rm{crit}} = 12 (v_{H}/v_{\rm{gas}})^3$ is replicated in our model for small cores and small particle radii. However, as can also be seen in the figure, there are numerous regimes covered in our modeling that aren't captured by extending the OK10 expressions. A prominent example is the feature in the lower righthand corner, where low mass core can accrete ``large" particles ($\sim 10^2-10^3$ cm) on rapid timescales. This features results from the fact that, for low core masses, the velocities of these larger particles are set by their interactions with the nebular gas. Because these large particles are not well coupled to the gas flow, they have low kinetic energies relative to the core, but high velocities relative to the gas, allowing a certain range of particle radii to dissipate energy during the encounter that is comparable to their kinetic energy. The shape of this feature depends on the transitions where $KE \sim W$ as well as where $R_{WS} \sim R_H$ and $v_{pk} \sim v_H$. In the presence of turbulence these transitions no longer need occur at similar regions of parameter space, leading to a more complex model of pebble accretion that is not well captured by extending the OK10 expressions.
	
\subsection{Comparison to Xu et al. (2017)} \label{xu_comp}
We close this section by comparing the results of our model to numerical simulations of pebble accretion in the presence of turbulence. Xu et al. (2017) performed magnetohydrodynamic (MHD) simulations of gas-assisted growth of turbulence due to the magnetorotational instability (MRI). These simulations provide an excellent point of comparison for our order of magnitude model, since they use much more detailed physics but apply over a narrower range of parameter space. As we will show below, our model agrees with the Xu et al. results to order of magnitude, as we would expect. 

Xu et al. perform six sets of simulations; they use two different values of core mass, at three different levels of turbulence for each core. The core masses are parameterized in terms of the ``thermal mass," which Xu et al. give as
\begin{align} \label{eq:m_therm}
M_T \approx 160 M_\oplus \left( \frac{a}{30\,\text{AU}} \right)^{3/4} \; .
\end{align}

\begin{figure} [h]
	\centering
	\includegraphics[width=\linewidth]{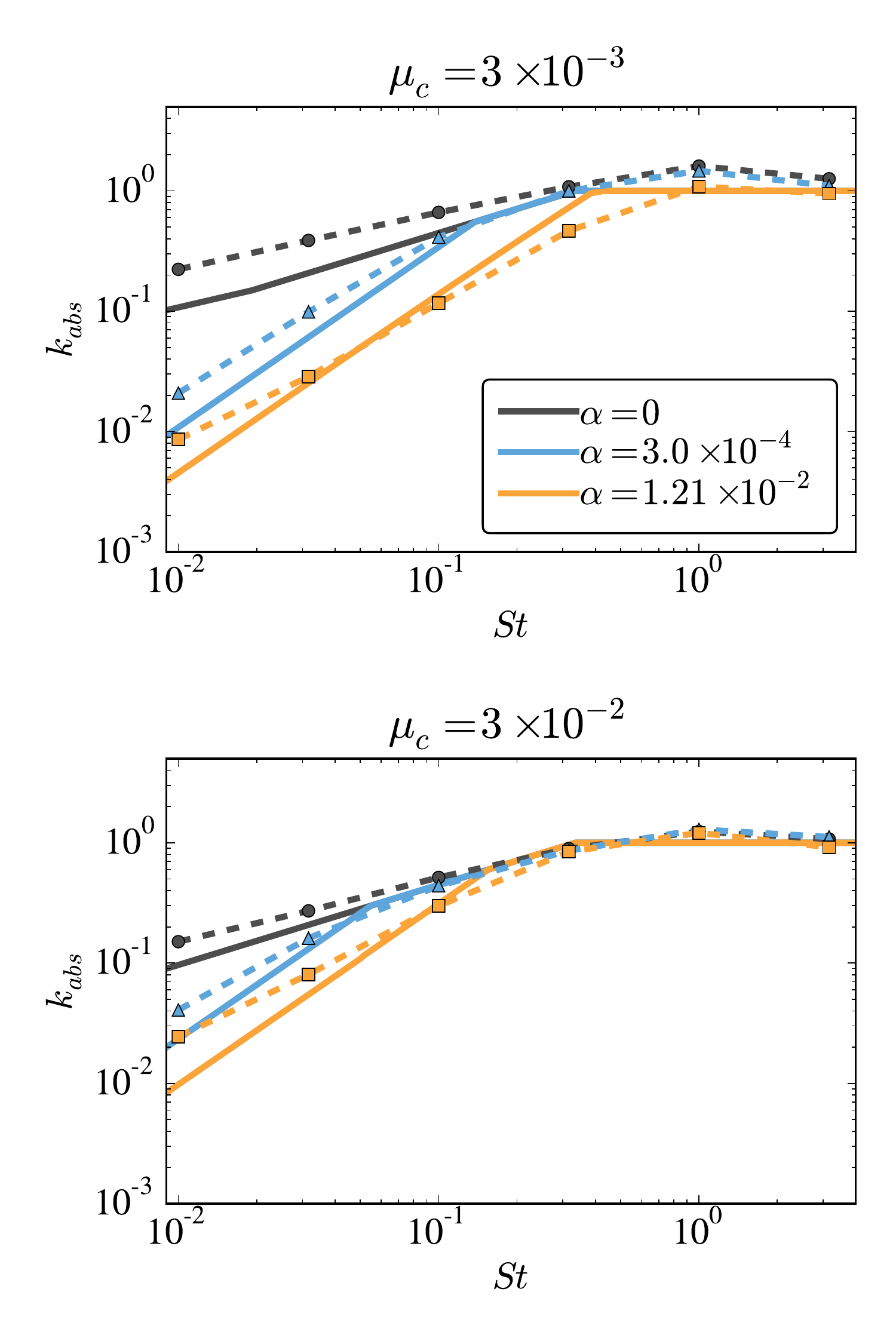}
	\caption{A comparison of the growth rate $k_{\rm{abs}}$ from our model and the MHD simulations of Xu et al. (2017), plotted as a function of the Stokes number of the particles the cores are accreting. Here $k_{\rm{abs}} \equiv \dot{M}/\dot{M}_{\rm{Hill}}$, where $\dot{M}= M/t_{\rm{grow}}$, and $\dot{M}_{\rm{Hill}} = M/t_{\rm{Hill}}$, i.e. $k_{\rm{abs}}$ represents the growth rate of the core in units of the accretion rate for growth at $t_{\rm{Hill}}$. The two panels depict growth for two different values of core mass -- $\mu_c \equiv M/M_T = 3 \times 10^{-3}$ and $\mu_c = 3 \times 10^{-2}$, where $M_T$ is the thermal mass, defined in Equation \eqref{eq:m_therm}. Each panel shows the growth rate for three different levels of turbulence, which are listed in the legend. The solid lines show the output from our model, while the data points show the Xu et al. results.}
	\label{fig:xue}
\end{figure}

Note that this is the same as the flow isolation mass (see Section \ref{flow_iso_mass}) for the values used in their work. The simulations are performed at core masses of $\mu_c \equiv M/M_T = 3 \times 10^{-3}$ and $\mu_c = 3 \times 10^{-2}$. The three levels of turbulence are achieved by performing a pure hydrodynamic simulation, a non-ideal MHD simulation with ambipolar diffusion, and an ideal MHD simulation. While thus far in our work we have assumed that $\alpha \approx \alpha_z$, where $\alpha_z$ is the diffusion coefficient for the particle scale height, Xu et al. are able to separately calculate $\alpha$ and $\alpha_z$. We also use their value for $v_z$ to calculate our $\alpha$ value, as we use $\alpha$ to parameterize the turbulent gas velocity. For the pure hydrodynamic run they have $v_z =  \alpha_z = 0$. For the ambipolar diffusion run the authors give $\braket{v_z^2/c_s^2} = 3.0 \times 10^{-4}$, $\alpha_z = 7.8\times10^{-4}$, while for the ideal MHD simulation they give $\braket{v_z^2/c_s^2} = 1.21\times10^{-2}$, and $\alpha_z = 4.4\times10^{-3}$.

\begin{figure*} [htbp]
	\centering
\includegraphics[width=7in]{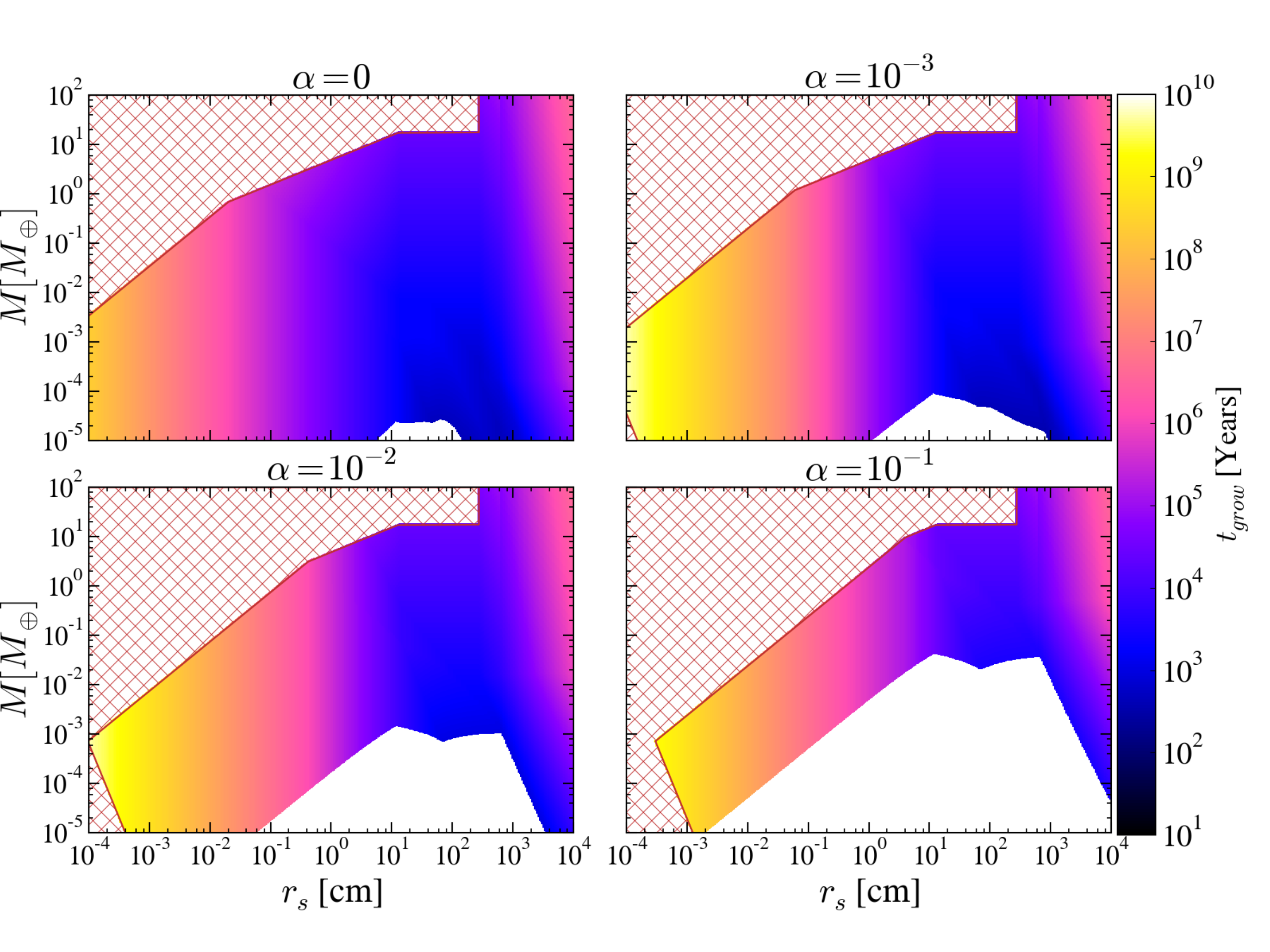}
	\caption{The growth timescale as a function of core mass, for $a = 5\,\text{AU}$. The red hatched region denotes where growth is completely shut off in our modeling, whereas the white regions show places where gas-assisted growth will not operate, but the core can still grow by other means (e.g. gravitational focusing). The feature in the upper right emerges for $R_b>R_H$ (see Section \ref{flow_iso_mass}).}
	\label{fig:heatmap_a1}
\end{figure*}

\begin{figure*} [htbp]
	\centering
	\includegraphics[width=7in]{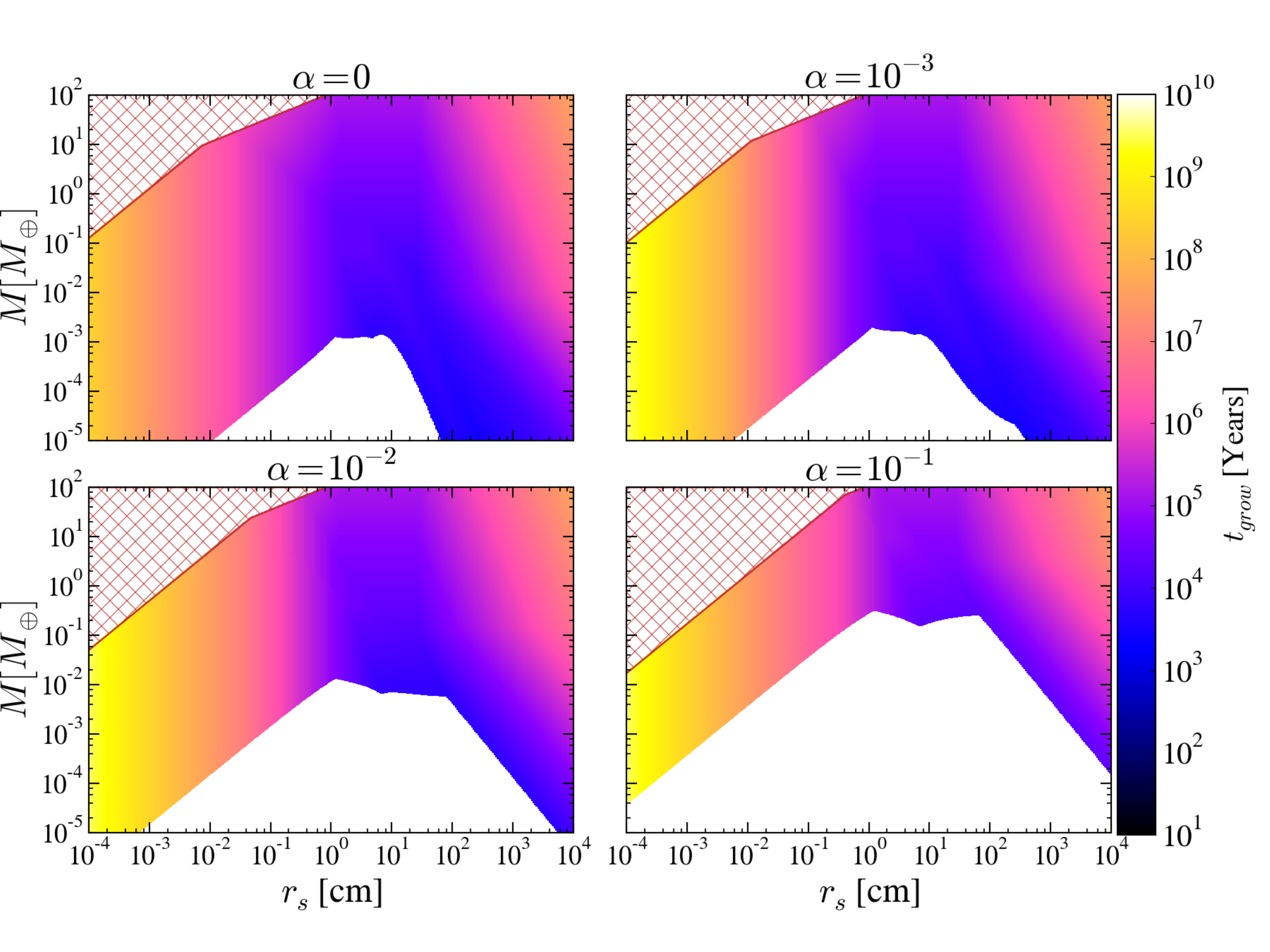}
	\caption{The growth timescale as a function of core mass, for $a = 50\,\text{AU}$. The red hatched region denotes where growth is completely shut off in our modeling, whereas the white regions show places where gas-assisted growth will not operate, but the core can still grow by other means (e.g. gravitational focusing).}
	\label{fig:heatmap_a10}
\end{figure*}

A comparison between our analytic model and their simulations is shown in Figure \ref{fig:xue}. This figure plots $k_{\rm{abs}} \equiv \dot{M}/\dot{M}_{\rm{Hill}}$ as a function of the particle Stokes number $St$. Here $\dot{M} = M/t_{\rm{grow}}$ is the growth rate of the core, and $\dot{M}_{\rm{Hill}} = M/t_{\rm{Hill}}$ is the growth rate for Hill accretion. The solid lines depict the model presented in this paper, while the data points show the results from the Xu et al. paper. For the purposes of matching their model, we've used a temperature profile consistent with their results, used their values of $\alpha_z$ for the calculation of $H_p$ (as opposed to using $\alpha$), and set the scale height in the laminar case to be $H_{p,\rm{lam}} = 0.01 H$ to be consistent with these authors' choice.

As can be seen in Figure \ref{fig:xue}, our model nicely achieves the intended goal of reproducing the trends found by the numeric results, as well as being accurate to within a factor of 2 for all of the values found in the numeric simulations. It is worth noting here that the simulations are carried out at quite large values of core mass, and serve mostly to confirm our prediction that the efficiency of pebble accretion is not hugely reduced by increasing turbulence. It would be interesting to investigate in future simulations whether the drop off in efficiency for lower core masses predicted in our model (see Section \ref{var_a_M}) is reproduced in the numeric simulations.

\section{Exploration of Parameter Space} \label{param_space}
In this section we give an overview of the effects of varying a few of the most important parameters in our model. In the first three sections we discuss varying parameters related to the disk in which growth is occurring. We begin by discussing the combined effects of varying the semi-major axis, $a$, and the core mass, $M$, both of which determine the importance of interactions between the small body and the gas relative to gravitational effects between the small body, growing core, and central star. We then investigate the effects of varying $T_0$, prefactor which sets the temperature profile, $\Sigma_g$, the local gas surface density, and $M_*$, the mass of the central star.

\subsection{Effects of Variation of Orbital Separation and Core Mass} \label{var_a_M}
In Section \ref{min}, we identified the minimal timescale $t_{\rm{Hill}}$ that pebble accretion can operate on. We also emphasized, however, that pebble accretion can operate on timescales that are substantially slower -- in many cases these timescales can exceed the lifetime of the disk, $\tau_{disk} \sim 2.5$ Myr, which implies that pebble accretion essentially won't occur. In this section we highlight how at low core masses at wide orbital separations only a small range of particle sizes, if any, accrete on timescales comparable to $t_{\rm{Hill}}$.

To illustrate how core mass and orbital separation effect the growth timescale, we plot $t_{\rm{grow}}$ as a function of both $r_s$ and $M$ for a core at $a =$ 5 AU in Figure \ref{fig:heatmap_a1}, and for a core at $a =$ 50 AU in Figure \ref{fig:heatmap_a10}. To begin, we note that growth is generally slower at wide orbital separation, since the dynamical time $t_{\rm{dyn}}\sim \Omega^{-1}$ is larger and the solid surface density is smaller. For low core mass there is also a gap in the particle sizes that can be accreted. For low turbulence this gap is small, but as $\alpha$ increases, the width of this gap increases as well. This effect is enhanced at wide orbital separation, where the gap extends to higher core masses and is overall wider. Thus, low mass cores can often only accrete small pebble radii. These small particles are accreted on slow timescales which are $\gtrsim \tau_{disk}$, as small pebbles can only be captured at low impact parameters, and have low densities since they can be easily lofted by turbulence. Thus we see that growth is much slower at low core mass and wide orbital separation, particularly as the strength of turbulence is increased. If there exists of population of $\gtrsim 10$ m ``boulders," then we see from the figure that there exists a narrow range of these larger particle sizes for which accretion can proceed on rapid timescales, even if accretion of pebble sized particles is slow.

\begin{figure*} [h]
	\centering
	\includegraphics[width=\linewidth]{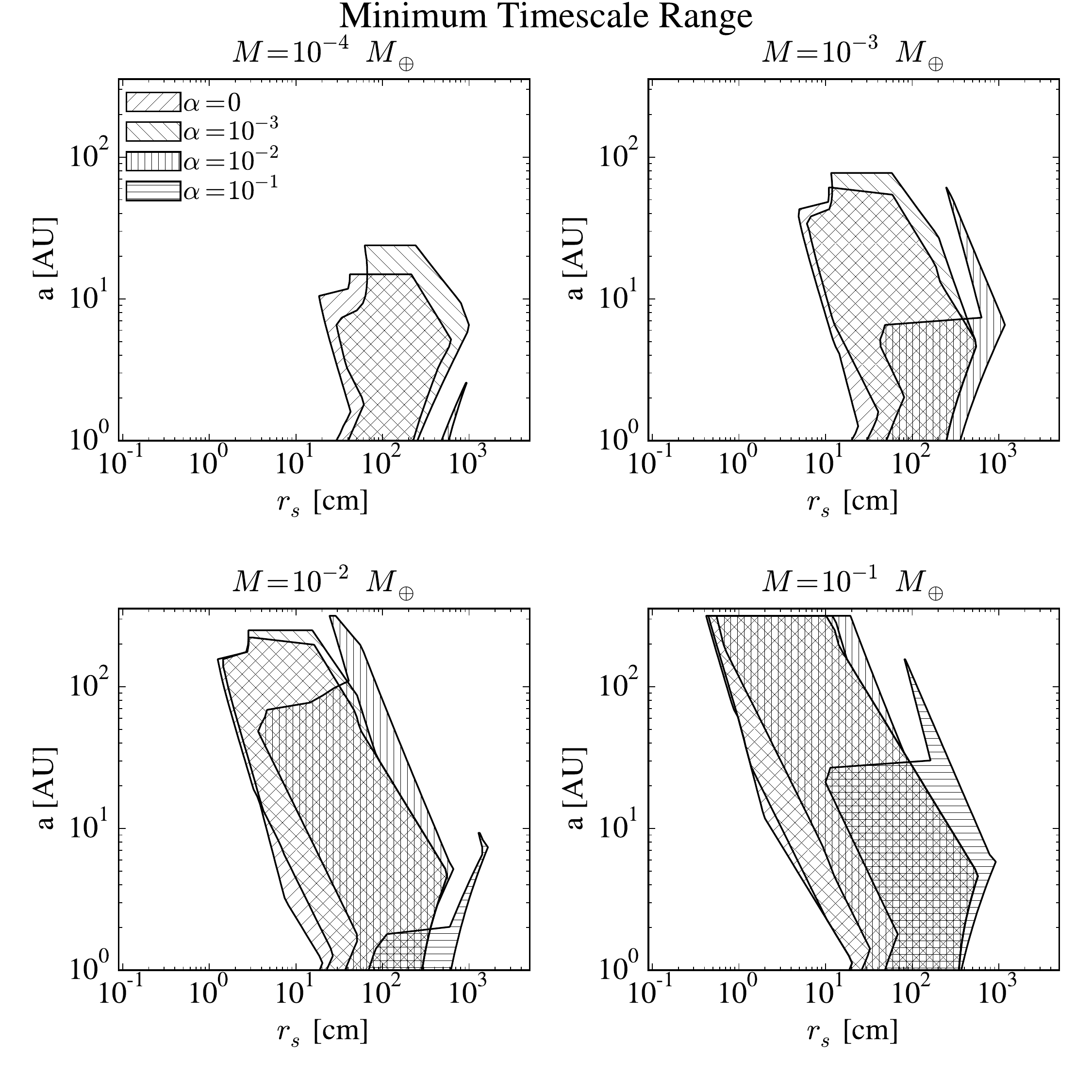}
	\caption{The range of small body radii which can accrete within a factor of two of $t_{\rm{Hill}}$ for a given semi-major axis and core mass. As indicated in the legend, different hatching styles indicate different levels of turbulence. If a given region is accessible for different levels of turbulence the various hatching styles are overlaid.}
	\label{fig:rad_range}
\end{figure*}

Once the core exceeds a certain critical mass, the gap disappears and accretion proceeds much more rapidly, allowing pebbles to accrete on timescales comparable to $t_{\rm{Hill}}$. To emphasize this effect in Figure \ref{fig:rad_range} we plot the range of particle sizes that can accrete on timescales within a factor of two of $t_{\rm{Hill}}$. Each panel depicts a different core mass, while different hatching patterns depict different levels of turbulence. As expected from the above discussion, for low core mass only close in cores with low levels of nebular turbulence can accrete particles on timescales comparable to $t_{\rm{Hill}}$. As core mass increases however, particles farther out in the disk can be accreted with growth timescales $\sim t_{\rm{Hill}}$, even when turbulence is strong.

We now discuss more quantitatively what sets the range of particle sizes that can be accreted, and how that range is affected by changing the core mass and orbital separation. For low mass cores accreting small particles, we expect $R_{WS}$ to set the impact parameter. Because the particles are so small, and therefore well coupled to the gas, we will have $v_{pk} \approx v_{\rm{gas}}$, and $v_{pg} \ll v_{\rm{kick}}$. Thus, in this regime we expect $v_\infty \approx v_{\rm{gas}}$ (since shear is negligible for low mass cores), and $v_{\rm{enc}} = v_{\rm{kick}}$ (since $v_{\rm{gas}} < v_{\rm{orbit}}$ in this regime). Plugging these values into our expressions for $KE$ and $W$ we can show that having $KE<W$ requires
\begin{align} \label{eq:st_crit_param_space}
St < 12 \frac{v_H^3}{v_{\rm{gas}}^3} \; ,
\end{align}
which was given previously in Equation \eqref{eq:st_crit_low}. This is one facet that makes accretion less efficient for low mass cores, high turbulence, and wide orbital separation -- since $v_H/v_{\rm{gas}} \propto M^{1/3} a^{-4/7}$ in the laminar regime (i.e. approximating $v_{\rm{gas}} \approx \eta v_k$) and $\propto M^{1/3} a^{-2/7} \alpha^{-1/2}$ in the turbulent regime ($v_{\rm{gas}} \approx \sqrt{\alpha} c_s$), decreasing core mass, increasing $\alpha$ or increasing orbital distance will all make the limit on $St$ more stringent.

As small body radius increases, $v_{\rm{kick}}$ decreases due to the increasing size of the core's WISH radius, while $v_{pg}$ rises as particles decouple from the gas. Eventually we reach the point that $v_{pg} > v_{\rm{kick}}$, meaning that now $v_{\rm{enc}} = v_{pg}$. In this regime the dependence of the energy on small body radius becomes more complex: throughout a large amount of parameter space $KE/W$ increases with particle size, which encapsulates the fact that it is more difficult for heavier particles to dissipate their kinetic energy. If $v_{\rm{enc}}=v_{pg}$ however, the work done during the encounter increases strongly with small body radius, which actually make it \textit{easier} for large particles to be accreted. Furthermore, if $v_\infty = v_{pk}$ then the incoming $KE$ of particles can also decrease with small body radius, making it even easier for heavier particles to be accreted.

\begin{figure} [h]
	\centering
\includegraphics[width=\linewidth]{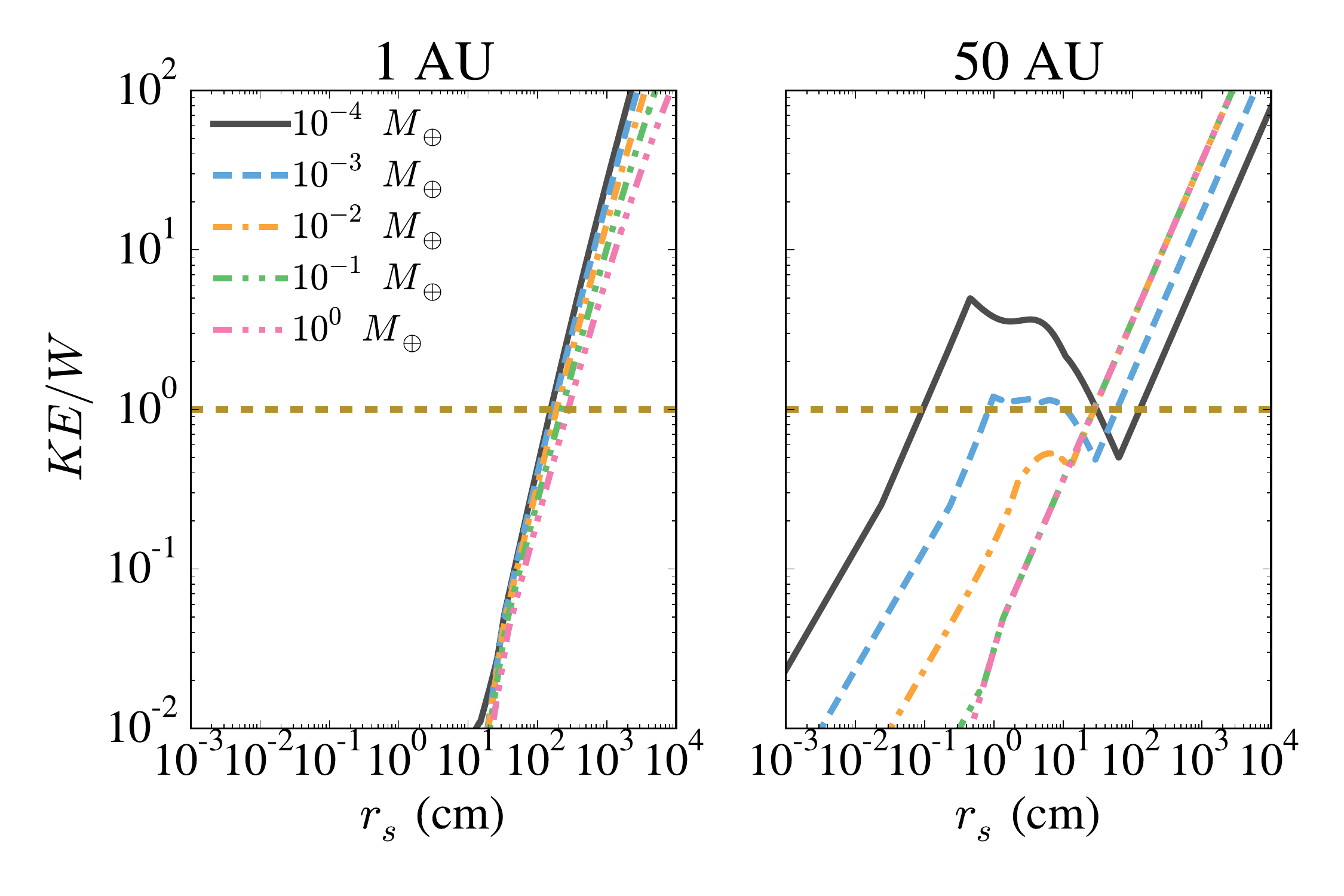}
	\caption{The ratio of kinetic energy relative to the core to work done by the gas on an incoming particle assuming $\alpha = 0$, plotted as a function of small body radius $r_s$. Curves are shown for a range of core masses. The left panel shows the situation at $a=1\,\text{AU}$, while the right panel is for $a=50\,\text{AU}$.} 
	\label{fig:en_ratio}
\end{figure}

This effect is illustrated in Figure \ref{fig:en_ratio}, which plots the ratio $KE/W$ as a function of $r_s$ for different core masses. The two different panels show cores at different orbital separations. For the low mass cores in the righthand panel, we see that once $r_s$ increases past a certain value the qualitative behavior of $KE/W$ changes -- instead of monotonically increasing, the slope flattens out or even decreases. However, for the very low mass cores this behavior is inconsequential, since particles are ruled out from accreting before we reach a large enough particle size that $v_{\rm{enc}} = v_{pg}$ and this more complex behavior starts. For larger core masses however this is no longer the case, and the range of particle sizes that can be accreted is greatly extended. This what causes the gap seen in the range of accreted particle sizes to disappear -- once particles with $v_{\rm{enc}} = v_{pg}$ become available accretion generally continues until the limit $St = 4 \sqrt{3}$ discussed in Equation \eqref{eq:st_crit_num} is reached.  From comparison of the two panels we also see that this effect is much more prominent at wide orbital separations -- at 1 AU none of the cores exhibit the change in slope seen at 50 AU.

Thus, we see that growth changes qualitatively once particles with $v_{pg} > v_{\rm{kick}}$ can be accreted. The critical value of mass where this occurs can be approximated by calculating the mass at which $v_{pg} > v_{\rm{kick}}$ is reached before the critical Stokes number given in Equation \eqref{eq:st_crit_param_space} is reached. If we keep the approximations that led to Equation \eqref{eq:st_crit_param_space} but take $v_{pg} \approx 2 v_{\rm{gas}} St$ in the laminar regime and $v_{pg} \approx v_{\rm{gas}} \sqrt{St}$ in the turbulent regime, we can solve for the Stokes number past which $v_{pg} > v_{\rm{kick}}$. Doing so, and setting the resulting Stokes number equal to \eqref{eq:st_crit_param_space} yields a mass limit, which, in both regimes, may be approximated by
\begin{align}
\frac{v_H}{v_{\rm{gas}}} \approx 48^{-1/3}
\end{align}

This inefficiency of pebble accretion at lower core masses is also identified by  \cite{vo_2016}, who calculate the growth timescale of planetesimals by numerically integrating the pebble equation of motion in the presence of two different laminar gas flow patterns. \citeauthor{vo_2016} find that pebble accretion is only faster than gravitational focusing  (which they refer to as the ``Safronov regime") once the growing planetesimal exceeds a certain radius, $R_{PA}$. They also find that this critical radius increases further out in the disk, which is again consistent with our results. It can be seen from our analytic expressions however, that while these effects are present in our model even in the laminar case, they are strongly amplified by the presence of turbulence.

\subsection{Effects of Varying $T_0$} \label{T_0}
In this section we consider the effects of varying the prefactor to the temperature profile, $T_0$. While changing the disk temperature will affect the properties of the small bodies, such as number density and composition, by changing what volatile species are present in solids at a given location, we neglect such effects in what follows. The consequences of changing the temperature profile are complex and depend on the local disk parameters as well as the core mass and strength of turbulence. Nevertheless, in general increasing the temperature is detrimental to the accretion rate when gas dominated processes set the scales relevant to growth. For growth at $t_{\rm{Hill}}$, however, none of the scales depend on temperature, so the most rapid growth timescales are unaffected by increasing the disk temperature. 

The primary effect of varying $T_0$ on growth timescales is due to the dependence of $c_s$ on $T$: $c_s \propto T^{1/2}$. Since $H_{KH} \propto c_s^2$, and $H_t \propto c_s$, increasing $T_0$ will increase $H_p$. Similarly, $v_{pk,\ell} \propto c_s^2$ and $v_{pk,t} \propto c_s$, so increasing $T_0$ will also increase $v_\infty$ in the drift dominated regime. Finally, from inspection of Equation \eqref{r_ws_exp}, we see that $T_0$ affects $R_{WS}$ by changing $\rho_g \, (\propto c_s^{-1}$), $v_{th}$, and $v_{\rm{gas}}$. By inspecting the scaling of $R_{WS}$ one can show that $R_{WS}$ is a decreasing function of $T_0$.  

While some the above effects increase $t_{\rm{grow}}$, while other decrease it, the general effect of increasing $T_0$ is to slow down growth. To see this, we can make the same approximations described in Equations \eqref{KH_vdisp} - \eqref{vdisp_approx}. As in Section \ref{min} we consider in the 2D and 3D regimes separately.

In the 2D regime where $H_p < R_{\rm{acc}}$, $H_{\rm{acc}}=R_{\rm{acc}}$, so we have $t_{\rm{grow}} \propto (R_{\rm{acc}} v_\infty)^{-1}$. Having $R_{\rm{acc}} > H_p$ implies that $v_{\rm{shear}} = R_{\rm{acc}} \Omega >v_{pk}$, which in turn implies that $R_{\rm{shear}} < R_{WS}$. Thus we expect that $t_{\rm{grow}}$ is independent of $T_0$ in this regime.

\begin{figure} [h]
	\centering
	\includegraphics[width=\linewidth]{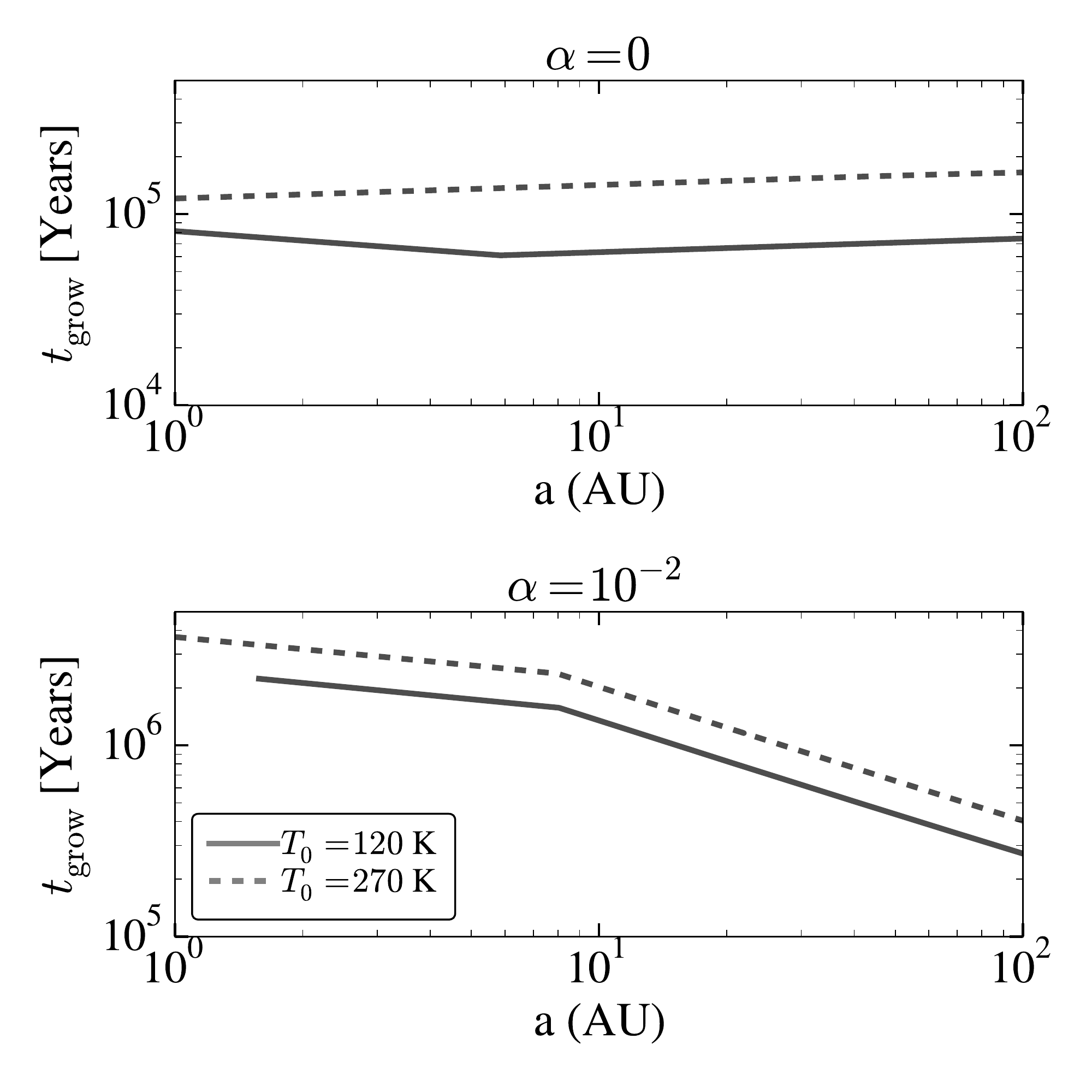}
	\caption{The growth timescale as a function of semi-major axis for two different values of the prefactor of the temperature profile, $T_0$. Both panels use the values $r_s = 0.25\,\text{cm}$, $M=10^{-1}M_\oplus$. The panel on the top is for $\alpha = 0$, while the panel on the bottom is for $\alpha = 10^{-2}$. The effect of increasing $T_0$ is more substantial in the laminar case, since $H_p$ and $v_{pk}$ both scale as $c_s^2$ in this regime, as opposed to $c_s$ in the turbulent case.}
	\label{T_0_comp}
\end{figure}

For 3D accretion, we have $H_p > R_{\rm{acc}}$, which implies that $v_\infty = v_{pk}$. Therefore we have $t_{\rm{grow}} \propto (R_{WS}^2 \Omega)^{-1}$, which is an increasing function of $T_0$. We therefore expect that (approximately) higher $T_0$ should lead to slower growth. An example of the difference caused by increasing $T_0$ is shown in Figure \ref{T_0_comp}. 

A secondary effect of increasing $T_0$ is that $t_s$ and $St$, which depend on $\rho_g$, $v_{th}$, and $\lambda$, are also affected. However, inspection of Equation \eqref{t_s_exp} shows that in the Epstein regime the effects of $T_0$ on $t_s$ cancel out. Therefore $T_0$ only affects $St$ in the fluid regime. This substantially diminishes the importance of this dependence, since for a large amount of parameter space the small body sizes which grow the most efficiently are in the Epstein regime. For particles that are in the fluid regime, higher values of $T_0$ will shift values of $St$ to lower values of $r_s$. 

Because of the approximate cancellation between $v_\infty$ and $H_p$ in the 3D regime, in general the maximal increase in timescale that can be provided by increasing $T_0$ is of order $t_{\rm{hot}}/t_{\rm{cold}} \sim R_{\rm{acc},\rm{cold}}^2/R_{\rm{acc},\rm{hot}}^2$, where hot and cold denote the higher and lower values of $T_0$ respectively. Since $R_H$ generally represents an upper limit on $R_{\rm{acc}}$ (ignoring accretion in the $R_{\rm{acc}} = R_b$ regime), the maximal increase is of order $(R_{WS,\rm{cold}}/R_{WS,\rm{hot}})^2$, which in the laminar Stokes and Ram regimes can be of order $(c_{s,\rm{hot}}/c_{s,\rm{cold}})^3 = (T_{0,\rm{hot}}/T_{0,\rm{cold}})^{3/2}$. In practice, this maximal value is rarely reached, since it requires simultaneously satisfying $r>9 \lambda /4$ to be in the fluid regime, while also having $St < v_{\rm{gas}}^3/(3 v_H^3)$ in order to have $R_{WS} < R_{\rm{shear}}$. In practice therefore, the increases increase growth timescale generally goes as $T_{0,hot}/T_{0,cold}$. Because accretion timescales at $t_{\rm{Hill}}$ are not dependent on temperature, increasing $T_0$ does not affect the rapid growth rates supplied by gas-assisted growth.

Changing $T_0$ will also have an effect on the range of particle sizes that the growing core is able to accrete. For small particle sizes, the accretion cut off is determined by the point where $R_b = R_{WS}$. Since $R_b \propto c_s^{-2}$, which is stronger than the dependence of $R_{WS}$ on $c_s$ regardless of drag regime, increasing $T_0$ will decreases the size of the Bondi radius relative to the WISH radius. This in turn will allow the core to accrete smaller sized bodies. For larger particle sizes, the effect of increasing $T_0$ is not as clear cut. For particles accreting at $t_{\rm{Hill}}$, the only effect of increasing $T_0$ is to increase the drag force on accreting small bodies, which allow the core to accrete larger sizes. In other regimes however, increasing $T_0$ can also increase the approach velocity of small bodies. In these cases the increase in kinetic energy of accreting particles outweighs the larger amount of work done, and the maximal size of small bodies accreted is reduced.

\subsection{Growth in a Gas Depleted Disk}
In this section, we explore growth in a disk where the gas density has been reduced by a factor of 100, but the solid surface density is unchanged; i.e we have $\Sigma = \Sigma_{g,0}(a/\text{AU})^{-1}$, where $\Sigma_{g,0} = \Sigma_0/100$, and $\Sigma_0 = 500 \, \text{g} \, \text{cm}^{-2}$ is the prefactor employed elsewhere in this paper for the gas surface density. However we keep $\Sigma_{p,0} = 5 \, \text{g} \, \text{cm}^{-2}$. We note that this may affect some of the expressions used in this work, which implicitly assume $\rho_g \gg \rho_p$, where $\rho_p$ is the volumetric mass density of the small bodies. We neglect in any such effects in what follows. These choices for gas and solid densities are made to emulate the conditions in the disk when the gas component of the disk is in the process of photoevaporating, which is an important stage in some theories of planet formation (e.g. \citealt{lee_sup_earth}, \citealt{fm_2017}). As we shall show below, the predominant effect of reducing the gas surface density is to shift the range of small body sizes where accretion occurs to lower values.

\begin{figure} [h]
	\centering
	\includegraphics[width=\linewidth]{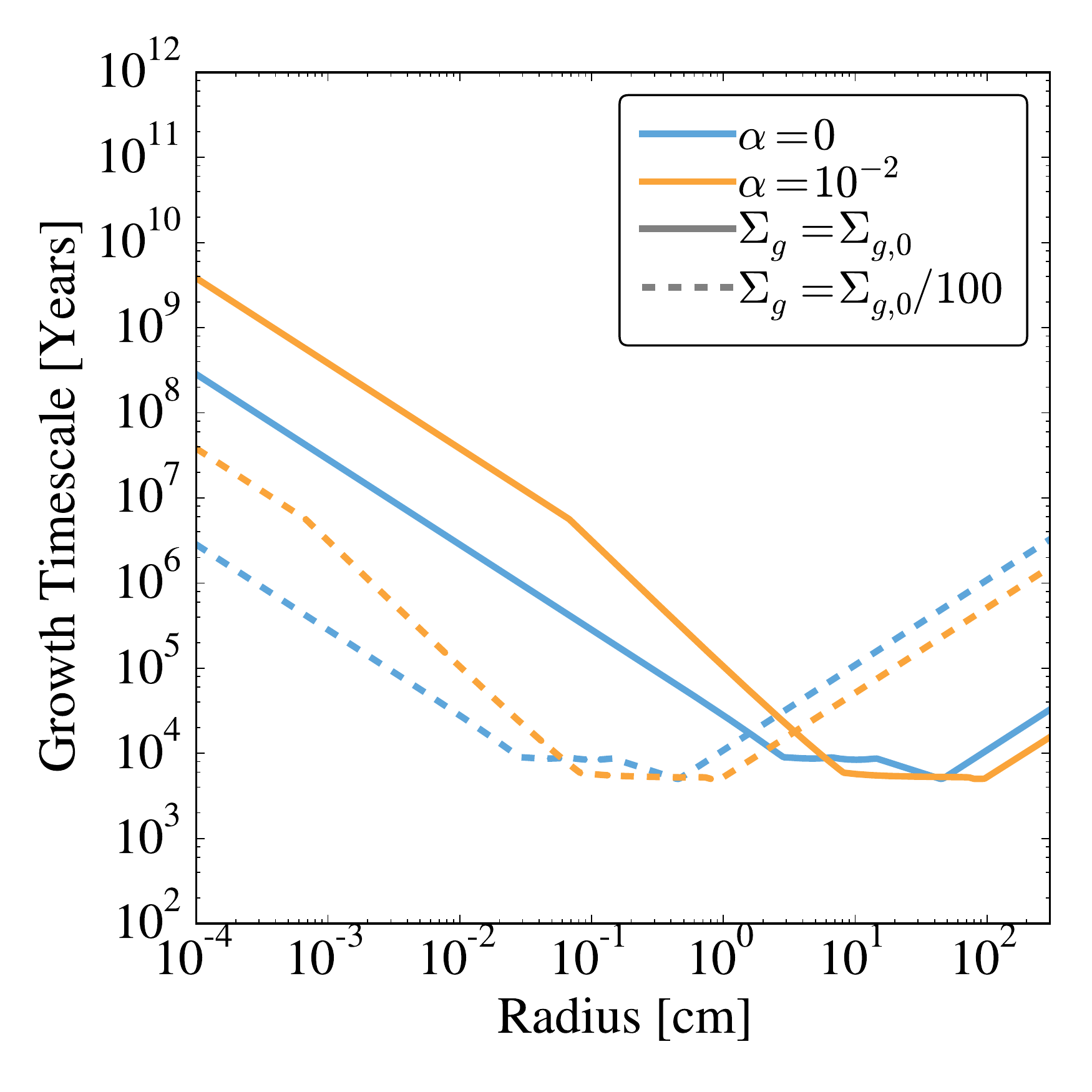}
	\caption{The gas-assisted growth timescale for a disk with its gas surface density depleted by a factor of 100. The values shown are for $a = 30 \, \text{AU}$, and $M = 1.5 \times 10^{-2} M_\oplus$. Both a laminar ($\alpha = 0$) and a strongly turbulent ($\alpha = 10^{-2}$) case are shown. The solid lines shown the values for the surface density used in the paper, while the dashed lines depict the effect of changing the surface density.}
	\label{fig:sig_0_comp}
\end{figure}

An example of growth in a depleted disk is shown in Figure \ref{fig:sig_0_comp}. As can be seen in the figure, the predominant effect of changing $\Sigma$ is to shift growth down to lower values of $r_s$. This is due to the fact that the quantities that go into calculating $t_{\rm{grow}}$, even the energy criteria, are functions of $r_s$ through their dependence on $St$ alone. Since in the Epstein regime $t_{s} \propto r_s/\rho_g$ and in the Stokes regime $t_{s} \propto r_s^2/\rho_g$, the radius corresponding to a given Stokes number decreases when the surface density, and correspondingly the volumetric density, are decreased. This is what causes the shift to lower radii seen in Figure \ref{fig:sig_0_comp}. Other than this shift however, there is essentially no change in the growth timescale. Said another way, when the timescale is viewed as a function of Stokes number, i.e. if we consider $t_{\rm{grow}} (St)$ as opposed to $t_{\rm{grow}}(r_s)$, then this function is independent of $\Sigma$. 

The sole caveat is that for particles not in a linear drag regime, the Stokes number of a particle is now dependent on the particle's velocity, which means that the Stokes numbers used for calculating different quantities might not be the same. For example we can express the WISH radius as $R_{WS} = \sqrt{G M St /(v_{\rm{gas}} \Omega)}$, but the $St$ value in that expression is defined with respect to $v_{\rm{rel}} = v_{\rm{gas}}$, meaning it will not be the same as the Stokes number for e.g, $v_{pk}$, which assumes of course that $v_{\rm{rel}} = v_{pk}$. In this case the simple argument that all quantities are solely dependent on $St$ breaks down. In practice if we use e.g. the Stokes number defined with respect to laminar drift velocities for comparison purposes, the discrepancy between the two surface densities is minor. Furthermore, if we write the critical radius dividing the fluid and diffuse drag regimes, $r_s = 9 \lambda /4$, in terms of Stokes number, it is straightforward to show the particle is in the fluid regime for

\begin{align}
St_{\rm{crit}} > 7.4 \times 10^{-2} \left( \frac{a}{\text{AU}} \right)^{23/7}  \left( \frac{\Sigma_{g,0}}{500 \, \text{g} \, \text{cm}^{-2}} \right)^{-2} \; .
\end{align}
Because of the strong scaling of $St_{\rm{crit}}$ with $a$, for $a\gtrsim5\,\text{AU}$ the differences between the two surface densities disappear entirely.

Thus, in general the range of small body sizes where accretion is effective will shift to lower values as the gaseous component of the disk dissipates. The subsequent effects of such a shift are quite sensitive to the underlying size distribution of the small bodies. A particularly salient issue here is whether radius or Stokes number controls the processes that produce the size distribution, as could be the case if e.g fragmentation near $St \sim 0.1$ generates an upper cutoff to growth (\citealt{blum_wurm_coll}). If the Stokes number is what matters, then the effects of dissipating the surface density would be rather minimal, provided the surface density of the disk evolves faster than the disk dissipation timescale. If the size distribution is determined by radius however, and there exists a small range of sizes where most of the mass of the disk is located, then the shift in effective accretion range could have substantial effects, either positive or negative, on the accretion rate of small bodies. In the outer disk, for example, the combination of growth and radial drift may set an important particle scale (e.g. \citealt{bke_2012}, \citealt{pms_2017}). The combination of these effects merits further investigation. See \cite{LJ14} for an example of pebble accretion in the presence of particle sizes determined by the interplay of growth and drift (note these authors use a full gas disk, not one depleted in gas surface density).

\subsection{Effects of Varying Stellar Mass} \label{vary_M_*}
In this section we discuss the dynamical effect of varying $M_*$. Clearly, higher mass stars will have a higher overall temperature, should on average have higher surface densities as well. Since we have discussed those effects in previous sections, in this section we consider only the effect of varying $M_*$, leaving the other properties of the disk unchanged.

Changing $M_*$ has an effect on the growth process solely though its effect on the local Keplerian orbital frequency $\Omega$ and the size of the Hill radius $R_H$. Higher mass stars have lower dynamical times, which tends to speed up growth. On the other hand, in the presence of more massive stars the Hill radius of a growing planet will shrink as it becomes more difficult to hold on to accreting material, which clearly inhibits growth. The interplay between these two factors will determine the net effect of varying the stellar mass -- which effect dominates, and therefore whether changing stellar mass is beneficial or detrimental to growth, depends on the other input parameters.

  \begin{figure} [h]
  	\centering
  	\includegraphics[width=0.9\linewidth]{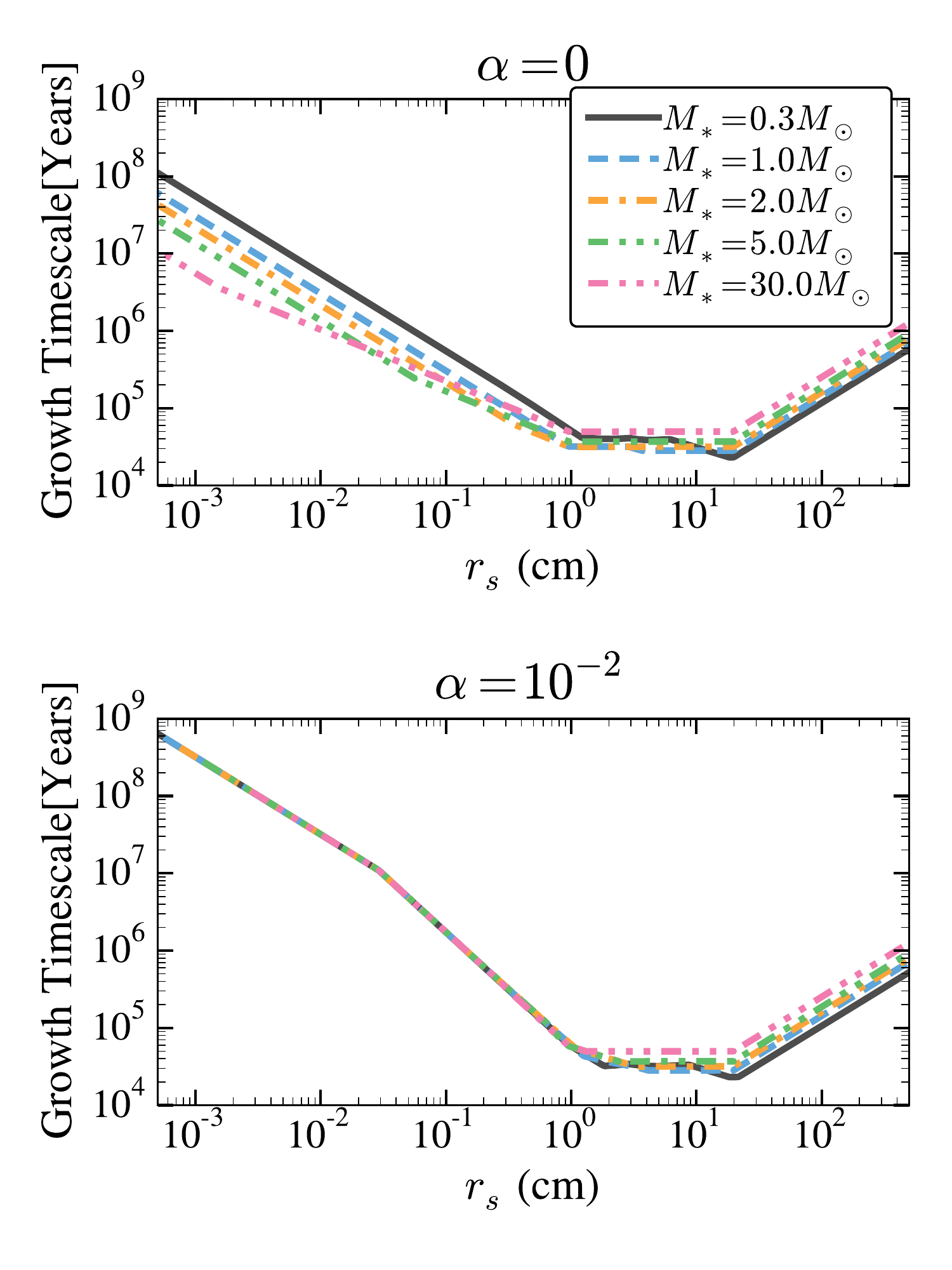}
  	\caption{The effect of varying the mass of the central star. The values shown here are at a distance of $a = 70 \, \text{AU}$, and $M = 0.5 \, M_\oplus$.  The panels show the grow timescale as a function of small body size for a laminar ($\alpha = 0$) and strongly turbulent ($\alpha = 10^{-2}$) disk. Each panel shows the timescale for five different stellar masses. \textit{Top Panel}: In the laminar case the small particles which accrete at $R_{\rm{acc}} = R_{WS}$ are accreted more rapidly around higher mass stars, as more massive stars have higher rates of shear and reduced small body scale height. As particle size increases the particles will begin accreting at $R_{\rm{acc}} = R_H$ -- the effect of stellar mass in this regime depends on the size of $H_{KH}$ relative to $R_H$ (see text). \textit{Bottom Panel}: The inclusion of turbulence allows the scale height of smaller particles to be set by turbulent diffusion instead of the Kelvin-Helmholtz shear instability, making the scale height independent of $M_*$ in this regime. In this case the increase in shear rate is balanced by the decrease in $R_{WS}$, causing the growth curves to merge for low values of $r_s$. For higher values of $r_s$ where $R_{\rm{acc}}  = R_H$ the situation is the same as in the top panel. }
  	\label{fig:M_*_plot}
  \end{figure}
  
An example is shown in Figure \ref{fig:M_*_plot}. The Figure shows a plot of $t_{\rm{grow}}$ vs. $r_s$ for five different stellar masses; the panels shows the growth timescale for two different levels of turbulence. Many of the main features of varying $M_*$ are visible in the figure. For the $\alpha = 0$ case, small particles ($r_s \lesssim 0.1 \, \text{mm}$) are accreted much more rapidly by the more massive stars. In this regime particles have $R_{\rm{acc}} = R_{WS}$ and accrete in 3D, with scale height $H_p = H_{KH}.$. Thus the primary effect here of varying stellar mass is to change $\eta v_k$ and $\rho_g$ along with $\Omega$. In the Epstein regime however, these effects on both $R_{WS}$ and $St$ cancel out, and because $H_{KH} \sim 2 \eta v_k / \Omega$, the increase in growth rate in this regime scales roughly as $\Omega \propto M_*^{1/2}$. Eventually, the particle size increases to the point that $R_{\rm{shear}} < R_{WS}$, which also implies that particles are now shear dominated. Because, in the laminar regime, this change roughly corresponds to where we shift from 3D to 2D accretion, the effect of increasing stellar mass in this regime is actually to \textit{increase} the growth timescale -- $t_{\rm{grow}} \propto (R_{\rm{shear}}^2 \Omega)^{-1} \propto  St^{-2/3} R_H^{-2} \Omega^{-1} \propto M_*^{1/6}$, since $St$ is independent of $M_*$ in the Epstein regime. This is what causes the overlap in $t_{\rm{grow}}$ curves as $r_s$ increases. Due to the fact that changing $M_*$ can switch the regime that determines one of our growth parameters (e.g. shear vs. dispersion dominated) the maximal change in growth rate for two stellar masses $M_{*,1}$ and $M_{*,2}$ in this regime can be complicated. In general, however, the change is of order $\Omega_{1}/\Omega_2 = (M_{*,1}/M_{*,2})^{1/2}$, as can be seen in Figure \ref{fig:M_*_plot}.

As we increase in particle size, the cores reach the point where they can accrete particles at $R_{\rm{acc}} = R_H$. As can be seen in Figure \ref{fig:M_*_plot}, the value of $r_s$ where this transition occurs is a decreasing function of $M_*$ due to the decreased size of the Hill radius at higher stellar mass. Figure \ref{fig:M_*_plot} also shows that increasing stellar mass initially decreases the growth timescale, but for higher $M_*$ further increasing the mass of the star actually increases the growth timescale. This is due to the fact that decreasing $M_*$ increases $H_{KH}$, affecting whether the largest size particles are accreting in the 2D or 3D regime. Low mass stars will accrete in 3D, in which case increasing $M_*$ will decrease the growth timescale by bringing the particle density down and the rate of shear up. Once the stellar mass becomes sufficiently large however, we will have $R_H > H_{KH}$, and the core will accrete in 2D. In this case increasing $M_*$ will slow down growth by decreasing the size of the Hill radius.

As turbulence increases, the difference between the various stellar masses for small particle size disappears. This effect occurs due to a balance between the decreased growth rate from decreasing encounter rate and an increased growth rate due to an increased particle density, since we generally accrete in 3D in this regime. If particles are drift-dispersion dominated, then $v_\infty = v_{pk} \approx v_{\rm{gas}}$ and $R_{\rm{acc}} = R_{WS}$. When $\alpha \neq 0$, $R_{WS}$ is no longer independent of $M_*$. If $\alpha > \eta$, then $v_{\rm{gas}}$ is approximately constant with respect to varying $\Omega$, meaning that the quantity in the denominator of $t_{\rm{grow}}$ -- $R_{WS}^2 v_{pk} \propto M_*^{-1/2}$. Furthermore, if $v_{\rm{shear}} > v_{pk}$, then $R_{\rm{acc}}=R_{\rm{shear}}$, and the denominator of $t_{\rm{grow}}$ is $\propto R_{\rm{shear}}^3 \Omega \propto M_*^{-1/2}$. For sufficiently strong turbulence that $H_{turb} > H_{KH}$, we have $H_p \propto \Omega^{-1} \propto M_*^{-1/2}$. Thus for 3D accretion in the strong turbulence regime these effects approximately cancel out, causing the growth curves to merge. 

For high core masses, the transition on the right side of the graphs where $t_{\rm{grow}}$ increases as a function of $r_s$ occurs at essentially the same value of small body radius, independent of $M_*$. As noted in Equation \eqref{eq:st_crit_num}, the energy criteria $KE < W$ can be rewritten as $St < 4\sqrt{3}$ for particles in this growth regime. For particles in the Epstein regime the Stokes number is independent of $M_*$, causing the cutoff radius to be approximately the same for all stellar masses. The timescale in this regime is weakly dependent on $M_{star}$ -- these large particles accrete at $t_{\rm{Hill}}$, with an increase in growth timescale $\propto St$, which again is essentially independent of $M_*$. Thus the growth timescale goes as $t_{\rm{grow}} \propto (R_H v_H)^{-1} \propto M_*^{1/6}$.

In the laminar regime the cutoff at small radii is also independent of $r_s$. This cutoff occurs when $R_{WS} < R_b$, and since, as previously noted, $R_{WS}$ is independent of $M_*$ in the laminar regime, this cutoff is independent of $M_*$ as well. As turbulence increases $R_{WS}$ becomes a decreasing function of $M_*$, causing the cutoff in growth to shift to higher values of $r_s$.

For low core masses, the low end cutoff still has the same dependence on $M_*$. However, growth will now cut off at the limit given by Equation \eqref{eq:st_crit_param_space}, which is an increasing function of $M_*$ in both the laminar and turbulent regimes. 

\section{Flow Isolation Mass} \label{flow_iso_mass}
In this section we discuss accretion in the regime $R_b > R_H$. In general, the core's atmosphere extends up to a scale $R_{\rm{atm}} = \min(R_b,R_H)$, as once $R_b > R_H$ $R_{\rm{atm}}$ is limited by gravitational effects from the central star, as opposed to the core's ability to bind nebular gas. Thus for core masses large enough that $R_b > R_H$ the energy criteria discussed in Section \ref{energy} are the same with $R_b \rightarrow R_H$.

\begin{figure} [h]
	\centering
	\includegraphics[width=\linewidth]{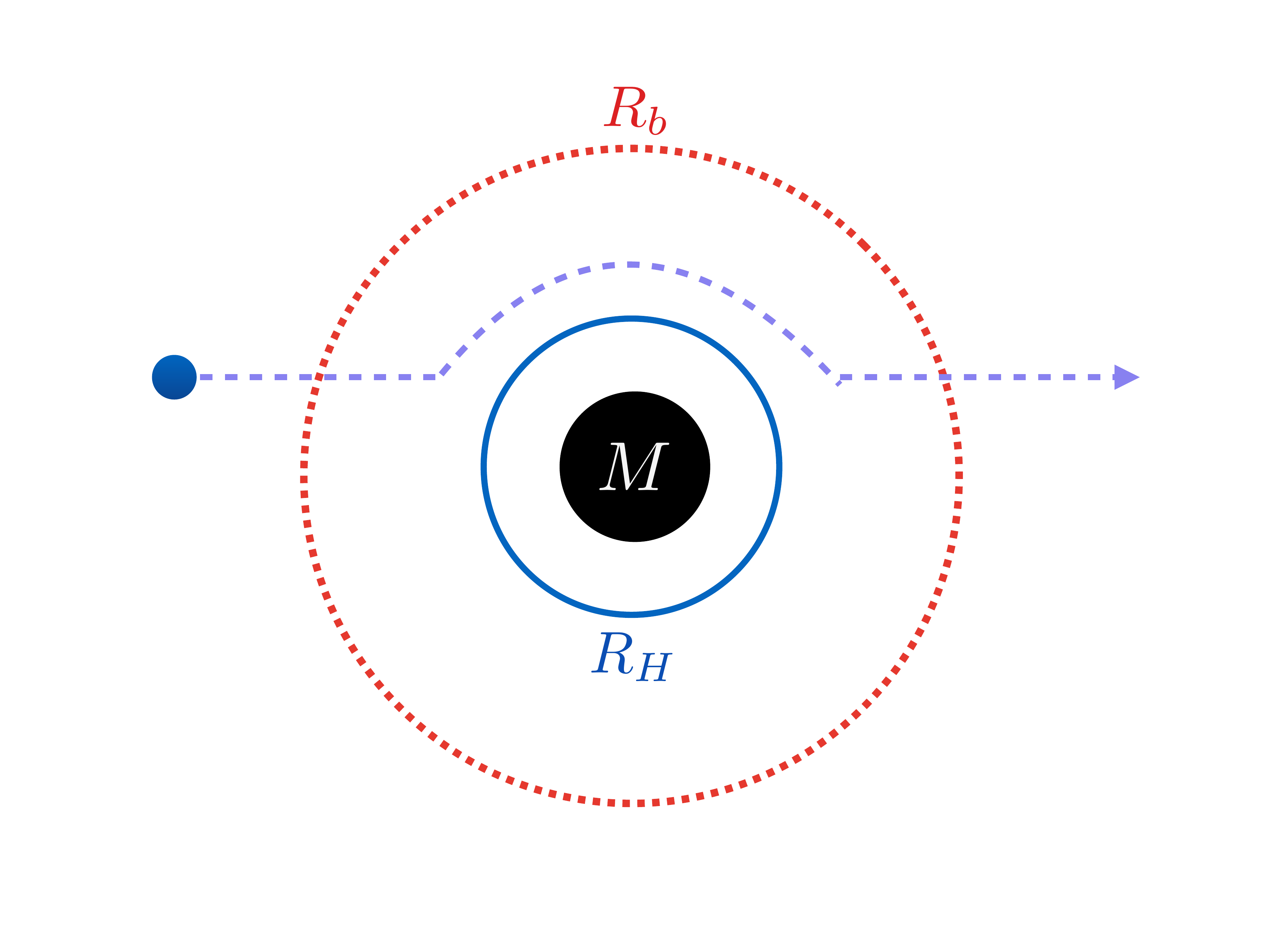}
	\caption{A schematic illustration of the trajectories of particles with $KE<W$ for $M>M_{\rm{flow}}$. The particle (blue circle), comes in from the left. Since the particle is able to dissipate its kinetic energy relative the core, it begins to follow the local gas flow. The core's atmosphere extends up to $R_H$, and the nebular gas flows around this obstacle. The particle is pulled along with the gas, causing it to flow around the core without being accreted.}
	\label{fig:flow_iso_illus}
\end{figure}

Figure \ref{fig:heatmap_a1} shows the emergence of a feature for high core masses where only larger sizes of particles can accrete. This occurs when the core reaches a mass $M_{\rm{flow}}$ such that $R_b = R_H$. As discussed in Section \ref{energy}, in the regime $R_H > R_b$, small bodies with $R_{\rm{stab}} < R_{b}$ will not be able to accrete if they dissipate their kinetic energy during their interaction with the core -- since the gas will flow around the core's approximately incompressible static atmosphere, particles that couple to the gas flow near the core will be pulled around without accreting. For lower mass cores, only particles that are small enough that $R_{WS} < R_b$ are restricted from accreting in this manner, i.e. this consideration dictates the lower limit on the particle size that can be accreted. Once the core's mass is large enough that $R_b>R_H$ however, we now have $R_{\rm{atm}} = R_H$, and therefore $R_{\rm{stab}} \lesssim R_{\rm{atm}}$ for all particle sizes. Thus, any particles that dissipate their kinetic energy relative to the core will be pulled around by the gas flow without accreting. Because $R_{\rm{atm}} = R_H$, we expect the gas flow to be around $R_H$, instead of $R_b$ as it was in the lower mass case. See Figure \ref{fig:flow_iso_illus} for an illustration. 

We note that the gas will be accelerated by the core's gravity as it passes through $R_b$; interior to $R_b$ the local orbital velocity exceeds the sound speed, which means the flow can accelerate to supersonic velocities.  In this case, it is less clear that the core's atmosphere will act as an incompressible obstacle.  However, there still should exist a scale $r$ past which the flow can no longer penetrate the core's atmosphere. To see this, we can compare the incoming kinetic energy of the flow to the binding energy of the core's atmosphere. In a time $\Delta t$ the mass of gas entering into the length scale $r$ is of order $\rho_{\rm{neb}} v_{\rm{app}} \Delta t r^2$, where $\rho_{\rm{neb}}$ is the volumetric mass density of the nebular gas and $v_{\rm{app}}$ is the velocity of the incoming gas relative to the core. Since $R_b \gtrsim R_H$ implies that $v_H \gtrsim c_s > v_{\rm{gas}}$ we have $v_{\rm{app}} \lesssim v_H$.  The timescale for gas to enter $r$ is $\Delta t \sim r/v_{\rm{app}}$. The binding energy of the atmosphere at scale $r$ is $\sim G M^2/r \sim \rho_{\rm{atm}} v_{esc}^2 r^3$. Thus, the ratio of the incoming kinetic energy to the binding energy is
\begin{align}
\frac{\rho_{\rm{neb}} v_{\rm{app}}^2 r^3}{\rho_{\rm{atm}} v_{\rm{esc}}^2 r^3} \leq \frac{\rho_{\rm{neb}} r}{3 \rho_{\rm{atm}} R_H} \; ,
\end{align}
where we've used the fact that $v_{\rm{app}} \leq v_H$ and that $v_H^2/v_{esc}^2 = r/(3 R_H)$. The quantity above is clearly $\ll1$ for $r \lesssim R_H$, particularly since close to the planet we expect $\rho_{\rm{atm}} \gg \rho_{neb}$. Thus there exists a scale $r \lesssim R_H$ where the incoming kinetic energy of the gas is much less than the binding energy of the core's atmosphere.

The core reaching $M_{\rm{flow}}$ signals a rapid cutoff in the accretion of pebbles: for masses just below $M_{\rm{flow}}$ there will always exist a wide range of particle sizes that dissipate their kinetic energy during the interaction with the core. Once $M > M_{\rm{flow}}$ accretion of these particle sizes, which generally represent the most rapid accretion rates, suddenly shuts off. We can demonstrate analytically that a broad range of particle sizes satisfying $KE<W$ will be present for $M=M_{\rm{flow}}$. To begin, we first show that at this mass scale, we never have $R_{\rm{stab}} = R_{WS}$. Firstly, $R_b = R_H$ implies that $v_H =c_s/\sqrt{3}$. The analytic criterion for the relative sizes of $R_b$, $R_{WS}$, and $R_{\rm{shear}}$ for $R_b = R_H$ can therefore be summarized as:
\begin{align*}
R_{\rm{shear}}<R_{WS}&:\;\;St>\sqrt{3}\left(\frac{v_{\rm{gas}}}{c_s}\right)^{3}
\\R_{WS}>R_{b}&:\;\;St>\frac{1}{\sqrt{3}}\left(\frac{v_{\rm{gas}}}{c_{s}}\right)
\end{align*}
Since $v_{\rm{gas}}/c_s \ll 1$, the above implies that $R_{\rm{stab}} = R_{\rm{shear}}$ occurs prior to accretion commencing for $R_{WS} > R_b$. Thus, accretion commences for $R_{\rm{shear}} > R_b$, which occurs at a Stokes number of
\begin{align}
St > 9 \left(\frac{v_H}{c_s}\right)^6 = \frac{1}{3} \; .
\end{align}
On the other hand, accretion will cease when $KE>W$. Since these particles are clearly shear dominated, we will simply have our criterion given in Equation \eqref{eq:st_crit_num} for particles which shear into $R_H$
\begin{align}
St < 4 \sqrt{3} \; .
\end{align}
Thus, any particle in the range $1/3 < St < 4 \sqrt{3}$ will be accreted for masses just below $M_{\rm{flow}}$, but for $M>M_{\rm{flow}}$ all particles in this range will no longer be accreted, i.e. $M > M_{\rm{flow}}$ represents a general point throughout parameter space where accretion of small bodies through pebble accretion cuts off.

Solving $R_b = R_H$ for $M$ gives $M_{\rm{flow}}$ as

\begin{align}
M_{\rm{flow}} =\frac{1}{\sqrt{3}} \frac{c_s^3}{G \Omega} \; .
\end{align}
Plugging in fiducial values of our parameters gives
\begin{align}
M_{\rm{flow}} &= 4.4 \left( \frac{ T_0 }{200 \, \text{K}} \right)^{3/2} \left( \frac{a}{\text{AU}}\right)^{6/7}  \left( \frac{M_*}{M_\odot}\right)^{-1/2}\,M_\oplus \; .
\end{align}

\begin{figure} [h]
	\centering
	\includegraphics[width=\linewidth]{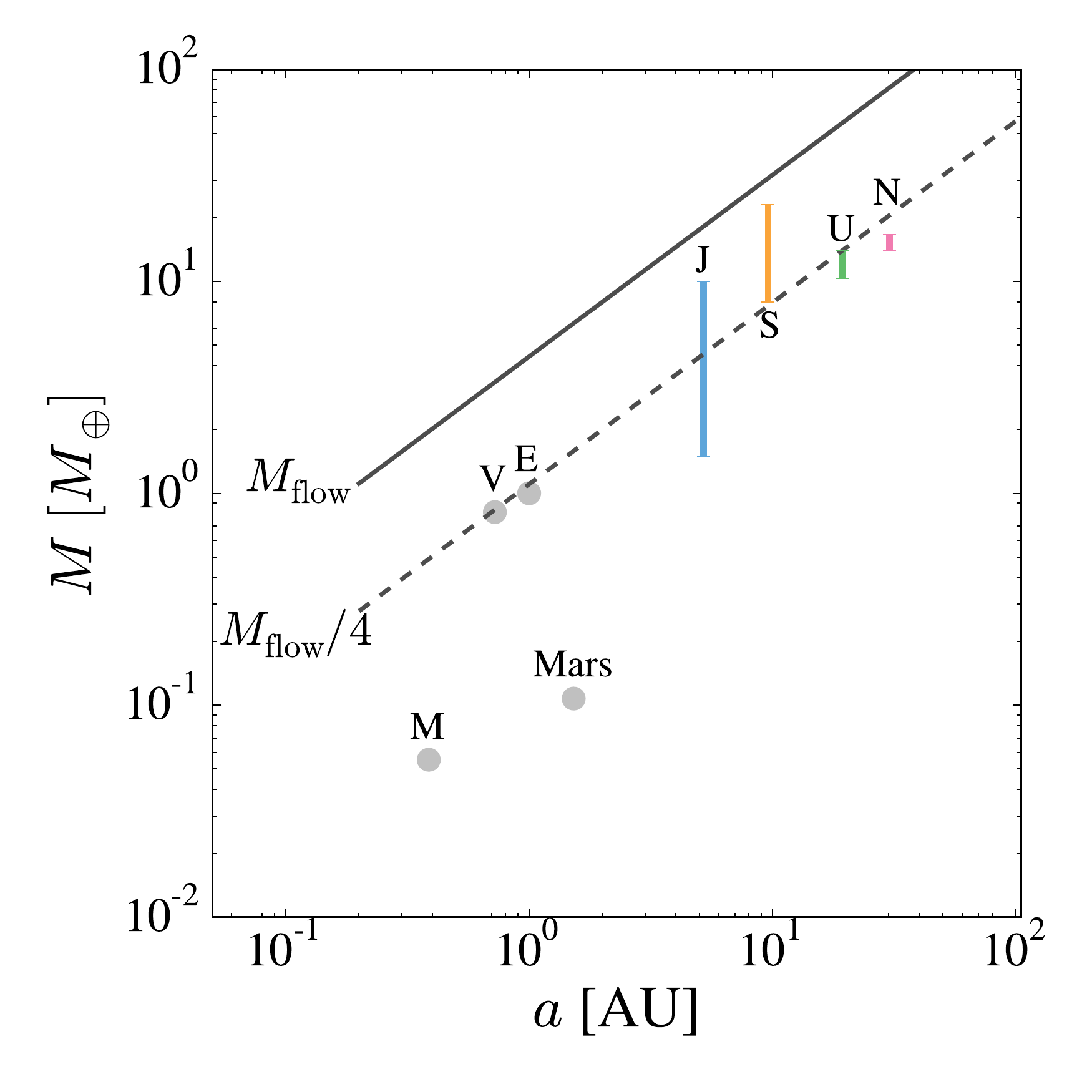}
	\caption{The value of the flow isolation mass (solid line), as well as the flow isolation mass scaled down by a factor of four (dashed line), both plotted as a function of semi-major axis. Also shown are the masses of the cores of the solar system planets. Values for the four giant planets are taken from \cite{g_2005}. For the terrestrial planets the total mass is shown. For the gas giant the mass of the core is used, whereas for the ice giants the total mass in solids is plotted.}
	\label{fig:flow_iso}
\end{figure}

For the temperature profile used in this work, the value of flow isolation mass is markedly similar to the distribution of solar system cores. Figure \ref{fig:flow_iso} shows the flow isolation mass scaled down by a factor of 4 as a function of semi-major axis. This corresponds to a cutoff in accretion for $R_b = 4^{-2/3} R_H \approx 0.4 R_H$, as opposed to $R_b = R_H$. For the terrestrial planets we have plotted the total mass of the planet. For the gas giants the bars indicate the range of possible masses, since these values are not as well constrained. For the gas giants the mass of the cores, as opposed to the total mass in solids, are shown, as once runaway gas accretion begins the amount of solids in the planet will not be set by the flow isolation mass. The values plotted are taken from Figures 7 and 8 of \cite{g_2005}, with the maximal range of core masses shown. For the ice giants we use the total mass in solids, since the flow isolation mass will more directly influence this number if runaway gas accretion does not occur. Again, these values are taken from \cite{g_2005} (Section 3.4). The correspondence between the flow isolation mass and the masses of the solar system cores may indicate that the flow isolation mass played a role in influencing the final masses of the solar system planets. 

We stress that this figure should not be over interpreted, as there a variety of factors that complicate the formation of the solar system planets. In particular, meteoritic dating (e.g. \citealt{yjy_2002}, \citealt{ktb_2009}) and dating of the Moon-forming impact (e.g. \citealt{bvm_2015}) provide strong evidence that the final assembly of the terrestrial planets occurred on timescales $\gtrsim 10 \, \rm{Myr}$ and that a period of ``giant impacts'' was important for setting the final masses of these planets. Because protoplanetary disks do not last longer than $10 \, \rm{Myr}$, nebular gas would not be present and gas-assisted growth would not occur during this phase. It is possible that the flow isolation mass played a role in setting in the initial embryo masses that underwent this phase of giant impacts. Any such scenario however, would have to explain the low masses of Mercury, Mars, and the inferred mass of the Moon-forming impactor (\citealt{c_2012}). It is conceivable that some cores reached flow isolation, while others stalled at low core mass due to the inefficiency of growth at low core masses described in Section \ref{var_a_M}. If all cores reached flow isolation, giant impacts could in principle remove mass from some, as has been proposed to explain the anomalously high density of Mercury (e.g. \citealt{bsc_1988}). Modeling by \cite{ls_2012} demonstrates that, for collision events typical of the final stages of terrestrial planet formation, the outcomes span the range from perfect accretion to erosion of the more massive target. Note that any scenario invoking flow isolation in the inner solar solar system is contrary to standard models of terrestrial planet formation, which rely on the giant impacts phase to increase the masses of the terrestrial planets above their isolation mass (e.g. \citealt{acl_1999}).

The flow isolation mass may be most relevant in the context of the ice giant planets, which were previously thought to form by growing to their local isolation mass, though note that the ice giants are thought to have migrated substantially from the location of their initial formation (reviewed e.g. in \citealt{m_2008}). In the context of pebble accretion however, planets are not limited by locally available material due to radial drift of solids. The flow isolation mass could provide a plausible mechanism that sets the mass of these planets in the context of pebble accretion (see \citealt{fm_2017} for a more in depth discussion).

We note here that a similar mass is identified by \cite{LJ14}, who refer to it as the ``Pebble Isolation Mass." We strongly emphasize, however, that the existence of the pebble isolation mass is based on different physics than our flow isolation mass: the pebble isolation mass is based on the gravity of the planet opening a gap in the pebble disk, as opposed to the planet altering the local flow of nebular gas. \cite{LJ14} calculate the pebble isolation mass by identifying the point where gravitational perturbations from the growing core's gravity render the gas velocity immediately outside of the planet's orbit ``super-Keplerian," which tends to push pebbles outwards rather than bring them in. \citeauthor{LJ14} perform numerical simulations at 5 AU in order to determine the pebble isolation mass at this orbital separation. They then calculate the dependence of $M_{\rm{flow}}$ on the disk aspect ratio, $H/a$, analytically, and determine that $M_{\rm{flow}} \propto (H/a)^3$. Combining these results, \citeauthor{LJ14} give the pebble isolation mass as

\begin{align}
M_{\rm{flow}} \approx 20 \left( \frac{a}{5 \, \text{AU}} \right)^{3/4} M_\oplus \; .
\end{align}
Despite the different physics used to calculate these mass scales, the flow isolation mass and the pebble isolation mass occur at roughly the same value -- to see this, we first note that since $H/a \propto c_s/(\Omega a) $, we have $(H/a)^3 \propto T^{3/2} a^{3/2}$. Thus our scaling agrees with the \citeauthor{LJ14} result. Furthermore, if we use a temperature in agreement with these authors' choice, $T \approx 270\,\text{K}\, (a/\text{AU})^{-1/2}$, then our criterion for pebble isolation mass, i.e. that $R_b = R_H$, gives  

\begin{align}
M_{\rm{flow}} = 23.2 \left( \frac{a}{5 \, \text{AU}} \right)^{3/4} M_\oplus \; ,
\end{align}
in rough agreement with the \citeauthor{LJ14} result. Finally, we note that the flow isolation mass is similar in scale to the ``Thermal mass" discussed in Section \ref{xu_comp}, past which the core's Hill radius exceeds the disk scale height (\citealt{lin_gap}). 

\section{Summary and Conclusions} \label{summary}

In this paper we have presented an order of magnitude model of pebble accretion, which accounts for the effects of turbulence on a variety of the timescale parameters. We calculate the growth timescale for a planet as
\begin{align}
t_{\rm{grow}} = \frac{M H_p}{2 f_s \Sigma v_\infty R_{\rm{acc}} H_{\rm{acc}}} \; .
\end{align}
We calculate $v_\infty$, $R_{\rm{acc}}$, $H_{\rm{acc}}$, and $H_p$ separately, allowing a number of different physical processes to set each of these velocity and length scales. Our model uses the wind-shearing radius of \cite{pmc11} to take into account the effects of gas drag on the stability of small bodies during the accretion process, which can set $R_{\rm{acc}}$ instead of $R_H$ or $R_b$. We also use the approximate formulae presented by \cite{oc07} for the RMS turbulent velocity of small bodies in a turbulent medium to incorporate the effects of turbulence into $v_\infty$. An incoming small body has its incoming kinetic energy compared to the work done by gas drag during the encounter, which determines the range of small body sizes that the core can accrete. The resulting model gives the growth timescale as well as a variety of other important parameters ($v_{pk},H_p,F_D,\ldots$). Due to its relative simplicity, our model can be applied over a large range of parameter space, and can be coupled with other physical processes, such as planetary migration. 

Studying the output of our model reveals many important of aspects of protoplanetary growth via pebble accretion in the presence of turbulence. Once protoplanets reach a large enough size (ranging from $10^{-4} - 10^{-1} M_\oplus$, depending on the strength of turbulence, as well as the location in the disk and the sizes of pebbles available), growth timescales become far shorter than the lifetime of the gaseous protoplanetary disk. This results remains true even for extremely strong ($\alpha \gtrsim 10^{-2}$) turbulence.  These enhanced growth rates are more than substantial enough to allow the formation of gas giants at wide orbital separations, where planetesimal accretion is inefficient, provided the cores can first reach this high mass.

Of equal importance however, are the regions where pebble accretion is not as efficient. We find that turbulence can substantially lower the growth rate at low core masses. For lower core masses and stronger turbulence, a smaller range of particle sizes are able to accrete at $t_{\rm{Hill}}$, and it becomes easier for turbulence to cut off gas-assisted growth for more massive particles entirely. These effects are exacerbated at wide orbital separations, where the detrimental effects of the gas on the accretion process are more substantial. Thus when studying the growth in pebble accretion it is important to consider not just the maximal accretion efficiency where the core accretes over its Hill radius, but to consider what sizes of small bodies are available and how these small bodies are affected by their interactions with the gas. These effects can have considerable ramifications for the predictions of any theory of planet formation by pebble accretion.

While we have used fiducial values of disk parameter in order to provide concrete numeric results in this paper, our model is quite flexible with regards to the disk parameters used. We briefly discussed the effects of modifying a few of these parameters, namely the temperature and surface density profiles and the stellar mass. While these effects are complicated, we can briefly summarize them as follows:

\begin{itemize}
	\item For higher temperatures, accretion generally slows down. The timescale for the most rapid accretion, where particles accrete over the entirety of the core's Hill sphere, is unaffected by disk temperature, though the range of particle sizes that can accrete at this rate may shrink in hotter disks.
	\item Depleting the gas surface density shifts the correspondence between Stokes number and small body radius, essentially shifting the curve of $t_{\rm{grow}}(r_s)$ to smaller values of particle size. The scale of $t_{\rm{grow}}$ is unchanged.
	\item In a laminar disk, increasing stellar mass makes accretion of smaller particle radii, which accrete at $R_{\rm{acc}} = R_{WS}$, more efficient due to the increased gas density and shear rate. In a turbulence disk the growth rate for small particle radii is insensitive to stellar mass. For larger particle sizes where $R_{\rm{acc}} = R_H$ the effect of increasing stellar mass is not as clear cut, but the overall effect is much less significant than in the small radius case.
\end{itemize}

Finally, we identified a natural upper limit to core mass in the context of pebble accretion, the ``Flow Isolation Mass." Past this mass the Bondi radius of the core will exceed its Hill radius, in which case the region where the core's gravity alters the gas flow exceeds the radius for stable orbits about the core, regardless of the small body size being accreted. In this regime the normal mechanism for pebble accretion will not be able to operate, since particles that dissipate their kinetic energy relative to the core will follow the flow of gas and be pulled around the core without accreting. This upper limit may set the critical mass that triggers run away gas accretion, since once cores reach this limit their accretion luminosity will drop, allowing them to accrete substantially more gas from the surrounding nebula. We note that the value of the flow isolation mass as a function of semi-major axis is quite similar to the distribution of the cores of the solar system planets. While this is an intriguing possibility, further study is needed to determine the importance of this mass scale.

An interesting and important extension to our model would be to consider lower mass cores, for which the effects of gas are more pronounced. This would require modeling not just the effects of the RMS random velocity of the small bodies, but also the particle-particle relative velocity between the small body and the core. Expressions for the turbulence induced relative velocity are given in e.g. OC07, but these formulae are much more complex than Equation \eqref{RMSturb}. Another effect that may be more important for lower masses is consideration of the full probability density function of the particle-particle relative velocity. In this work we have used only the RMS value for velocity, but considering particles in the low and high velocity tails of the distribution can have important effects, as has been noted for early stages of growth (\citealt{wbod_2012}). Our preliminary investigations of gas-assisted growth at planetesimal sizes show that many novel features appear in this regime that are not present in the higher mass case -- for example the range of particle sizes that planetesimals can accrete efficiently can be much narrower than the range for protoplanets. This effect could lead to stratification in the composition of planetesimals, which could be observed in our solar system. Furthermore, for low masses the actual collision velocity between the core and small body can be smaller than their initial relative velocity, due to the inspiral of the particle, which can have important ramifications for whether a given collision results in growth, bouncing, or fragmentation.

There are a large number of applications for our model to address issues in planet formation. In Paper II (submitted) we present one such application: we apply our theory to the question of formation of gas giants at wide orbital separation. We are particularly interested in how the strength of turbulence can help place restrictions on when gas giant formation is possible, which may help us understand why wide orbital separation gas giants are so uncommon even though pebble accretion timescales are so rapid.

\vspace{2mm}

\noindent The authors wish to thank Diana Powell, Renata Frelikh, and John McCann for their useful discussion, and Eugene Chiang for his thoughtful suggestions on the manuscript. We also thank the anonymous referee for their helpful comments, which improved the quality of the manuscript. MMR and RMC acknowledge support from NSF CAREER grant number AST-1555385. HBP is supported by the Israel science foundation I-CORE grant 1829/12 and the MINERVA center for life under extreme planetary conditions.

\appendix

\section{Summary of Calculation Algorithm}\label{app:sum}

In this appendix, we summarize the recipe for calculating the growth timescale.

Our model takes in five input parameters: $M_*,a,M,r_s,$ and $\alpha$. Once these parameters are specified we can calculate the parameters of the protoplanetary disk at the given orbital separation (see Table \ref{tab:param_table}). We then need to  calculate the Stokes number of the small body $St \equiv t_s \Omega$. Particles with $r_s < 9 \lambda /4$ (which applies over most of parameter space) are in the Epstein regime, which allows us to immediately calculate their stopping time:
\begin{align}
t_{s,\rm{Eps}} = \frac{\rho_s}{\rho_g} \frac{r_s}{v_{th}}  \; .
\end{align}
For $r_s > 9 \lambda /4$ the stopping time is calculated numerically. We begin by setting $v_{pg} = v_{\rm{gas}} \equiv \sqrt{\alpha c_s^2 + \eta^2 v_k^2}$. We then calculate the drag force on the particle using
\begin{align}
F_D = \frac{1}{2} C_D  (Re) \pi r_s^2 \rho_g v_{pg}^2 \; ,
\end{align}
where
\begin{align}
C_D(Re) = \frac{24}{Re} \left ( 1 + 0.27 Re \right)^{0.43} + 0.47 \left [ 1 - \exp \left( - 0.04 Re^{0.38} \right) \right] \; ,
\end{align}
and $Re = 4 r_s v_{pg} / (v_{th} \lambda)$. The stopping time is then given by
\begin{align}
t_s = \frac{m v_{pg}}{F_D} \; ,
\end{align}
where $m$ is the mass of the small body. Using this stopping time we can recalculate $v_{pg}$. The relevant equations are
\begin{align} \label{eq:app_v_pg_ell}
v_{pg,\ell} = \eta v_k St \frac{ \sqrt{4+St^2} }{1+St^2} \; ,
\end{align}
and
\begin{align} \label{eq:app_v_pg_turb}
v_{pg,t}=\begin{dcases*}
\sqrt{\alpha} c_s\left(\frac{St^{\prime 2} \left(1-Re_t^{-1/2} \right)}{(St^\prime+1) \left(St^\prime+Re_t^{-1/2} \right)}\right)^{1/2}, & $St^\prime < 10$\\
\sqrt{\alpha} c_s \sqrt{\frac{St^\prime}{1+St^\prime}}, & $St^\prime \geq 10$
\end{dcases*}
\end{align}
Here $Re_t \equiv \alpha c_s H_g / (v_{th} \lambda)$ and $St^\prime \equiv t_s/t_{\rm{eddy}}^\prime$, where
\begin{align} \label{eq:app_eddy_cross}
t_{\text{eddy}}^\prime &=\frac{\Omega^{-1}}{\sqrt{1+\left(\frac{v_{pg,\ell}}{\sqrt{\alpha} c_s}\right)^2}}  \; .
\end{align}
The total velocity relative to the gas is
\begin{align} \label{eq:app_comb_pg}
v_{pg} = \sqrt{v_{pg,\ell}^2 + v_{pg,t}^2} \; .
\end{align}
Using this new velocity we can recalculate $F_D$ and obtain a new value of $t_s$. We then iterate this process until we obtain the desired accuracy for $St$.

Once $St$ is known, we calculate the length scales needed to determine the growth timescale. We first calculate $R_{\rm{stab}}$:
\begin{align}
R_{\rm{stab}} = \min(R_{WS},R_{\rm{shear}},R_H)
\end{align}
Here $R_H$ is the planet's Hill radius:
\begin{align}
R_H = a \left( \frac{M}{3 M_*} \right)^{1/3} \; .
\end{align}
$R_{WS}$ is the planet's WISH radius:
\begin{align}
R_{WS}=\sqrt{\frac{G M m}{F_D(v_{\rm{gas}})}} \; .
\end{align}
For a particle in a linear drag regime, there is a simple analytic expression for $R_{WS}$:
\begin{align}
R_{WS} = R_H \sqrt{3 St \left( \frac{v_H}{v_{\rm{gas}}} \right)} \; .
\end{align}
$R_{\rm{shear}}$ is the shearing radius, which is the solution to the equation
\begin{align} \label{eq:app_r_shear}
R_{\rm{shear}} = \sqrt{\frac{G M m}{F_D(R_{\rm{shear}} \Omega)}} \; .
\end{align}
In non-linear drag regimes, we solve this equation numerically. For a particle in a linear drag regime, the above equation has the analytic solution
\begin{align} \label{eq:app_r_shear_guess}
R_{\rm{shear}} = R_H \left(3 St \right)^{1/3} \; .
\end{align}

Using $R_{H}$ and the planet's Bondi radius
\begin{align}
R_b = \frac{G M}{c_s^2}
\end{align}
we calculate $R_{\rm{atm}}$:
\begin{align}
R_{\rm{atm}} = \min(R_b,R_H) \; ,
\end{align}
which in turn tells us the impact parameter for accretion:
\begin{align}
R_{\rm{acc}} = \max(R_{\rm{stab}},R_{\rm{atm}}) \; .
\end{align}

The scale height of the small bodies is determined by the Kelvin-Helmholtz scale height
\begin{align}
H_{KH} = \frac{2 \eta v_k}{\Omega} \min(1,St^{-1/2}) \; ,
\end{align}
and turbulent scale height
\begin{align}
H_t = H_g\min\left(\sqrt{ \frac{\alpha}{St} },1\right) \; .
\end{align}
We take $H_p$ to be:
\begin{align}
H_p = \max(H_t,H_{KH}) \; .
\end{align}
Using $R_{\rm{acc}}$ and $H_{p}$ we can determine $H_{\rm{acc}}$:
\begin{align}
H_{\rm{acc}} = \min\left(R_{\rm{acc}},H_{p}\right) \; .
\end{align}

We now calculate the approach velocity of the small bodies, $v_\infty$. The laminar and turbulent components of the velocity due the particle's interactions with the gas are given by
\begin{align} \label{eq:app_v_pk_ell}
v_{pk,\ell} = \eta v_k \frac{ \sqrt{1+4 St^2} }{1+St^2} \; ,
\end{align}
and
\begin{align} \label{eq:app_v_pk_turb}
v_{pk,t}=\begin{dcases*}
\sqrt{\alpha} c_s\left(1-\frac{St^{\prime 2} \left(1-Re_t^{-1/2} \right)}{(St^\prime+1) \left(St^\prime+Re_t^{-1/2} \right)}\right)^{1/2},& $St^\prime < 10$\\
\sqrt{\alpha} c_s \sqrt{\frac{1}{1+St^\prime}}, & $St^\prime \geq 10$
\end{dcases*}
\end{align}
where again $St^\prime \equiv t_s/t_{\rm{eddy}}^\prime$, with $t_{\rm{eddy}}^\prime$ given by Equation \eqref{eq:app_eddy_cross}.
The total velocity is again given by
\begin{align} \label{eq:app_comb_pk}
v_{pk} = \sqrt{v_{pk,\ell}^2 + v_{pk,t}^2} \; ,
\end{align}
The value of $v_\infty$ is then given by 
\begin{align} \label{eq:app_v_infty}
v_\infty = \max(v_{pk},v_{\rm{shear}}) \; ,
\end{align}
where $v_{\rm{shear}} = R_{\rm{acc}} \Omega$.

We now have enough information to calculate the growth timescale $t_{\rm{grow}}$:
\begin{align} \label{eq:t_grow_app}
t_{\rm{grow}} = \frac{M H_p}{2 f_s\Sigma v_{\infty} R_{\rm{acc}} H_{\rm{acc}}} \; .
\end{align}
In order to determine whether a particle can accrete, we calculate its incoming kinetic energy
\begin{align}
KE = \frac{1}{2} m v_\infty^2 \; ,
\end{align}
and the work done by gas drag
\begin{align}
W = 2 F_D(v_{\rm{enc}}) R_{\rm{acc}} \; .
\end{align}
Here $v_{\rm{enc}}$ is the velocity of the particle during its interaction with the core
\begin{align} \label{eq:v_enc}
v_{\rm{enc}} = \begin{dcases}
\max\left(v_{\rm{orbit}}, v_{pg}\right), & v_\infty < v_{\rm{orbit}} \\
\max\left(v_{\rm{kick}}, v_{pg}\right), & v_\infty > v_{\rm{orbit}}
\end{dcases} 
\end{align} 
where $v_{\rm{orbit}} = \sqrt{G M / R_{\rm{acc}}}$, and $v_{\rm{kick}} = G M / (R_{\rm{acc}} v_\infty)$. Particles can accrete if
\begin{center}
\begin{minipage}{0.4\textwidth}
\begin{enumerate}
	\item $R_{\rm{stab}} > R_b$ and $W>KE$
	\item  $R_{\rm{stab}} < R_b$ and $W<KE$
    \item $R_{\rm{stab}} = R_H$ and $v_\infty = v_H$    
\end{enumerate}
\end{minipage}
\end{center}
If the particle does not fall into any of these regimes then we set $t_{\rm{grow}} = \infty$. In case 3 the growth timescale is given by
\begin{align}
t_{\rm{grow}} = \frac{t_{\rm{grow}}^\prime}{\min(1,W/KE)} \; .
\end{align}
where $t_{\rm{grow}}^\prime$ is given by Equation \eqref{eq:t_grow_app}.

\section{Derivation of Velocity Formulae in Different Frames}
 
 In this appendix we give detailed derivations of the equations given in Section \ref{rey_avg}. In what follows we assume that the turbulence is ergodic, i.e. that time averaging the system is equivalent to ensemble averaging, and that the turbulence is a stationary process.
 
 The methods for calculating turbulent velocity in e.g \cite{volk} begin by decomposing the total velocity $\boldsymbol{v}$ into time averaged and fluctuating components, a technique known as ``Reynolds averaging." (See \citealt{cuzzi93}, Appendix A). That is, we take $\boldsymbol{v} = \bar{\boldsymbol{v}} + \boldsymbol{\delta v}$, such that $\braket{\boldsymbol{\delta v}} =0$, where $\braket{\ldots}$ denotes ensemble averaging. $\boldsymbol{\bar{v}}$ is associated with the laminar component of velocity while $\boldsymbol{\delta v}$ is associated with the turbulent velocity. We can use this same decomposition to determine how to combine the laminar and turbulent components as well as how to compute the velocity after changing reference frames.
 
 We first note that decomposing the velocity as above, taking the dot product of each side of the equation and time averaging gives
 \begin{align}
 \braket{\boldsymbol{v} \cdot \boldsymbol{v}} &= \bar{v}^2 + \braket{\delta v^2} + 2\braket{\boldsymbol{\bar{v}} \cdot \boldsymbol{\delta v} } 
 \\v^2&=  \bar{v}^2 + \braket{\delta v^2} \; ,
 \end{align}
which is the same as Equation \eqref{turb_lam}, and is used to combine the turbulent and laminar components of the velocity, which are calculated separately.

 For the purposes of this problem, we are concerned with velocities relative to two frames: velocities relative to the total gas velocity are needed for the calculation of drag forces, while velocities relative to the local Keplerian velocity are needed to determine the rate that small bodies encounter large ones. If the subscripts $p,g,$ and $k$ denoted the velocity of the small bodies, the gas, and the Keplerian velocity respectively, then we may write
 
 \begin{align} \label{rel_v_unavg}
 \vp=\vrel+\vg \; .
 \end{align}
 Reynolds averaging \eqref{rel_v_unavg} gives
 \begin{align} \label{rel_v_avg_2}
 \avvp = \avvrel + \avvg \; .
 \end{align}
So the laminar component of the particle's velocity can be changed from one frame to another in the usual manner independent of the turbulent velocity, as is done in the main text (c.f. Equation \ref{rel_v_avg}).
 
 To compute how the turbulent velocity changes between frames, we require the equation of motion for the grains. Following \cite{volk}, for a particle in a linear drag regime ($F_D \propto v$) we can write
 \begin{align} \label{basic_EOM}
 \frac{d\vp}{dt}=\mathbf{a}_{g}-\frac{\vrel}{t_{s}} \; ,
 \end{align}
 where $\mathbf{a}_{g}$ is the acceleration due to forces other than gas drag, such as gravity. Reynolds averaging and subtracting the result from \eqref{basic_EOM} gives
 \begin{align} \label{fluc_EOM}
 \frac{d}{dt}\left(\boldsymbol{\delta v}_{pk}\right)=-\frac{\boldsymbol{\delta v}_{pg}}{t_{s}} \; ,
 \end{align}
 where we've assumed that $\boldsymbol{a}_g$ only varies on large spatial scales, so $\braket{\mathbf{a}_{g}} = 0$. Now subtracting equation \eqref{rel_v_avg_2} from Equation \eqref{rel_v_unavg} and rearranging slightly gives
 \begin{align}
 \boldsymbol{\delta v}_{gk} = \boldsymbol{\delta v}_{pk} - \boldsymbol{\delta v}_{pg}  \; .
 \end{align}
 Taking the dot product and time averaging gives
 \begin{align}
 \braket{\dvg^{2}}=\braket{\dvp^{2}}+\braket{\dvrel^{2}}-2\braket{\boldsymbol{\delta v}_{pk} \cdot\boldsymbol{\delta v}_{pg} } \; .
 \end{align}
 Plugging in Equation \eqref{fluc_EOM} into the last term on the righthand side gives
 \begin{align}
 \braket{\dvg^{2}}	&=\braket{\dvp^{2}}+\braket{\dvrel^{2}}+2t_{s}\braket{\boldsymbol{\delta v}_{pk}  \cdot\frac{d}{dt}\left( \boldsymbol{\delta v}_{pk}  \right)}
 \\&=\braket{\dvp^{2}}+\braket{\dvrel^{2}}+t_{s}\frac{d}{dt}\braket{\dvp^{2}} \; .
 \end{align}
For a stationary process the last term will be zero, so we have
 \begin{align}
 \braket{\dvg^{2}}=\braket{\dvp^{2}}+\braket{\dvrel^{2}} \; ,
 \end{align}
 which is used in our model to convert the turbulent component of the velocity from the frame relative to the gas to the frame relative to Keplerian (c.f. Equation \ref{turb_frames}).
 
 \section{Canonical Core Accretion Timescale} \label{GF_sec}
 In this appendix we give a short summary of how growth proceeds for cores accreting particles for which the gas drag force is negligible. See \cite{gold} for a more in depth review of these processes. Since small bodies in this regime cannot disspate their kinetic energy through gas drag, in order for the accretion to occur the impact parameter of the incoming particle must be small enough that it collides with the core. The maximum impact parameter at which a particle will be gravitationally focused into a collision with the core is given by
 
 \begin{align} \label{eq:r_focus}
 R_{\rm{focus}}  = R \left( 1  + \frac{v_{esc}^2}{v_{\infty}^2} \right )^{1/2} \; ,
 \end{align}
 where $R$ is the radius of the core, and $v_{esc}=\sqrt{2 G M/R}$ is the escape velocity from the core. The scale height of small bodies is given by the vertical component of their velocity dispersion -- $H_p \sim v_z/\Omega$. 
 
 While there are a number of gas free growth regimes, and therefore timescales, we confine our attention to the regime where the velocity of dispersion of the small bodies is approximately the hill velocity, $v_H = R_H \Omega$. Here $R_H$ is the core's Hill radius (see Section \ref{len_scales}), and $\Omega$ is the local Keplerian orbital frequency. This regime gives the highest possible growth rate without invoking some external mechanism to bring the velocity dispersion below the Hill velocity, since interactions with the core will drive bodies up to the Hill velocity. In this regime the core can accrete over the entirety of $R_{\rm{focus}}$ in the vertical direction, so $\sigma \approx 4 R_{\rm{focus}}^2$. If we set $v_\infty \approx v_z \approx v_H$, and note that $v_{esc}/v_H \sim (R_H/R)^{1/2}$, then we see that $t_{GF}$ is of order
 
 \begin{align} \label{eqn:t_GF}
 t_{GF,v=v_H} \sim \frac{M}{2 f_s \Sigma  R_{H}^2 \Omega} \left( \frac{R_H}{R} \right) \; ,
 \end{align}
 Scaled to our fiducial values (see Section \ref{params}), the timescale is given by
 
 \begin{align} \label{eqn:t_GF_fid}
 t_{GF} \approx 7 \times 10^5  \left( \frac{a}{\text{AU}} \right)^{3/2} \left( \frac{M}{\,M_\oplus} \right)^{1/3} \text{years} \; .
 \end{align}
 
 If there are $\sim \, \text{km}$ sized objects available, then cores may grow by gravitational focusing in addition to gas-assisted growth. Furthermore, if for a given set of parameters, the gravitational focusing timescale is shorter than the gas-assisted growth timescale, then gravitational focusing will be the dominant mechanism of growth, which can cause cores to still grow in regimes where gas-assisted growth is slow.
 
 \section{Details of Turbulent Velocity Calculation}
In this appendix we describe some of the more minor details of the calculation of the velocity of small bodies due to turbulence. 

 \begin{figure} [htbp]
 	\centering
 	\includegraphics[width=6in]{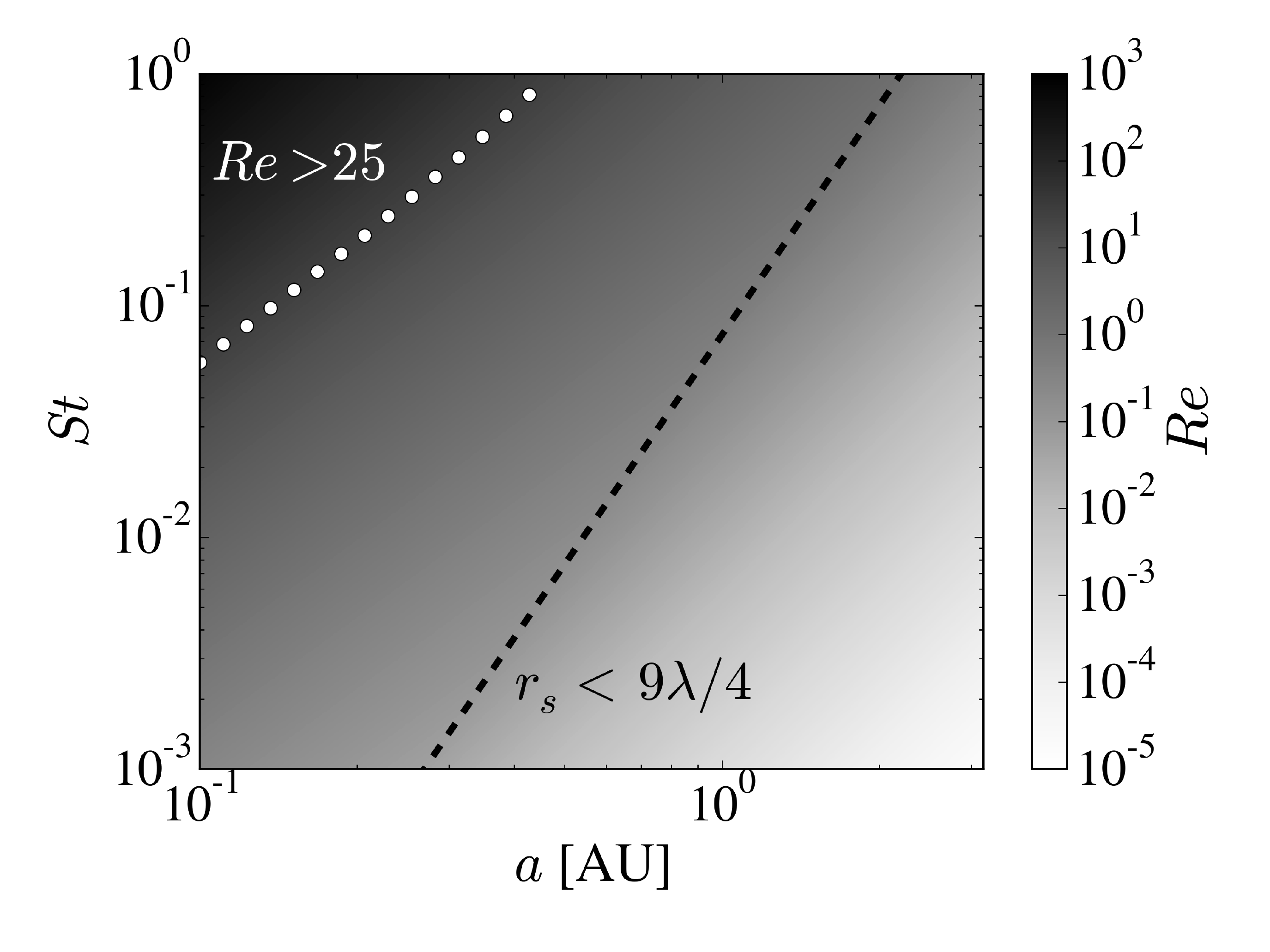}
 	\caption{The Reynolds number of a small body as a function of Stokes number and semi-major axis, for the case $\alpha = 10^{-1}$ . The region where $Re>25$ and the assumptions made in our model begin to be violated is indicated, as is the region where $r_s < 9 \lambda/4$ and the particle enters the diffuse regime. For lower values of $\alpha$ the region where $Re>25$ shrinks rapidly. }
 	\label{fig:st_check}
 \end{figure}

The results of \cite{volk} and all work derived from these results, in particular Equation \eqref{RMSturb}, rest on the assumption that there is a well-defined stopping time for the particle, independent of the particle's velocity. Thus Equation \eqref{RMSturb} may not hold when the particle enters the Ram pressure gas drag regime. However, Equation \eqref{RMSsimple} holds whenever the particle's stopping time is large enough that it receives many ``kicks" from the largest scale eddies before a stopping time has elapsed. In other words, Equation \eqref{RMSsimple} holds when $St \gg 1$. Since for $Re \gg 1$ the particle will be in the Ram pressure regime, which is quadratic in velocity, we need only be concerned about the validity of our approximation if we have $Re\gg1$ before $St\gg1$.  Figure \ref{fig:st_check} shows a plot of the Reynolds number of particles as a function of semi-major axis and Stokes number. We have restricted the figure to show $St<1$, since for larger values of $St$ we expect Equation \eqref{RMSsimple} to hold to reasonable accuracy. In the plot we indicate the region where $Re>25$, which we take as the approximate region where accuracy of our model is in question.  The plot is for $\alpha = 10^{-1}$, which is the most restrictive case; for lower values of $\alpha$ the region where $St<1$ and $Re>25$ shrinks.

The laminar velocity effects the turbulent velocity of the particle as well. This effect can be qualitatively understood by considering the fact that a laminar component to the particle's velocity decreases the amount of time that the particle interacts with a turblent eddy of a given wave number $k$. The original \cite{volk} result is dependent on the value of what they call $k^*$, which is the divide between eddies which are large enough that the particle comes into equilibrium with the eddy, and eddies which either decay or are traversed by the particle before they have a substantial frictional effect. To order of magnitude, the relative velocity between the particle and the gas is simply equal to the velocity of the eddy that the particle is marginally coupled to, i.e. the velocity of the eddy with wavenumber $k_*$. Since the presence of a laminar component to the velocity will affect which eddies the small bodies can drift through over a stopping time, introducing a laminar velocity will change the value of $k^*$, which in turn has important effects on the RMS turbulent velocity as a function of particle size. OC07 therefore refer to the effects of a laminar component of velocity as ``eddy-crossing effects."
 
OC07 neglect the effect of a laminar component on the particle velocity, noting that it is only important in the weakly turbulent regime for small Stokes numbers. As we are interested here in varying the strength of turbulence and determining its effect of planetary growth rates, the weakly turbulent regime is of interest to us. 
 
\cite{orb} note that, in the regime where eddy crossing effects are non-negligible, we can approximate these effects by using an effective large eddy turnover time, $t_{\text{eddy}}'$, given by
\begin{align}
 t_{\text{eddy}}'&=\frac{t_L}{\sqrt{1+(v_{pg,\ell}/v_t)^2}} \label{eddy_cross_eqn} \; ,
 \end{align}
 where $v_{pg,\ell}$ is the laminar component of the particle's velocity, measured relative to the RMS gas velocity.
 
In conclusion, we calculate the turbulent component of the small body's velocity by combining equations \eqref{RMSsimple} and \eqref{RMSturb}, with the large eddy turnover time $t_{\rm{eddy}}$ modified by equation \eqref{eddy_cross_eqn}.

\bibliographystyle{yahapj}

\end{document}